\documentclass[twocolumn,prd,superscriptaddress]{revtex4-1}

\usepackage{color}
\usepackage{amsmath}
\usepackage{xspace}
\usepackage{graphicx}
\usepackage{hyperref}

\newcommand\mysec[1]{\section{#1}}

\def\prd{Phys.~Rev.~D}       

\newcommand{\beq}{\begin{equation}}
\newcommand{\eeq}{\end{equation}}

\newcommand{\be}{\begin{equation}}
\newcommand{\ee}{\end{equation}}
\newcommand{\bea}{\begin{eqnarray}}
\newcommand{\eea}{\end{eqnarray}}
\newcommand{\bes}{\begin{subequations}}
\newcommand{\ees}{\end{subequations}}

\newcommand{\Caltech}{\affiliation{Theoretical Astrophysics 350-17, California Institute of Technology, Pasadena, CA 91125, USA}}
\newcommand{\CITA}{\affiliation{Canadian Institute for Theoretical Astrophysics, 60 St.~George Street, University of Toronto, Toronto, ON M5S 3H8, Canada}}
\newcommand{\AEI}{\affiliation{Max Planck Institute for Gravitational Physics (Albert Einstein Institute), Am M{\"u}hlenberg~1, 14476 Potsdam-Golm, Germany}}

\begin{document}

\title{Targeted numerical simulations of binary black holes for GW170104}
\author{J.~Healy}
\author{J.~Lange}
\author{R.~O'Shaughnessy}
\author{C.O.~Lousto}
\author{M.~Campanelli}
\author{A. R.~Williamson}
\author{Y.~Zlochower}
\affiliation{Center for Computational Relativity and Gravitation,
School of Mathematical Sciences,
Rochester Institute of Technology, 85 Lomb Memorial Drive, Rochester,
New York 14623}
\author{J.~Calder\'on Bustillo}
\author{J.A.~Clark}
\author{C.~Evans}
\author{D.~Ferguson}
\author{S.~Ghonge}
\author{K.~Jani}
\author{B ~Khamesra}
\author{P.~Laguna}
\author{D. M.~Shoemaker}
\affiliation{Center for Relativistic Astrophysics and School of Physics,
Georgia Institute of Technology, Atlanta, GA 30332, USA}
\author{M.~Boyle}
\affiliation{Center for Astrophysics and Planetary Science, Cornell University, Ithaca, New York 14853, USA}
\author{A.~Garc\'{i}a}
\affiliation{Gravitational Wave Physics and Astronomy Center, California State University Fullerton, Fullerton, California 92834, USA}
\author{D. A.~Hemberger}
\Caltech
\author{L. E.~Kidder}
\affiliation{Center for Astrophysics and Planetary Science, Cornell University, Ithaca, New York 14853, USA}
\author{P.~Kumar}
\affiliation{Center for Astrophysics and Planetary Science, Cornell University, Ithaca, New York 14853, USA}
\CITA
\author{G.~Lovelace}
\affiliation{Gravitational Wave Physics and Astronomy Center, California State University Fullerton, Fullerton, California 92834, USA}
\author{H. P.~Pfeiffer}\AEI\CITA
\author{M. A.~Scheel}
\Caltech
\author{S.A.~Teukolsky}
\affiliation{Center for Astrophysics and Planetary Science, Cornell University, Ithaca, New York 14853, USA}
\Caltech

\begin{abstract}
In response to LIGO's observation of GW170104, we performed a series of full numerical simulations of binary black
holes, each designed to replicate likely realizations of its dynamics and radiation.   These simulations have been
performed at multiple resolutions and with two independent techniques to solve Einstein's equations.   For the nonprecessing and precessing
simulations, we demonstrate the two techniques agree mode by mode, at a precision substantially in excess of statistical uncertainties in current
 LIGO's observations.  Conversely, we demonstrate our full numerical solutions contain information which is not
accurately captured with the approximate phenomenological models commonly used to infer compact binary
parameters.  
To quantify the impact of these differences on parameter inference for GW170104 specifically, we compare the predictions
of our simulations and these approximate models to LIGO's observations of GW170104.  


\end{abstract}
\maketitle

\section{Introduction}
The LIGO-Virgo Collaboration (LVC) has reported the confident discovery of three binary black hole (BBH) mergers via
gravitational wave (GW) radiation: GW150914\cite{LIGOVirgo2016a} and GW151226\cite{Abbott:2016nmj} from the first observing run O1\cite{TheLIGOScientific:2016pea}, and GW170104 \cite{2017PhRvL.118v1101A}, GW170608 \cite{Abbott:2017gyy}, 
and GW170814 \cite{Abbott:2017oio} from
the second observing run.  The
parameters of these detections were inferred by comparing the data to state-of-the-art approximate models
\cite{2014PhRvD..89f1502T,Purrer:2014, Hannam:2013oca}.
A reanalysis of GW150914 implementing full
numerical relativity (NR)\cite{NRPaper} simulations helped to better constrain the mass ratio of the system.
This is due to the fact that NR waveforms include physics omitted by current approximate models, notably 
higher order modes and accurate precession effects.
A full description of this methodology, including detailed tests of systematic
errors and parameter estimation improvements, can be found in \citet{Lange2017}.

This paper is organized as follows.
In Section \ref{sec:FN}, we describe the two independent techniques we use to solve Einstein's equations numerically for
the evolution of binary black hole spacetimes.
In Section \ref{sec:TargetedFollowupAndComparisons}, we describe the binary's parameters selected for detailed
followup, our simulations of these proposed initial conditions, and detailed  comparisons between 
our paired results, for both nonprecessing and precessing simulations.  We also contrast our simulations' radiation with the
corresponding results derived from the approximate phenomenological models used by LIGO for parameter inference. 
In Section \ref{sec:NR+GW170104}, we directly compare our simulations to GW170104.  These comparisons provide both a
scalar measure of how well each simulation agrees with the data (a marginalized likelihood), as well as the best-fitting
reconstructed waveform in each instrument \cite{NRPaper,Lange2017}.  
Using our reconstructed waveforms, we graphically
demonstrate that our simulations agree with each other and the data, with simulation differences far smaller than the
residual noise in each instrument.   
Using these real observations as a benchmark for model quality, we then quantify how effectively our simulations
reproduce the data, compared to the results of approximate and phenomenological models at the same parameters.  
Since our simulations parameters were selected using these approximate and phenomenological models, we also have the opportunity to
assess how effectively they identified the optimal binary  parameters.   
In Section \ref{sec:Conclusions} we discuss the prospects for future targeted simulations in followup of LIGO
observations. 

\section{Full Numerical Evolutions}\label{sec:FN}

The breakthroughs~\cite{Pretorius:2005gq,Campanelli:2005dd,Baker:2005vv}
in numerical relativity allowed for detailed predictions for the
gravitational waves from the late inspiral, plunge, merger and
ringdown of black hole binary systems. Catalogs of the simulated waveforms
are publicly available \cite{Mroue:2013xna, Jani:2016wkt, Healy:2017psd}
for its use for BBH parameter estimation \cite{Lange:2017wki},
as well as for determining how the 
individual masses and spins of the orbiting binary relate to the
properties of the final remnant
black hole produced after merger. This relationship~\cite{Healy:2014yta}
can be used as a consistency check for the observations of the inspiral
and, independently, the merger-ringdown signals as tests of general relativity
\cite{Ghosh:2016qgn,TheLIGOScientific:2016src,TheLIGOScientific:2016pea}.

\subsection{Simulations using finite-difference, moving-puncture methods}\label{sec:RIT}

In order to make systematic studies and build a data bank of
full numerical simulations, e.g.,~\cite{Healy:2017psd},
it is crucial to develop efficient
numerical algorithms, since large computational resources are required.
The Rochester Institute of Technology (RIT) group evolved the  BBH data sets described below using the {\sc
LazEv}~\cite{Zlochower:2005bj} implementation of the moving puncture
approach~\cite{Campanelli:2005dd,Baker:2005vv} with the conformal
function $W=\sqrt{\chi}=\exp(-2\phi)$ suggested by
Ref.~\cite{Marronetti:2007wz}.  For those runs, they used
centered, sixth-order finite differencing in
space~\cite{Lousto:2007rj} and a fourth-order Runge Kutta time
integrator (the code does not upwind the advection terms)
and a 5th-order Kreiss-Oliger dissipation operator.

The {\sc LazEv} code uses 
the {\sc EinsteinToolkit}~\cite{Loffler:2011ay,
einsteintoolkit} / {\sc Cactus}~\cite{cactus_web} /
{\sc Carpet}~\cite{Schnetter-etal-03b}
infrastructure.  The {\sc
Carpet} mesh refinement driver provides a
``moving boxes'' style of mesh refinement. In this approach, refined
grids of fixed size are arranged about the coordinate centers of both
holes.  The {\sc Carpet} code then moves these fine grids about the
computational domain by following the trajectories of the two BHs.

To compute the initial low eccentricity orbital parameters
RIT used the post-Newtonian techniques described
in~\cite{Healy:2017zqj} and then generated the initial data based on
these parameters using
approach~\cite{Brandt97b} along with the
{\sc TwoPunctures}~\cite{Ansorg:2004ds} code implementation. 

The {\sc LazEv} code uses {\sc AHFinderDirect}~\cite{Thornburg2003:AH-finding} to locate
apparent horizons, and measures the magnitude of the horizon spin using
the {\it isolated horizon} (IH) algorithm detailed in
Ref.~\cite{Dreyer02a} and as implemented in Ref.~\cite{Campanelli:2006fy}.
The horizon
mass  is calculated via the Christodoulou formula 
${m_H} = \sqrt{m_{\rm irr}^2 + S_H^2/(4 m_{\rm irr}^2)}\,,$
where $m_{\rm irr} = \sqrt{A/(16 \pi)}$, $A$ is the surface area of
the horizon, and $S_H$ is the spin angular momentum of the BH (in
units of $M^2$). 

The radiated energy,
linear momentum, and angular momentum, were measured in terms of the radiative Weyl
Scalar $\psi_4$, using the formulas provided in
Refs.~\cite{Campanelli:1998jv,Lousto:2007mh},
Eqs. (22)-(24) and (27) respectively. However, rather than
using the full $\psi_4$, it was decomposed it into $\ell$ and $m$ modes and
dropping terms with $\ell >
6$.  The formulas in Refs.~\cite{Campanelli:1998jv,Lousto:2007mh} are
valid at $r=\infty$.  To obtain the waveform and radiation quantities
at infinity, the perturbative extrapolation described in
Ref.~\cite{Nakano:2015pta} was used.

For the RIT simulations, different resolutions are denoted by NXXX where XXX is either 100, 118, or 140 for low, medium, and high resolutions, respectively.  This number is directly related to the wavezone resolution
 in the simulation.  For instance, N100 has a resolution of $M/1.0$ in the wavezone (where observer extraction takes place, preliminary to
perturbative extrapolation to infinity via \cite{Nakano:2015pta}),
and N140 has $
M/1.4$.  In each case UID\#1-5, there are 10 levels of refinement in all
and the grids followed a pattern close to those described in \cite{Healy:2017mvh}.

Other groups using the moving punctures \cite{Campanelli:2005dd,Baker:2005vv}
formalism with finite difference methods are Georgia Institute of Technology (GT) \cite{Jani:2016wkt}
and those based on BAM \cite{Bruegmann2006}.
The GT \cite{Jani:2016wkt} simulations were obtained with the \textsc{Maya} code \cite{Herrmann:2007ex,Herrmann:2007ac,Hinder:2007qu,Healy:2008js,Hinder:2008kv,Healy:2009zm,Healy:2009ir,Bode:2009mt}, which is also based on the BSSN formulation
with moving punctures. 
The grid structure for each run consisted of  10 levels of refinement 
provided by \textsc{Carpet} \cite{Schnetter-etal-03b}, a 
mesh refinement package for \textsc{Cactus} \cite{cactus_web}. 
Each successive level's resolution decreased by a factor of 2.
Sixth-order spatial finite differencing was used with the BSSN equations 
implemented with Kranc \cite{Husa:2004ip}. 

\subsection{Simulations using pseudospectral, excision methods\label{sec:SXS}}


Simulations labeled SXS are carried out using the Spectral Einstein
Code (SpEC)~\cite{SpECwebsite} used by the Simulating eXtreme Spacetimes Collaboration (SXS).  Given initial BBH parameters, a
corresponding weighted superposition of two boosted, spinning
Kerr-Schild black holes~\cite{Lovelace2008} is constructed, and then
the constraints are solved~\cite{York1999,Pfeiffer2003,Ossokine:2015yla} by a
pseudospectral method to yield
quasi-equilibrium~\cite{Caudill-etal:2006,Lovelace2008} initial data.
Small adjustments in the initial orbital trajectory are made
iteratively to produce initial data with low
eccentricity~\cite{Pfeiffer-Brown-etal:2007,Buonanno:2010yk,Mroue:2012kv}.

The initial data are evolved using a first-order
representation~\cite{Lindblom2006} of a generalized harmonic
formulation~\cite{Friedrich1985, Garfinkle2002, Pretorius2005c} of
Einstein's equations, and using damped harmonic
gauge~\cite{Lindblom2009c,Choptuik:2009ww,Szilagyi:2009qz}.  The
equations are solved pseudospectrally on an
adaptively-refined~\cite{Lovelace:2010ne,Szilagyi:2014fna} spatial grid that
extends from pure-outflow excision boundaries just inside apparent
horizons~\cite{Scheel2009,Szilagyi:2009qz,Hemberger:2012jz,Ossokine:2013zga,Scheel2014}
to an artificial outer boundary.  
Adaptive time-stepping automatically achieves
 time steps of approximately the Courant limit.

On the Cal State Fullerton cluster, ORCA, the simulation
achieved a typical evolution speed of $O(100 M)/{\rm day}$ for the highest
resolution (here we measure simulation
time in units of $M$, the total mass of the
binary).
After the holes merge, all variables
are automatically interpolated onto a new grid with a single excision
boundary inside the common apparent
horizon~\cite{Scheel2009,Hemberger:2012jz}, and the evolution is
continued.  Constraint-preserving boundary
conditions~\cite{Lindblom2006, Rinne2006, Rinne2007} are imposed on
the outer boundary, and no boundary conditions are required or imposed
on the excision boundaries.

We use a pseudospectral fast-flow algorithm~\cite{Gundlach1998} 
to find apparent horizons, and we compute spins on these apparent horizons
using the approximate Killing vector formalism of Cook, Whiting, and
Owen~\cite{Cook2007, OwenThesis}.

Gravitational wave extraction is done by three independent methods:
direct extraction of the Newman-Penrose quantity $\Psi_4$ at finite
radius~\cite{Scheel2009,Boyle2007,Pfeiffer-Brown-etal:2007},
extraction of the strain $h$ by matching to solutions of the
Regge-Wheeler-Zerilli-Moncrief equations at finite
radius~\cite{Buchman2007,Rinne2008b}, and Cauchy-Characteristic
Extraction~\cite{Bishop:1997ik,Winicour2005,Gomez:2007cj,Reisswig:2012ka,Handmer:2014}.
The latter method directly provides gravitational waveforms at future
null infinity, while for the former two methods the waveforms are
computed at a series of finite radii and then extrapolated to
infinity~\cite{Boyle-Mroue:2008}.  Differences between the different
methods, and differences in extrapolation algorithms, can be used to
as error estimates on waveform
extraction. These waveform extraction errors are important for
the overall error budget of the simulations, and are typically on the
order of, or slightly larger than, the numerical truncation
error~\cite{Taylor:2013zia,Chu:2015kft}. In this paper, the waveforms
we compare use Regge-Wheeler-Zerilli-Moncrief extraction and
extrapolation to infinity. We have verified that our choice of extrapolation
order does not significantly affect our results. We have also checked
that corrections to the wave modes~\cite{Boyle2015a} to account for
a small drift in the coordinates of the center of mass have a negligible
effect on our results.

\section{Simulations of GW170104}
\label{sec:TargetedFollowupAndComparisons}

We  extracted the maximum a posteriori (MaP) parameters from (preliminary) Bayesian posterior inferences performed by the
LIGO Scientific Collaboration and the Virgo Collaboration, using different waveform models \cite{Abbott:2017vtc,Veitch:2015}.  As described in Appendix
\ref{ap:DifferentTargets}, this point parameter estimate is one of a few well-motivated and somewhat different choices for followup parameters;
however, as described in that appendix, we estimate that the specific choice we adopt will not significantly change our
principal results.   Table \ref{tab:NRfollow} shows parameters  simulated with numerical relativity.

\begin{table*}
\caption{Follow-up Parameter Table
}\label{tab:NRfollow}
\begin{ruledtabular}
\begin{tabular}{l|cccccc}
Run              & $M_{total}/M_{\odot}$ & $f_{ref}$ [Hz] & $q=m_1/m_2$   & $\chi_1$ & $\chi_2$ &Approximant\\
\hline
\#1 & 58.49 & 24 & 0.8514 & ( 0, 0, 0.7343)              & ( 0, 0, -0.8278) & SEOBNRv4ROM \\
\#2 & 58.72 & 24 & 0.5246 & ( 0.1607, -0.1023, -0.0529 ) & ( -0.3623, 0.5679, -0.3474 ) & SEOBNRv3 \\ 
\#3 (*) & 62.13 & 20 & 0.4850 & ( 0.0835, -0.4013, -0.3036 ) & ( -0.3813, 0.7479, -0.1021 ) & IMRPhenomPv2 \\
\#4 & 53.46 & 20 & 0.7147 & ( 0, 0, 0.2205)              & ( 0, 0, -0.7110) & SEOBNRv4ROM \\
\#5 & 59.11 & 20 & 0.4300 & ( 0, 0, -0.3634)             & ( 0, 0, -0.1256) & IMRPhenomD \\
\end{tabular}
\end{ruledtabular}
\end{table*}

Spin Conventions: $(\chi_{1}^x, \chi_{1}^y, \chi_{1}^z)$ are specified in a frame where (i) $\hat{L}=(0,0,1)$, i.e. the Newtonian orbital angular momentum is along the z-axis. (ii) the vector $\hat{n}$ pointing from $m_2$ to $m_1$ is the x-axis, (1,0,0). Note that the orientation of $\hat{n}$ is essentially undetermined by parameter estimation (PE) methods, so the choice (ii) is meant to break this degeneracy to arrive at concrete parameters. In other words, the spin-components given below are those consistent with Eqs. (9)-(11) of
\footnote{https://dcc.ligo.org/DocDB/0123/T1500606/006/\\
  NRInjectionInfrastructure.pdf}

The label, UID\#1-5, of the simulation identifies which parameters we are using in following up as
given by the initial data from Table \ref{tab:NRfollow}.   
For aligned spin runs, where spin-vectors are preserved, the initial orbital frequency
may be smaller than $f_{ref}/2$. For precessing runs, because we target a certain spin configuration at $f_{ref}$, this forces the initial NR-frequency to be identical (or only somewhat smaller than) the reference frequency.
$D/M$ is the initial orbital separation of the NR run in geometric units and the mass ratio and
intrinsic spins $(\chi_i=S_i^2/m_i^2)$ are denoted by $(q, \chi_1^x, \chi_1^y,...,)$.
Due to the way NR simulations are set-up the initial parameters can change due to the presence of junk radiation and/or imperfections in setting up initial data, therefore
these quantities should be reported as after-junk masses/spins, ideally extracted at the reference frequency.
For precessing runs, in particular, the spin-components should be specified at the reference frequency, following the convention $\chi_i^x=\vec{\chi_i}\cdot\hat n, \chi_i^z=\vec{\chi_i}\cdot\hat{L}$, with $\hat n$ and $\hat L$ computed at the reference frequency, too. We also provide $e$, the orbital eccentricity.
For instance, the actual initial data as measured for the RIT's followup simulations are described in Table \ref{tab:ID}

\begin{table*}
\caption{Initial data parameters for the quasi-circular
configurations with a smaller mass black hole (labeled 1),
and a larger mass spinning black hole (labeled 2). The punctures are located
at $\vec r_1 = (x_1,0,0)$ and $\vec r_2 = (x_2,0,0)$, with momenta
$P=\pm (P_r, P_t,0)$, mass parameters
$m^p/M$, horizon (Christodoulou) masses $m^H/M$, total ADM mass
$M_{\rm ADM}$, dimensionless spins $a/m_H = S/m_H^2$, and eccentricity, $e$.
}\label{tab:ID}
\begin{ruledtabular}
\begin{tabular}{l|cccccccccccc}
Run              & $x_1/M$ & $x_2/M$ & $P_r/M$   & $P_t/M$ & $m^p_1/M$ & $m^p_2/M$ & $m^H_1/M$ & $m^H_2/M$ & $M_{\rm ADM}/M$ & $a_1/m_1^H$ & $a_2/m_2^H$ & $e$ \\
\hline
\#1 & -7.9168 & 6.7407 & -2.829e-4 & 0.07467 & 0.3196 & 0.3056 & 0.4599 & 0.5401 & 0.9928 &  0.7343 & -0.8267 & 6e-4 \\
\#2 & -7.8211 & 4.1029 & -4.837e-4 & 0.07837 & 0.3277 & 0.4400 & 0.3441 & 0.6559 & 0.9922 &  0.1977 &  0.7580& 1e-3 \\
\#3 & -8.7720 & 4.2543 & -3.316e-4 & 0.07160 & 0.2796 & 0.3584 & 0.3266 & 0.6734 & 0.9930 &  0.5101 &  0.8445& 1e-3 \\
\#4 & -8.4742 & 6.0567 & -2.918e-4 & 0.07421 & 0.3991 & 0.4219 & 0.4168 & 0.5832 & 0.9928 &  0.2205 & -0.7110 & 2e-4 \\
\#5 & -9.4395 & 4.0593 & -2.718e-4 & 0.06698 & 0.2753 & 0.6865 & 0.3007 & 0.6993 & 0.9933 & -0.3634 & -0.1256 & 5e-4\\
\end{tabular}
\end{ruledtabular}
\end{table*}

\begin{table*}
\begin{ruledtabular}
\begin{tabular}{rcccccc}
$\#$  & $m^r_1/M$ & $m^r_2/M$ & $q^r$ & $\chi^r_1$ & $\chi^r_2$ & $mf^r$\\
\hline
UID1-N100 & 0.459757 & 0.539780 & 0.851718 & (0,0,0.737493) & (0,0,-0.829064) & 0.002586\\
UID1-N118 & 0.459758 & 0.539801 & 0.851718 & (0,0,0.737473) & (0,0,-0.829040) & 0.002587\\
UID1-N140 & 0.459758 & 0.539801 & 0.851718 & (0,0,0.737464) & (0,0,-0.829030) & 0.002587\\
UID1-L2   & 0.459913 & 0.540183 & 0.851401 & (0,0,0.734254) & (0,0,-0.827442) & 0.003135 \\
UID1-L3   & 0.459902 & 0.540171 & 0.851400 & (0,0,0.734138) & (0,0,-0.827657) & 0.003139 \\
UID1-L4   & 0.459907 & 0.540176 & 0.851402 & (0,0,0.734106) & (0,0,-0.827690) & 0.003139 \\
\hline
UID2-N100 & 0.344090 & 0.655693 & 0.524773 & ( 0.119494, 0.144676, -0.063539 ) & ( -0.675724, 0.181910, -0.293427 ) & 0.004010\\
UID2-N118 & 0.344091 & 0.655693 & 0.524774 & ( 0.118992, 0.144981, -0.063537 ) & ( -0.676089, 0.180798, -0.293141 ) & 0.004010\\
UID2-N140 & 0.344091 & 0.655694 & 0.524774 & ( 0.118754, 0.145124, -0.063541 ) & ( -0.676262, 0.180273, -0.293005 ) & 0.004010\\
UID2-L1   & 0.344082 & 0.655876 & 0.524614 & ( 0.101171, 0.156396, -0.066135 ) & ( -0.693687, 0.096203, -0.290389 ) & 0.004001 \\
UID2-L2   & 0.344069 & 0.655966 & 0.524523 & ( 0.100390, 0.156863, -0.066508 ) & ( -0.693803, 0.095200, -0.289578 ) & 0.004013 \\
UID2-L3   & 0.344072 & 0.655958 & 0.524534 & ( 0.100596, 0.156736, -0.066591 ) & ( -0.693727, 0.095442, -0.289439 ) & 0.004010 \\
\hline
UID3-N100 & 0.326605 & 0.672963 & 0.485325 & ( 0.360515, -0.132028, -0.338700 ) & ( -0.685165, 0.491281, -0.068777 ) & 0.003308\\
UID3-N118 & 0.326606 & 0.672963 & 0.485326 & ( 0.360458, -0.131982, -0.338637 ) & ( -0.685154, 0.491237, -0.068765 ) & 0.003308\\
UID3-N140 & 0.326607 & 0.672964 & 0.485326 & ( 0.360429, -0.132025, -0.338591 ) & ( -0.685106, 0.491276, -0.068763 ) & 0.003308\\
UID3-L1   & 0.326576 & 0.673382 & 0.484978 & ( 0.358221, -0.095241, -0.350438 ) & ( -0.749729, 0.387741, -0.052141 ) & 0.003330 \\
UID3-L2   & 0.326598 & 0.673451 & 0.484962 & ( 0.358434, -0.094398, -0.350478 ) & ( -0.750204, 0.386567, -0.051977 ) & 0.003339 \\
UID3-L3   & 0.326583 & 0.673462 & 0.484931 & ( 0.352293, -0.114550, -0.350800 ) & ( -0.727233, 0.427282, -0.059549 ) & 0.003308 \\
\hline
UID4-N100 & 0.416817 & 0.583036 & 0.714908 & (0,0,0.221608) & (0,0,-0.712189) & 0.002629\\
UID4-N118 & 0.416819 & 0.583037 & 0.714911 & (0,0,0.221602) & (0,0,-0.712184) & 0.002630\\
UID4-N140 & 0.416820 & 0.583037 & 0.714912 & (0,0,0.221600) & (0,0,-0.712182) & 0.002630\\
UID4-L1   & 0.416790 & 0.583195 & 0.714666 & (0,0,0.220660) & (0,0,-0.710930) & 0.002635 \\
UID4-L2   & 0.416809 & 0.583214 & 0.714675 & (0,0,0.220453) & (0,0,-0.710967) & 0.002635 \\
UID4-L3   & 0.416817 & 0.583203 & 0.714703 & (0,0,0.220427) & (0,0,-0.710930) & 0.002619 \\
\hline
UID5-N100 & 0.300721 & 0.699282 & 0.430043 & (0,0,-0.366067) & (0,0,-0.125360) & 0.002932 \\
UID5-N118 & 0.300722 & 0.699284 & 0.430043 & (0,0,-0.366031) & (0,0,-0.125352) & 0.002932 \\
UID5-N140 & 0.300723 & 0.699285 & 0.430043 & (0,0,-0.366015) & (0,0,-0.125348) & 0.002932 \\
UID5-L1   & 0.300697 & 0.699261 & 0.430021 & (0,0,-0.363371) & (0,0,-0.125641) & 0.002900 \\
UID5-L2   & 0.300708 & 0.699223 & 0.430061 & (0,0,-0.363424) & (0,0,-0.125550) & 0.002864 \\
UID5-L3   & 0.300722 & 0.699178 & 0.430107 & (0,0,-0.363345) & (0,0,-0.125616) & 0.002855 \\
\end{tabular}
\end{ruledtabular}
\caption{
Values of the individual masses, mass ratio, dimensionless spins, 
and frequency, given at a time after the gauge settles.  For the nonprecessing cases (1, 4, and 5), this time is $t^r/M=200$ for RIT and $t^r/M = 640 $ for SXS.  For the precessing systems (2 and 3), the values are given such that the relaxed frequency $mf^r$ is the same between RIT and SXS.
}
\label{tab:relaxed}
\end{table*}

For followup \#1, the initial spurious burst contains a non-negligible kick which
causes a center of mass drift of approximately 0.65M over the 4000M of evolution.
Because of this, information from the dominant $\ell=2,
m=\pm2$ modes leak into the other modes, particularly the $m=$odd modes.
To reduce this effect, we can recalculate the modes by finding the
{\it average} rest frame of the binary. We calculate the
average velocity of the center of mass of the binary (from $\psi_4$)
over the inspiral and then boost the waveform in the opposite
direction. This is done using  Eqs. (4-5) in \cite{Boyle:2015nqa} and
Eqs. (7-8) in \cite{Kelly:2012nd}.
Note that this does not change the physical waveforms, only how they
are spread over modes.

\begin{figure}
  \includegraphics[angle=270,width=\columnwidth]{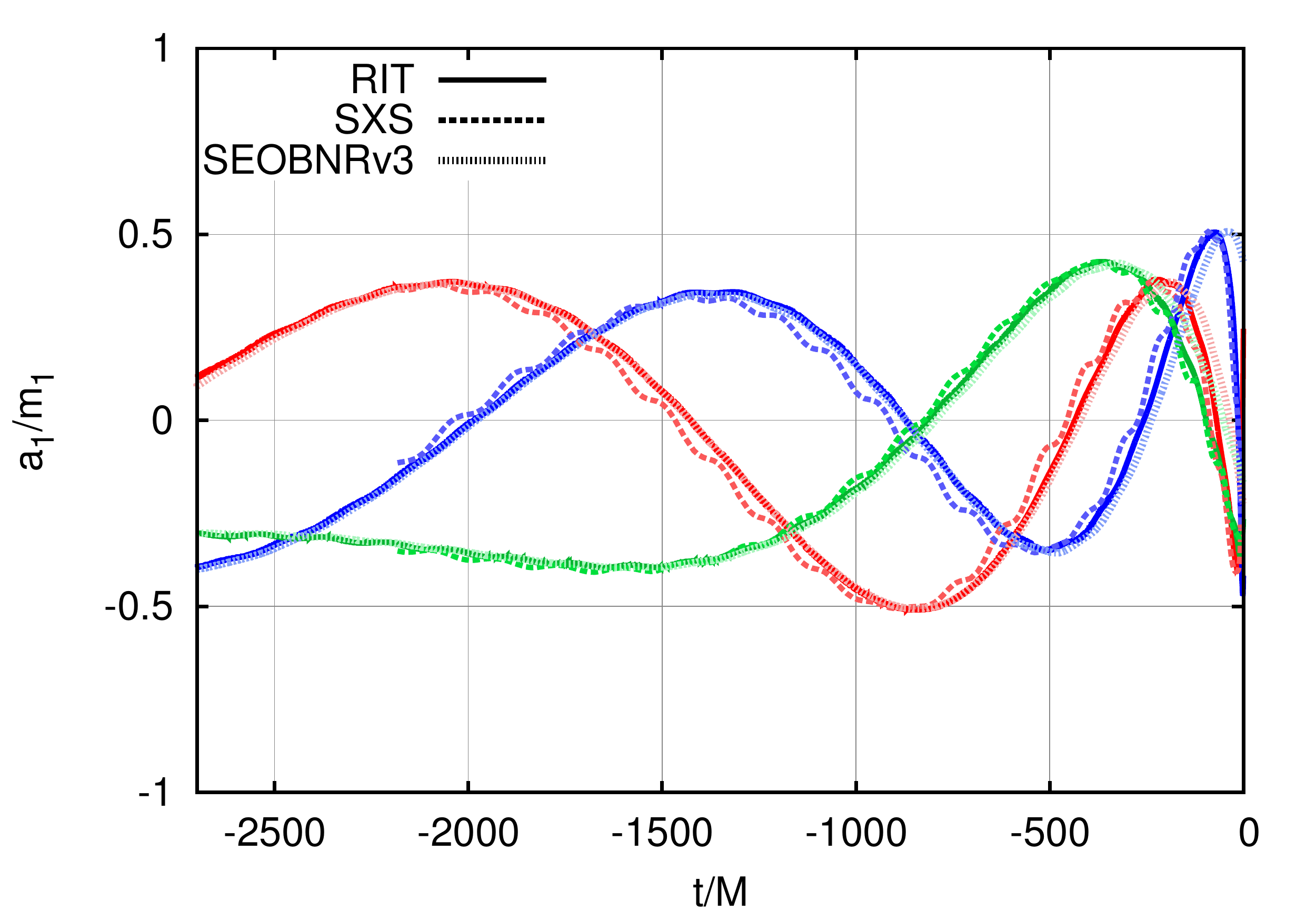}
  \includegraphics[angle=270,width=\columnwidth]{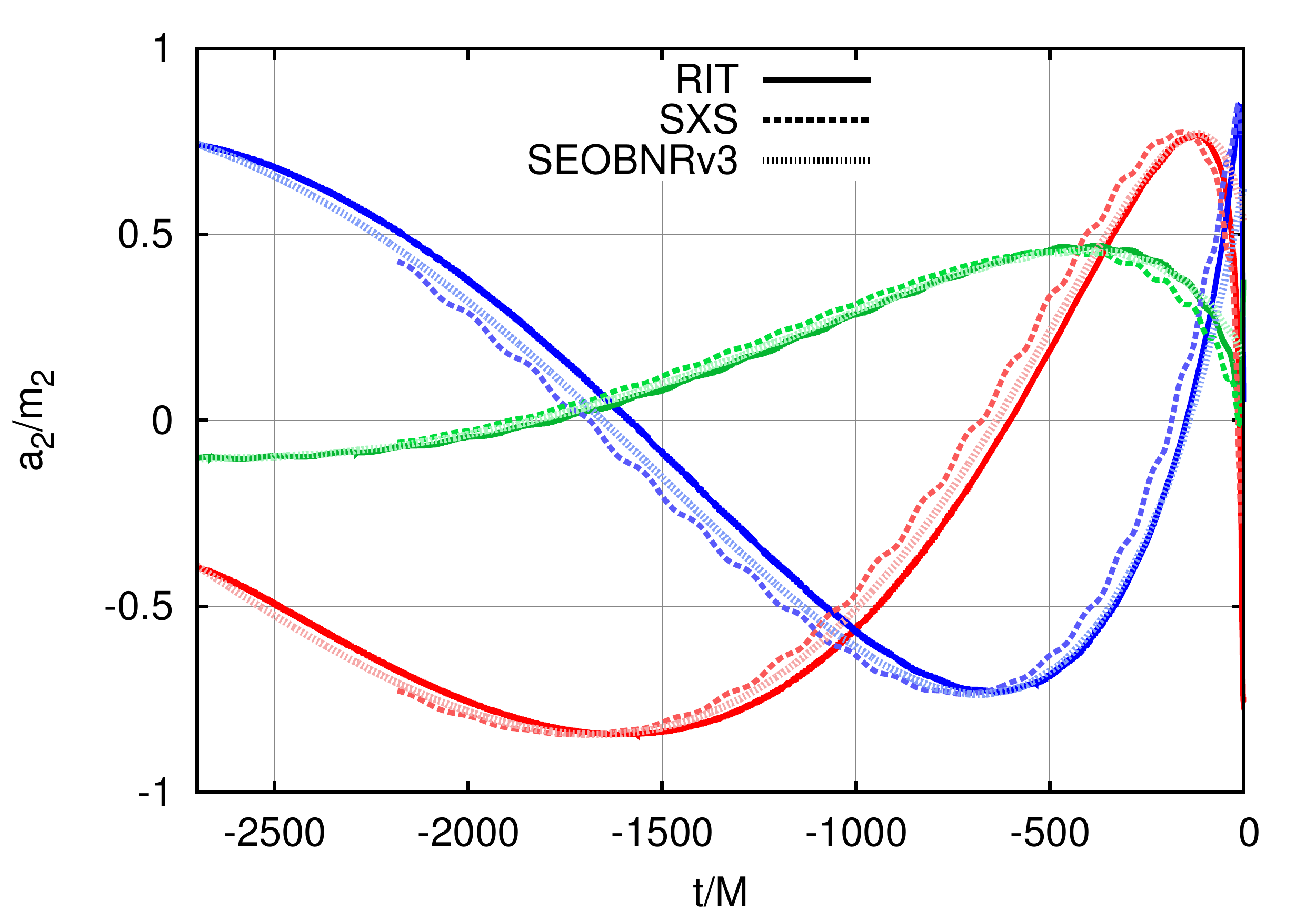}
  \caption{ The small BH (top) and large BH (bottom) spins for followup \#3.  
 RIT's simulation has solid lines; SXS's has dashed lines; and spin evolution as predicted by
   SEOBNRv3 is shown with a wide-dashed line.
   The spin components, $x$, $y$, and $z$, are red, blue, and green respectively.
  \label{fig:uid3_spins}}
\end{figure}


For the RIT simulations, the initial data parameters in Table \ref{tab:ID} for the nonprecessing systems 
1, 4, and 5 were determined by choosing the starting frequency
just below the reference frequency.  This gives the gauge time to settle, and since the
spins do not change, this gives us a cleaner waveform once we hit the reference frequency.
For the precessing simulations 2 and 3, since the spins will now evolve, 
we determine the initial data parameters by choosing the initial spins at the specified reference frequency.  
The initial data used by SXS, being not conformally flat have less spurious radiation content than the
Bowen-York data and hence produce a different set of masses and spins after settling down. See Table \ref{tab:relaxed} for the specific values of each simulation by the two kind of initial data families.
This process can be iterated to get closer initial parameters for each approach, although it requires
some extra evolution time and coordination to reach a fractional agreement below $10^{-3}$.
This process has been followed for UID\#1, but not for the other cases, in particular the two precessing ones
\#2 and \#3, and hence the differences, for instance, displayed in Fig.~\ref{fig:uid3_spins} for the precessing case \#3.

The SXS simulations used in this work have been assigned SXS catalog numbers BBH:0626 (UID1), BBH:0627 (UID2), BBH:0628
(UID3), BBH:0625 (UID4), and BBH:0631 (UID5).

Each simulation has an asymptotic frame relative to which we extract   $rh_{lm}(t)$.  In all cases used here, this axis
corresponds to the $\hat{z}$ ($=\hat{L}$) axis of the simulation frame.   For all simulations, this axis also agrees with the orbital angular momentum axis $\hat{L}$ at the start of the evolution.

\subsection{Outgoing radiation very similar for different NR methods}

\begin{figure}
  \includegraphics[angle=270,width=\columnwidth]{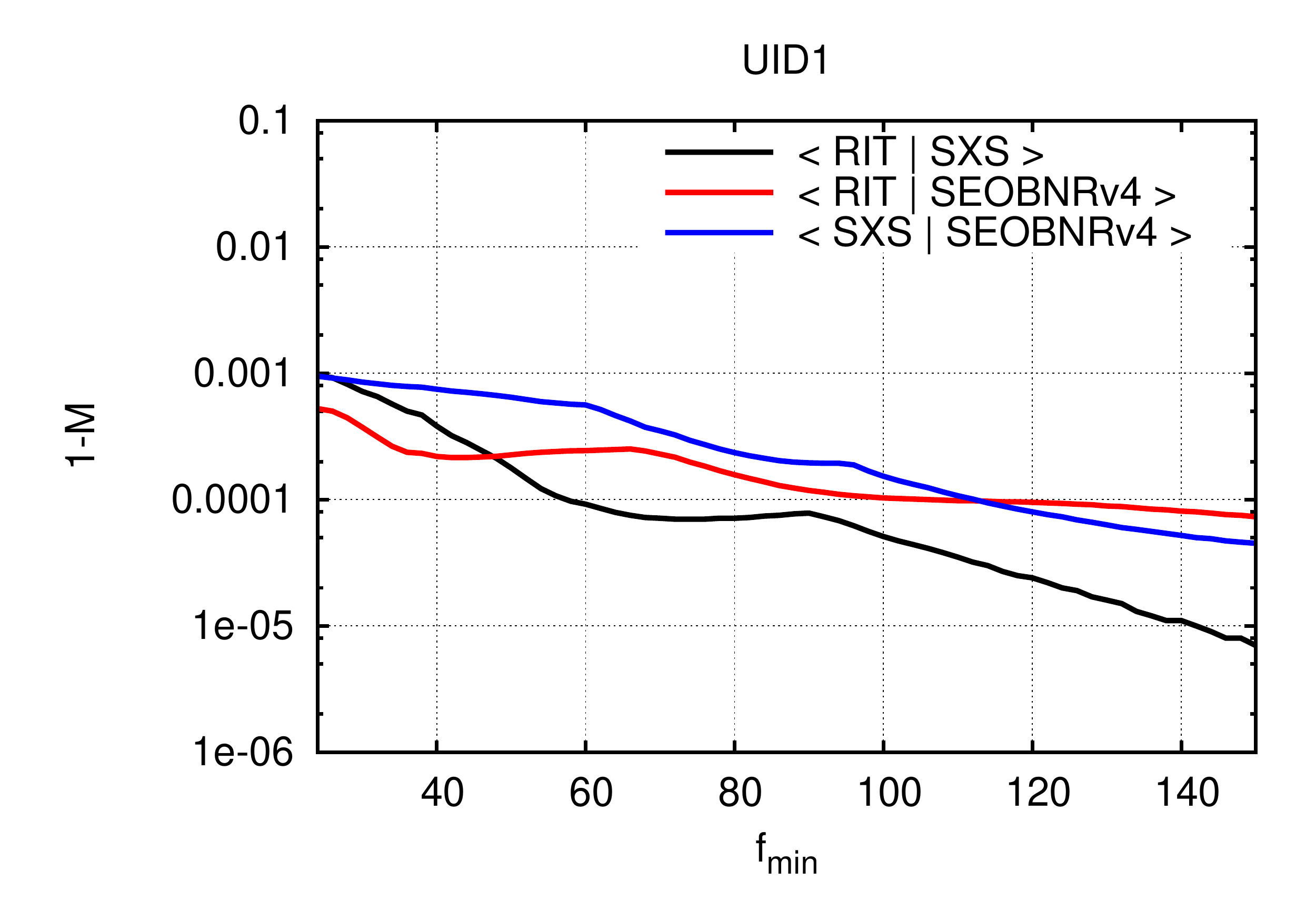}
  \includegraphics[angle=270,width=\columnwidth]{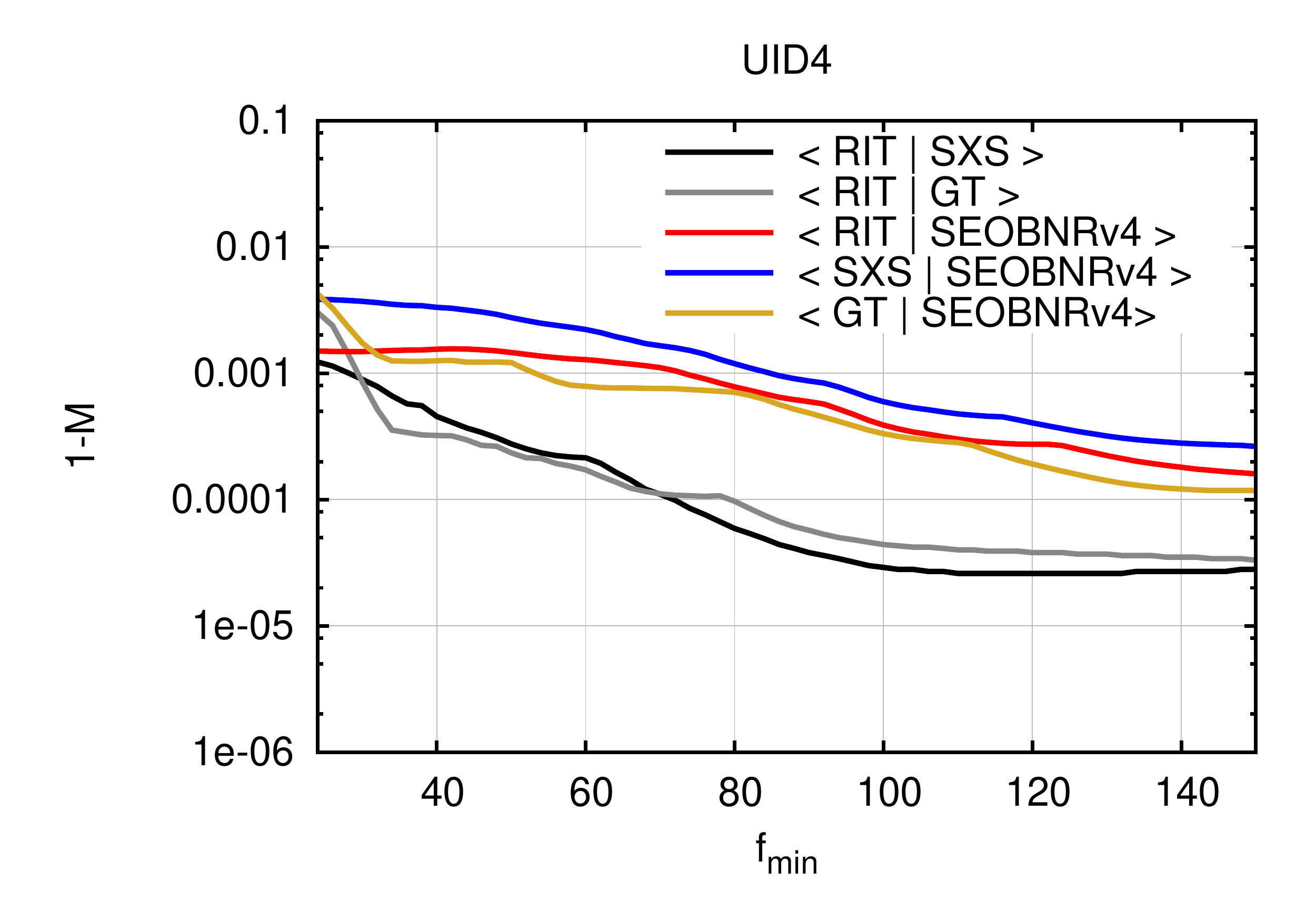}
  \includegraphics[angle=270,width=\columnwidth]{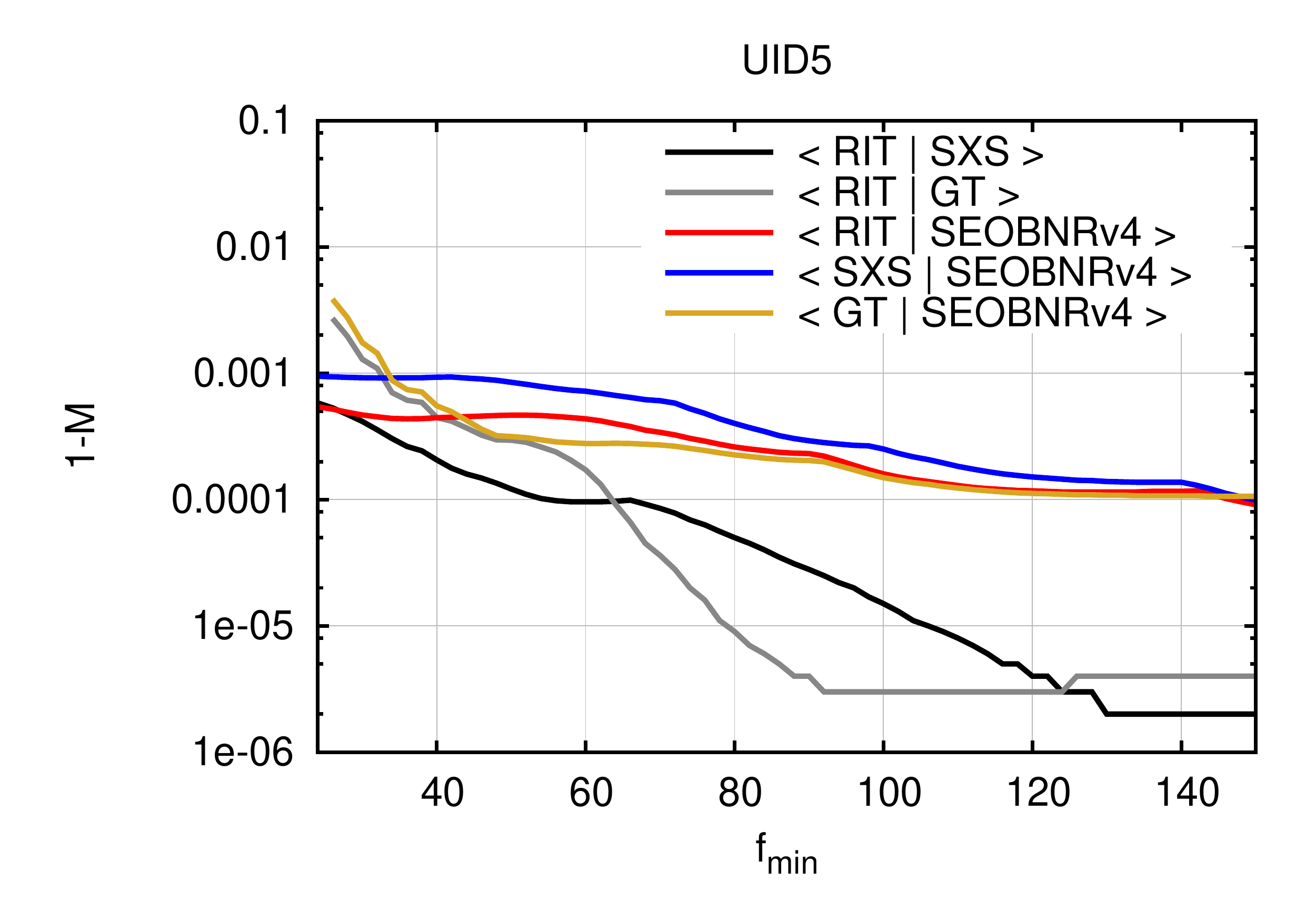}
  \caption{For the three nonprecessing UIDs \# 1,4,5 in Table~\ref{tab:ID}, matches between SXS, RIT, and SEOBNRv4 (2,2) modes
    as a function of $f_{\rm min}$, 
using the H1 PSD characterizing data near GW170104. We also compare with GT runs for UIDs \# 4,5.
    Compare to  also to similar plots for GW150914 \cite{LovelaceLousto2016}.
  \label{fig:UID1_match}  
}
\end{figure}

\begin{figure}
  \includegraphics[angle=270,width=\columnwidth]{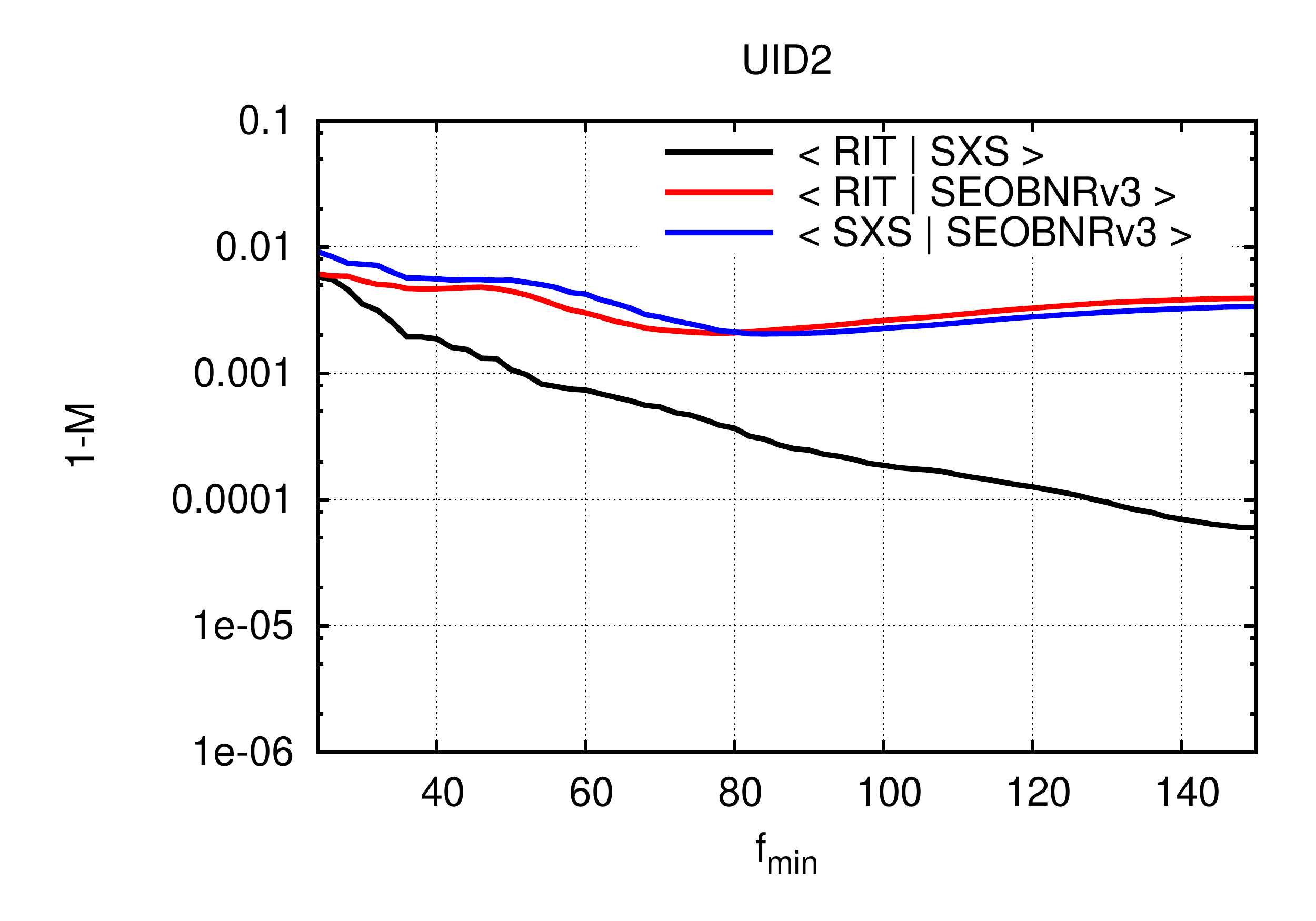}
  \includegraphics[angle=270,width=\columnwidth]{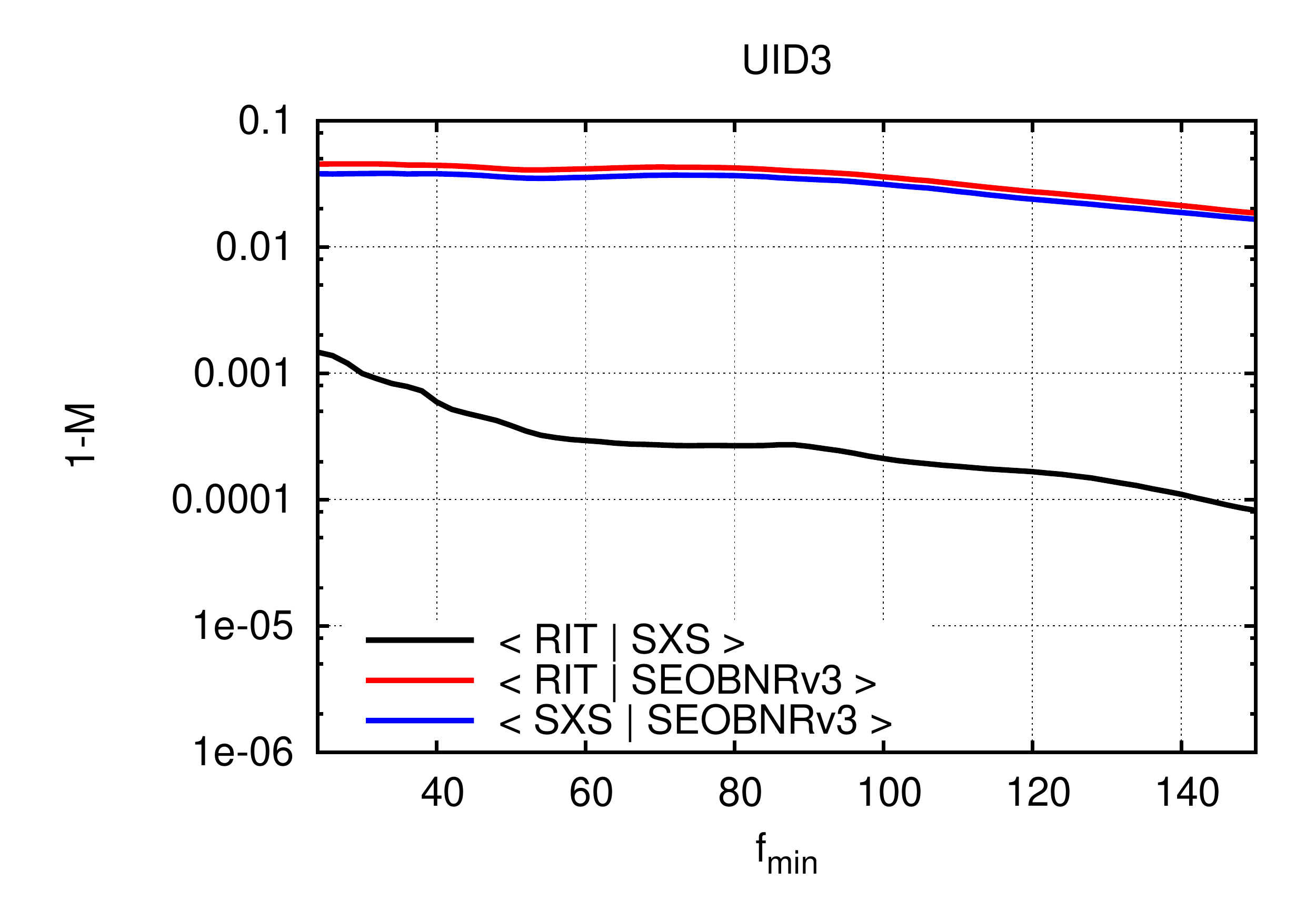}
  \caption{For the two precessing UIDs\#2,3 in Table~\ref{tab:ID}, matches between SXS, RIT, and SEOBNRv3 (2,2) modes
as a function of $f_{\rm min}$
    as a function of $f_{\rm min}$, 
using the H1 PSD characterizing data near GW170104.   In this comparison, the $(2,2)$ mode of all three simulations and
SEOBNRv3 are extracted relative to the $\hat{L}$ axis, identified from their common initial orbital parameters.
While these frame identifications are coordinate-dependent for precessing binaries -- implying our comparisons here
could include both intrinsic disagreement and systematic error due to (say) overall misalignment -- the good agreement
shown in Figure \ref{fig:uid3_spins} for the equally coordinate-dependent spins suggests that   convention-dependent
sources contribute little to the mismatches illustrated here. 
  \label{fig:UID2_match}}
\end{figure}

Following previous studies \cite{LovelaceLousto2016}, we compare the outgoing radiation mode by mode, using an
observationally-driven measure: the overlap or \emph{match}.  
The black and grey lines in Figures \ref{fig:UID1_match} and \ref{fig:UID2_match} show the match between the two simulations' (RIT-SXS and RIT-GT respectively)
(2,2) modes, as a function of the minimum frequency used in the match.  In this calculation, we use a detector noise
power spectrum appropriate to GW170104, and a total mass $M_\odot=$ as given in Table \ref{tab:NRfollow}.   By increasing the minimum frequency, we
increasingly omit the earliest times in the signal, first eliminating transient startup effects associated due to finite
duration and eventually comparing principally the merger signals from the two black holes.  
For comparison, the red, blue, and yellow lines show the corresponding matches between RIT, SXS, and GT simulations respectively and effective one body
models with identical parameters (faithfulness study).   In Figure \ref{fig:UID1_match}, which illustrates only nonprecessing simulations,
these comparisons are made to the nonprecessing model SEOBNRv4 \cite{Bohe:2016gbl}.  In Figure \ref{fig:UID2_match}, which targets the two
precessing UIDs, we instead compare to SEOBNRv3, which approximates some precession effects.   
For both nonprecessing and precessing simulations, these figures show that the different NR groups' simulations produce similar
radiation, with mismatches $\le 10^{-3}$ even at the longest durations considered.  By contrast, comparisons with SEOBNRv4
and SEOBNRv3 show that these models do not replicate our simulations' results, particularly for precessing binaries.  

To demonstrate good agreement beyond the (2,2) mode for precessing simulations, for multiple resolutions, Tables
\ref{tab:match_uid3} and \ref{tab:match_n100vL3} systematically compare all modes between RIT and SXS.  The match
calculations in this Table are performed using a strain noise power spectral densities (PSD)  characterizing data near
GW170104.  Following \cite{LovelaceLousto2016}, one phase- and time-shift is computed by maximizing the overlap of the (2,2) mode; this phase- and time-shift is then applied to all other modes without any further maximization.
Table \ref{tab:match_uid3} shows a resolution test: the match between RIT and SXS simulations, as a function of RIT
simulation resolution.  As the most challenging precessing case, UID3 is shown by default
Except for the $m=0$ modes, all the simulations show good agreement mode-by-mode, for all resolutions.
Based on Figures  \ref{fig:UID1_match},  \ref{fig:UID2_match} and Table  \ref{tab:match_uid3} we anticipate the lowest
production-quality NR resolution (N100 for RIT; L3 for SXS) will usually be sufficient to go well beyond the accuracy of approximate and 
phenomenological models.  To elaborate on this hypothesis,  Table   \ref{tab:match_n100vL3} shows the mode-by-mode
overlaps between these two lowest NR resolutions.   The $\ell=2$ modes agree without exception.  Good agreement also
exists for the most significant modes up to $\ell\le 4$. On the other hand, the last columns of Table \ref{tab:match_uid3}
shows that the rough agreement between NR and models for the modes (2,2) displayed in Fig.~\ref{fig:UID2_match}, notably
worsens when looking at other than the leading modes.

In Table \ref{tab:srem_uid2}, we also provide a comparison of the remnant properties,
ie. final mass, spin and recoil velocity of the final, merged, black hole,
as computed by the two NR methods and for a set of three increasing resolutions. We observe good agreement and convergence of their values.
In the case of the RIT runs, a nearly 4th order convergence with resolution
for the recoil velocity of the remnant is displayed. The same convergence
properties are shown for the final mass and spin, despite these quantities being over-resolved
at these resolutions. Those remnant
quantities are also important to model fitting formulae 
\cite{Healy:2016lce,Jimenez-Forteza:2016oae} to be used to infer the final black hole properties from the binary
parameters and thus serve as a test of the general theory of relativity,
as in \cite{Ghosh:2016qgn}. The phenomenological approximate waveform models \cite{Bohe:2016gbl,Blackman:2017pcm,London:2017bcn} also benefit from information
of the final remnant properties as one of their inputs and can thus
produce a more accurate precessing and including higher modes model.

In conclusion we see that the typical production NR simulations are well into the convergence regime and produce accurate enough waveforms for all practical applications of the current generation of gravitational wave observations. This includes different NR approaches, modes and remnant properties. The distinction with the current models is also clear and those still show signs of systematic errors with respect to the most accurate NR waveforms.

\begin{table}
\begin{ruledtabular}
\begin{tabular}{ccccccr}
$\ell$ & $m$ & $ N100 $ & $ N118 $ & $ N140 $ & $\mathcal{M}_{v3}$ & $\mathcal{O}_{N140}$ \\
\hline
2 & -2 & 0.9989 & 0.9990 & 0.9990 & 0.9347 & 244.54 \\
2 & -1 & 0.9965 & 0.9972 & 0.9968 & 0.6257 & 96.12 \\
2 & 0  & 0.9972 & 0.9973 & 0.9966 & 0.3091 & 56.06 \\
2 & 1  & 0.9982 & 0.9983 & 0.9983 & 0.5797 & 102.66 \\
2 & 2  & 0.9986 & 0.9986 & 0.9986 & 0.9600 & 215.48 \\
\hline
3 & -3 & 0.9901 & 0.9902 & 0.9912 & - & 29.63 \\
3 & -2 & 0.9887 & 0.9913 & 0.9902 & - & 17.14 \\
3 & -1 & 0.9785 & 0.9811 & 0.9801 & - & 8.98 \\
3 & 0 & 0.9803 & 0.9814 & 0.9834 & - & 5.57 \\
3 & 1 & 0.9848 & 0.9845 & 0.9847 & - & 9.17 \\
3 & 2 & 0.9867 & 0.9864 & 0.9862 & - & 17.01 \\
3 & 3 & 0.9899 & 0.9896 & 0.9901 & - & 28.87 \\
\hline
4 & -4 & 0.9921 & 0.9927 & 0.9938 & - & 11.99 \\
4 & -3 & 0.9800 & 0.9798 & 0.9814 & - &6.61 \\
4 & -2 & 0.9830 & 0.9851 & 0.9838 & - & 4.17 \\
4 & -1 & 0.9856 & 0.9871 & 0.9868 & - & 2.30 \\
4 & 0 & 0.9317 & 0.9341 & 0.9377 & - & 1.55 \\
4 & 1 & 0.9854 & 0.9862 & 0.9861 & - & 2.32 \\
4 & 2 & 0.9825 & 0.9845 & 0.9836 & - & 4.26 \\
4 & 3 & 0.9827 & 0.9825 & 0.9835 & - & 6.93 \\
4 & 4 & 0.9906 & 0.9911 & 0.9919 & - & 10.13 \\
\hline
5 & -5 & 0.9703 & 0.9819 & 0.9848 & - & 3.19 \\
5 & -4 & 0.9646 & 0.9681 & 0.9735 & - & 1.72 \\
5 & -3 & 0.9641 & 0.9674 & 0.9708 & - & 1.09 \\
5 & -2 & 0.9575 & 0.9743 & 0.9765 & - & 0.66 \\
5 & -1 & 0.9657 & 0.9722 & 0.9734 & - & 0.36 \\
5 & 0 & 0.8730 & 0.8897 & 0.9013 & - &0.25 \\
5 & 1 & 0.9636 & 0.9695 & 0.9710 & - &0.37 \\
5 & 2 & 0.9541 & 0.9728 & 0.9765 & - &0.67 \\
5 & 3 & 0.9688 & 0.9718 & 0.9738 & - &1.13 \\
5 & 4 & 0.9643 & 0.9692 & 0.9725 & - & 1.71 \\
5 & 5 & 0.9657 & 0.9796 & 0.9825 & - & 2.73 \\
\end{tabular}
\end{ruledtabular}
\caption{Match between individual spherical harmonic modes $(\ell,m)$
  of the SXS and RIT UID3 waveforms, using the H1 PSD 
 characterizing data near GW170104.
 Following \cite{LovelaceLousto2016}, rather than
maximize over time and phase for each independently, our
mode-by-mode comparisons fix the event time and overall phase using one mode (here, the (2,2) mode).  
The successively higher resolution simulations from RIT, labeled as
$N100,N118,N140$ are compared to the L3 (highest) resolution run from SXS. The minimal frequency is taken as $f_{min}=30m\mbox{ Hz}$ for $m>1$ and $f_{min}=30 Hz$ for $m=0,1$ for a fiducial
total mass of $M=58.73M_\odot$. The column labeled $\mathcal{M}_{v3}$ shows the match between RIT N140 
and the corresponding SEOBNRv3 mode.  Rows with a ``-" are not modeled by SEOBNRv3.  
The column labeled $\mathcal{O}_{N140}$ shows the overlap of N140 with itself, with a minimum frequency
of $30\mbox{ Hz}$ in all cases, to indicate the significance of the mode.
}
\label{tab:match_uid3}
\end{table}

\begin{table*}
\begin{ruledtabular}
\begin{tabular}{cccccccccccc}
$\ell$ & $m$ & $\mathcal{M}_1$ & $\mathcal{O}_1$ & $\mathcal{M}_2$ & $\mathcal{O}_2$ & $\mathcal{M}_3$ & $\mathcal{O}_3$ & $\mathcal{M}_4$ & $\mathcal{O}_4$ & $\mathcal{M}_5$ & $\mathcal{O}_5$ \\
\hline
2 & -2 & 0.9993 & 282.02 & 0.9940 & 220.13 & 0.9993 & 259.04 & 0.9991 & 241.44 & 0.9996 & 237.10 \\
2 & -1 & 0.9985 & 28.16 & 0.9933 & 127.06 & 0.9973 & 96.60 & 0.9859 & 24.19 & 0.9991 & 19.24 \\
2 & 0 & 0.9294 & 5.81 & 0.9145 & 72.81 & 0.9975 & 54.88 & 0.9795 & 6.72 & 0.9816 & 5.14 \\
2 & 1 & 0.9986 & 28.13 & 0.9946 & 130.33 & 0.9985 & 103.84 & 0.9859 & 24.19 & 0.9991 & 19.24 \\
2 & 2 & 0.9993 & 282.02 & 0.9965 & 201.91 & 0.9990 & 226.98 & 0.9991 & 241.44 & 0.9996 & 237.10 \\
\hline
3 & -3 & 0.9463 & 6.49 & 0.9870 & 22.17 & 0.9904 & 31.88 & 0.9363 & 15.23 & 0.9978 & 37.33 \\
3 & -2 & 0.9993 & 7.80 & 0.9827 & 17.40 & 0.9915 & 17.89 & 0.9947 & 6.06 & 0.9986 & 6.15 \\
3 & -1 & 0.8844 & 1.17 & 0.9850 & 9.31 & 0.9788 & 9.30 & 0.7229 & 1.95 & 0.9099 & 0.73 \\
3 & 0 & 0.7757 & 0.78 & 0.8437 & 5.46 & 0.9792 & 5.73 & 0.8199 & 0.69 & 0.8720 & 0.28 \\
3 & 1 & 0.8879 & 1.21 & 0.9766 & 9.99 & 0.9838 & 9.51 & 0.7228 & 1.95 & 0.9099 & 0.73 \\
3 & 2 & 0.9993 & 7.80 & 0.9930 & 18.29 & 0.9861 & 17.81 & 0.9947 & 6.06 & 0.9986 & 6.15 \\
3 & 3 & 0.9463 & 6.49 & 0.9843 & 22.66 & 0.9898 & 30.67 & 0.9363 & 15.23 & 0.9978 & 37.33 \\
\hline
4 & -4 & 0.9922 & 12.17 & 0.9810 & 9.20 & 0.9926 & 12.97 & 0.9908 & 10.32 & 0.9956 & 13.75 \\
4 & -3 & 0.9930 & 2.24 & 0.9824 & 7.66 & 0.9794 & 7.11 & 0.9869 & 1.68 & 0.9967 & 1.94 \\
4 & -2 & 0.5968 & 0.69 & 0.9874 & 4.90 & 0.9841 & 4.37 & 0.9150 & 0.53 & 0.9068 & 0.49 \\
4 & -1 & 0.6361 & 0.14 & 0.9743 & 2.51 & 0.9857 & 2.39 & 0.7761 & 0.15 & 0.6647 & 0.10 \\
4 & 0 & 0.3514 & 0.49 & 0.6951 & 1.45 & 0.9355 & 1.60 & 0.2246 & 0.52 & 0.2578 & 0.28 \\
4 & 1 & 0.6533 & 0.13 & 0.9606 & 2.43 & 0.9848 & 2.45 & 0.7761 & 0.15 & 0.6647 & 0.10 \\
4 & 2 & 0.5967 & 0.69 & 0.9887 & 4.89 & 0.9830 & 4.51 & 0.9150 & 0.53 & 0.9068 & 0.49 \\
4 & 3 & 0.9930 & 2.24 & 0.9851 & 7.75 & 0.9830 & 7.37 & 0.9869 & 1.68 & 0.9967 & 1.94 \\
4 & 4 & 0.9922 & 12.17 & 0.9814 & 8.28 & 0.9912 & 10.78 & 0.9908 & 10.32 & 0.9956 & 13.75 \\
\hline
5 & -5 & 0.8662 & 0.64 & 0.9614 & 1.85 & 0.9709 & 3.29 & 0.9028 & 1.38 & 0.9807 & 3.83 \\
5 & -4 & 0.9758 & 0.58 & 0.9666 & 1.81 & 0.9648 & 1.85 & 0.9615 & 0.44 & 0.9893 & 0.61 \\
5 & -3 & 0.7287 & 0.12 & 0.9775 & 1.26 & 0.9631 & 1.16 & 0.7838 & 0.13 & 0.8748 & 0.19 \\
5 & -2 & 0.4258 & 0.22 & 0.9592 & 0.74 & 0.9552 & 0.71 & 0.6812 & 0.12 & 0.7639 & 0.13 \\
5 & -1 & 0.1107 & 0.07 & 0.9494 & 0.37 & 0.9658 & 0.38 & 0.6209 & 0.03 & 0.4149 & 0.04 \\
5 & 0 & 0.3735 & 0.10 & 0.5100 & 0.25 & 0.8761 & 0.27 & 0.3827 & 0.08 & 0.2603 & 0.06 \\
5 & 1 & 0.0553 & 0.04 & 0.9128 & 0.34 & 0.9639 & 0.40 & 0.6208 & 0.03 & 0.4152 & 0.04 \\
5 & 2 & 0.4258 & 0.22 & 0.9711 & 0.72 & 0.9509 & 0.74 & 0.6812 & 0.12 & 0.7639 & 0.13 \\
5 & 3 & 0.7286 & 0.12 & 0.9804 & 1.30 & 0.9683 & 1.22 & 0.7838 & 0.13 & 0.8748 & 0.19 \\
5 & 4 & 0.9758 & 0.58 & 0.9764 & 1.87 & 0.9661 & 1.86 & 0.9615 & 0.44 & 0.9893 & 0.61 \\
5 & 5 & 0.8662 & 0.64 & 0.9650 & 1.77 & 0.9663 & 2.76 & 0.9028 & 1.38 & 0.9807 & 3.83 \\
\end{tabular}
\end{ruledtabular}
\caption{Matches ($\mathcal{M}_i$) between individual spherical harmonic modes $(\ell,m)$
of the SXS and RIT waveforms, using the H1 PSD
 characterizing data near GW170104.
The lowest resolution simulations from RIT, labeled N100, are compared to the L3 resolution run from SXS. 
The minimal frequency is taken as $f_{min}=30m\mbox{ Hz}$ for $m>1$ and $f_{min}=30 Hz$ for $m=0,1$.
The columns labeled $\mathcal{O}_i$ show the overlap of N100 with itself,  
$\left<h_{\ell m}^{N100}|h_{\ell m}^{N100}\right>$, 
to indicate the significance of the mode.
}
\label{tab:match_n100vL3}
\end{table*}

\begin{table}
\begin{ruledtabular}
\begin{tabular}{rccc}
$\#$ & $m_{rem} / M $ & $\chi_{rem}^z$ & $V_{rem}^{xy} (km/s)$ \\
\hline
UID1-N100 & 0.955294 & 0.619052 & 402.78  \\
UID1-N118 & 0.955310 & 0.619079 & 404.40 \\
UID1-N140 & 0.955311 & 0.619100 & 405.63 \\
UID1-L2   & 0.955782 & 0.618905 &  \\
UID1-L3   & 0.955813 & 0.618893 &  \\
UID1-L4   & 0.955829 & 0.618899 &  \\
\hline
UID2-N100 & 0.963445 & 0.581627 & 962.55 \\
UID2-N118 & 0.963418 & 0.581480 & 996.49 \\
UID2-N140 & 0.963405 & 0.581392 & 1016.06 \\
UID2-L1   & 0.963768 & 0.579124 &  \\
UID2-L2   & 0.964063 & 0.579988 &  \\
UID2-L3   & 0.964063 & 0.579958 &  \\
\hline
UID3-N100 & 0.961903 & 0.659634 & 614.70 \\
UID3-N118 & 0.961920 & 0.659725 & 598.96 \\
UID3-N140 & 0.961927 & 0.659781 & 587.63 \\
UID3-L1   & 0.962123 & 0.658707 &  \\
UID3-L2   & 0.962388 & 0.657731 &  \\
UID3-L3   & 0.962401 & 0.657599 &  \\
\hline
UID4-N100 & 0.962020 & 0.529128 & 312.81  \\
UID4-N118 & 0.962028 & 0.529129 & 313.65  \\
UID4-N140 & 0.962030 & 0.529130 & 314.05 \\
UID4-L1   & 0.962114 & 0.528897 &  \\
UID4-L2   & 0.962184 & 0.529023 &  \\
UID4-L3   & 0.962174 & 0.529117 &  \\
\hline
UID5-N100 & 0.968160 & 0.531761 & 171.57  \\
UID5-N118 & 0.968171 & 0.531837 & 175.35  \\
UID5-N140 & 0.968173 & 0.531873 & 177.81 \\
UID5-L1   & 0.967872 & 0.531920 &  \\
UID5-L2   & 0.968041 & 0.531934 &  \\
UID5-L3   & 0.968051 & 0.531917 &  \\
\end{tabular}
\end{ruledtabular}
  \caption{Remnant results for spinning binaries.
    We show the remnant mass $m_{rem}$ in units of the total initial mass $M \equiv m_1+m_2$,
    the remnant dimensionless spin $\chi_{rem}^z \equiv J_{rem}^z / m_{rem}^2 $, and the
    remnant velocity in the x-y plane $V_{rem}^{xy}$. We show results for different LazEv
    resolutions (N100, N118, and N140) and different SpEC resolutions (L0, L2, L4, and L6).
}
\label{tab:srem_uid2}
\end{table}

\begin{table}
\begin{ruledtabular}
\begin{tabular}{l|rrrrrc|l}
UID & \multicolumn{3}{c}{$\ln {\cal L}$(RIT)} & $\ln {\cal L}$(SXS) & $\ln {\cal L}$(GT) & $\ln {\cal L}$(SEOB) & Model \\
 & N100 & N118 & N140 & L3 & M120 & (at NR) & \\ \hline
\#1 & 60.4 & 61.0 & 61.0 & 60.9 & - & 62.7 & v4 \\
\#2 & 61.0 & 60.9 & 60.6 & 60.9 & - & 61.4 & v3\\
\#3  & 60.4 & 60.5 & 60.7 & 60.7 & - & 60.4 & v3 \\
\#4 & 60.6 & 60.7 & 60.8 & 60.3 & 60.4 & 62.2 & v4\\
\#5 & 60.0 & 60.0 & 60.1 & 60.0 & 59.8 & 61.2 & v4\\
\end{tabular}
\caption{\label{tab:lnLmargPeak}\textbf{Marginalized likelihood of the data}: This table shows the results for the 5
  simulations when directly compared to the data. For these results, we use the same PSD adopted in all other
  calculations, with   $f_{\rm min}=30$Hz (i.e. low-frequency cutoff). The first column is the UID.  The second column  is the estimated peak log marginalized likelihood $\ln {\cal L}$, maximized over binary total mass, for the NR followup simulation.  The third column is the corresponding
  log marginalized likelihood, using exactly the same intrinsic parameters (e.g., masses and spins) as maximize the
  likelihood in the second column, evaluated using a phenomenological approximate model instead of numerical relativity. The fourth
  column is the specific model used: either SEOBNRv3 (for precessing simulations) or SEOBNRv4 (for nonprecessing
  simulations).  To see more on this parameter estimation method, see \cite{NRPaper,Lange2017}.}
  \end{ruledtabular}
\end{table}

\begin{table*}
\begin{ruledtabular}
\begin{tabular}{l|l|l|c|c|c|r|c|l}
NR & Label & Sim. ID & $q=\frac{m_1}{m_2}$ &$\chi_1$ & $\chi_2$ & $\ln {\cal L}$ & $\ln \mathcal{L}(SEOB)$ & Model\\
Group & & & & & & &(at NR)& \\
\hline
RIT & a & \verb|d0_D10.52_q1.3333_a-0.25_n100| & 0.7500 & ( 0, 0, 0 ) & ( 0, 0, -0.25 ) & 63.0 & 62.5 & v4\\
GT & b & \verb|(0.0,1.15)| & 0.8696 & ( 0, 0, 0 ) & ( 0, 0, 0 ) & 62.2& 61.5 & v4\\
RIT & c & \verb|q50_a0_a8_th_135_ph_30| & 0.5000 & ( 0, 0, 0 ) & ( 0.490, 0.283, -0.566 ) & 62.5 & 60.7 & v3\\
BAM & d & \verb|BAM150914:24| & 0.8912 & ( -0.278, -0.605, -0.085 ) & ( 0.151, 0.396, 0.017 ) & 62.7 & 61.0 & v3\\
SXS & e & \verb|SXS:BBH:0052| & 0.3333 & ( 0.001, 0.008, -0.499 ) & ( 0.494, 0.073, 0.001 ) & 62.3 & 60.4 & v3\\

\end{tabular}
\caption{\label{tab:lnLmargPeak:OtherSims}\textbf{Marginalized likelihood of the data: Selected other simulations}: This
  table shows the results for several other simulations that particularly match the data well and the SEOB model results at those parameter points.  These simulations are part of the top 15 simulations in $\ln {\cal L}$. When comparing the NR $\ln {\cal L}$ values here to the ones in Table \ref{tab:lnLmargPeak}, one can see these to be generally higher i.e. better match the data. When comparing the NR $\ln {\cal L}$ values to the SEOB at the same points, one sees a consistent lower SEOB $\ln {\cal L}$ value This implies that these points were not picked for NR Followup due to the lower SEOB $\ln {\cal L}$ value.}
  \end{ruledtabular}
\end{table*}

\section{Comparing NR simulations with observations of GW170104}\label{sec:NR+GW170104}

The comparisons above demonstrate that our simulations agree with one another, but differ from approximate and phenomenological
models oft-used to describe precessing mergers.   Fortunately, nature has provided us with a natural benchmark with
which to assess the efficacy of our calculations and the significance of any discrepancies: observations of BH-BH mergers.
We use standard techniques \cite{NRPaper,Lange2017} to directly compare GW170104 to our simulations.  For context, we also compare these
observations to the corresponding predictions of approximate and phenomenological models that purport to describe the same event.  

\begin{figure*}
  \includegraphics[width=1.7\columnwidth]{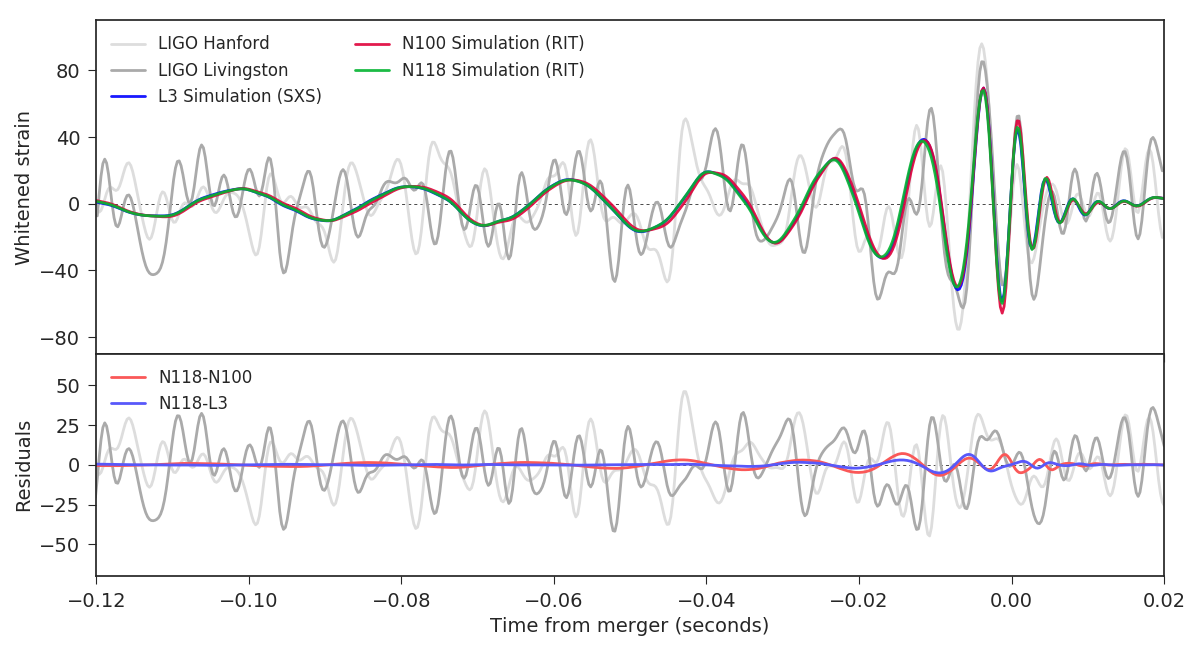}
  \includegraphics[width=1.7\columnwidth]{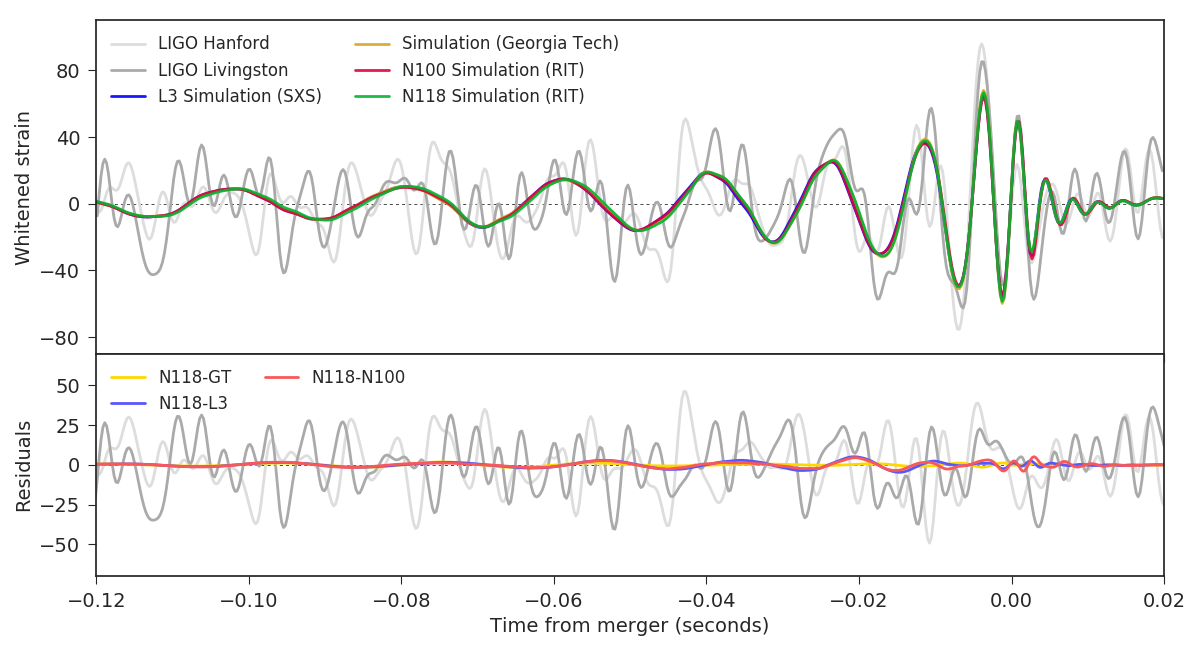}
  \includegraphics[width=1.7\columnwidth]{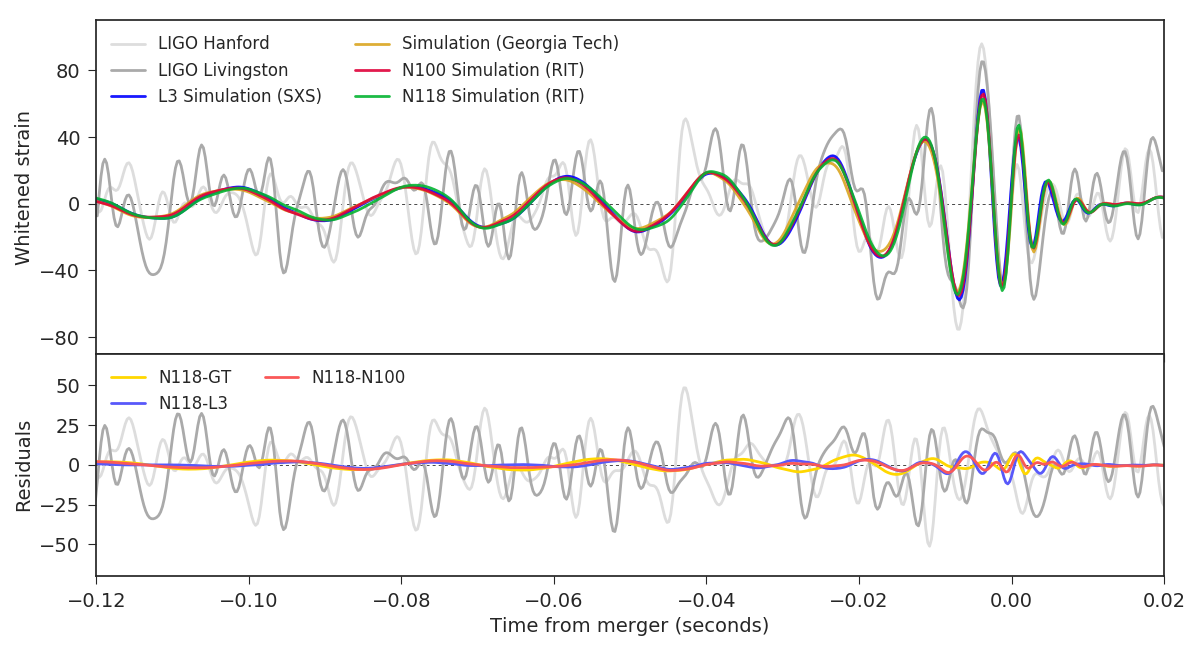}
  \caption{Comparison of the GW170104 signal seen by LIGO detectors
    H1 and L1 (in grey and dark grey) with the computer simulations of
    black hole mergers from SXS, RIT, and GT approaches for the
    nonprecessing cases labeled as $\#1$, $\#4$, and $\#5$ in Table~\ref{tab:ID}.
  \label{fig:UID145}}
\end{figure*}

Figure \ref{fig:UID145} displays the direct comparison of the
nonprecessing simulations
by RIT and SXS complementary approaches for the configurations
UID\#1, UID\#4, and UID\#5, as given in Table \ref{tab:ID}.
They directly compare to the signals as observed by LIGO H1 and L1
and with each other. The lower panel shows the residuals of the
signals with respect to the RIT N118 simulation and compares it
with the direct difference of the two approaches and also with the
difference of the N118 and N100 resolutions, that measure the
finite difference error of the N118, given the observed
near 4th order convergence seen when included the N140 run into the analysis.
For all three cases we note that the differences in any of the simulations
is much smaller than the residuals and hence typical noise of the
observations. This shows that the fast response
runs performed to simulate BBH (low-medium resolution)
are in an acceptable good agreement with the expected
higher resolution ones at the required level of errors.

\begin{figure*}
  \includegraphics[width=1.7\columnwidth]{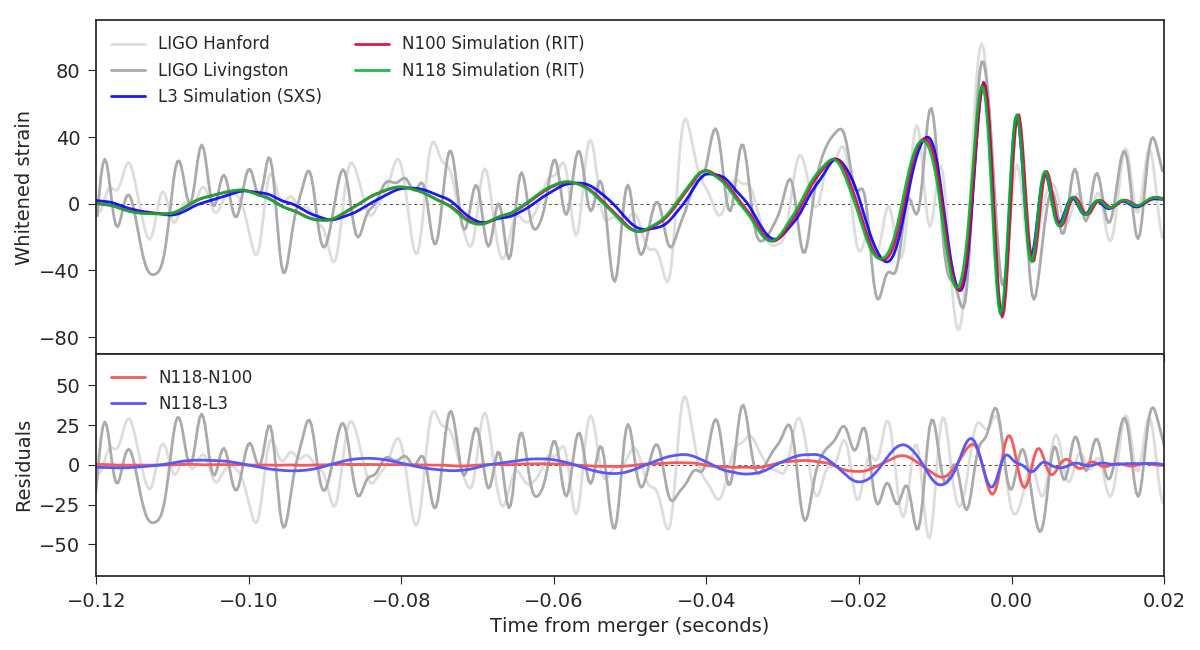}
  \includegraphics[width=1.7\columnwidth]{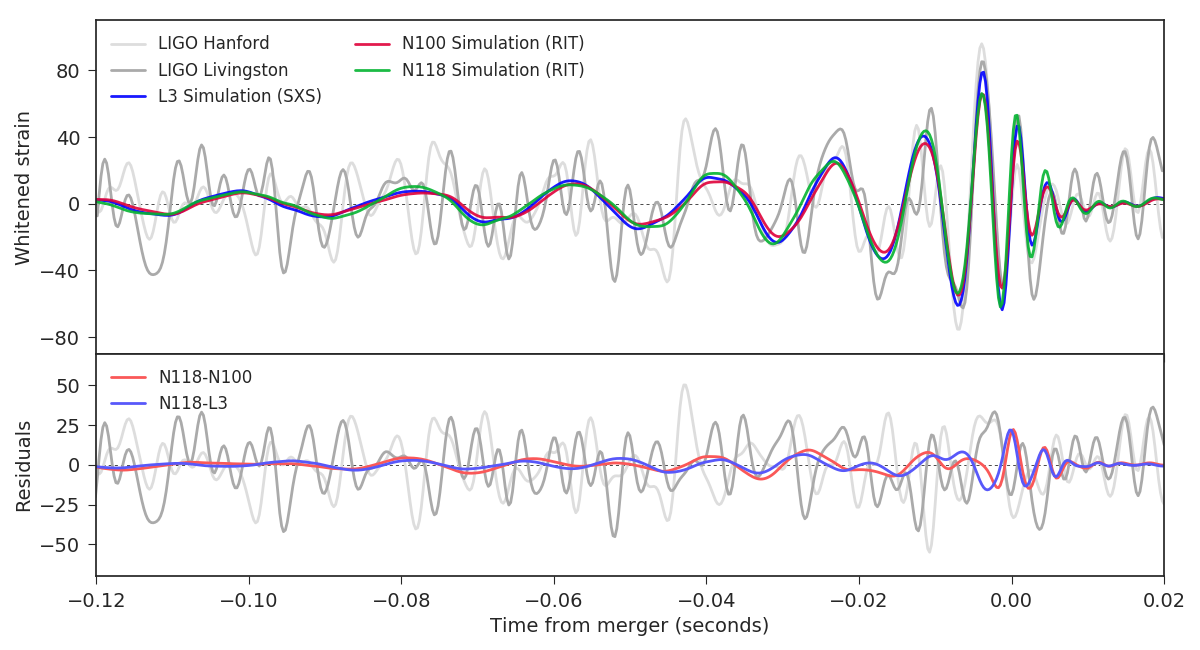}
  \caption{Comparison of the GW170104 signal seen by LIGO detectors
    H1 and L1 (in grey and dark grey) with the computer simulations of
    black hole mergers from SXS and RIT approaches for the
    nonprecessing cases labeled as $\#2$ and $\#3$ in Table~\ref{tab:ID}.
  \label{fig:UID23}}
\end{figure*}

Figure \ref{fig:UID23} displays the two precessing targeted simulations
for GW170104 studied in this paper. We compare them with the L1 and H1 signals
in grey and light grey in the plots. Here we also perform a double test
of the accuracy of the simulations by considering the two main approaches
to solve BBHs by the RIT and SXS groups and by considering the internal
consistency of convergence of the waveforms with increasing resolutions.
The waveforms again show a good agreement among themselves and their differences, shown in the lower panels are smaller than the residuals of the signals
with respect to the N118 simulations. They are larger than in the aligned
cases due to the choice of the initial spin configurations at at slightly
different reference orbital frequencies. Note also that this comparisons
do not align the peak of the waveforms and hence if independently fit to data would show much smaller differential residuals.

\subsection{Residuals versus resolution}

For each UID, direct comparison of our simulations to the data selects a fiducial total mass which best fits the
observations, as measured by the marginalized likelihood of the data assuming our simulations.   Using the same mass for all
simulations performed for that UID (e.g., by all groups and for all resolutions), we can for each simulation select the binary extrinsic
parameters like event time and sky location which maximize the likelihood of the
data, given our simulation and mass.   Then, using these extrinsic parameters, we evaluate the expected detector
response in the LIGO Hanford (H1) and Livingston (L1) instruments.  This procedure has been used to reconstruct the gravitational
wave signal for GW150914 \cite{NRPaper} and GW170104 events. 

\begin{figure*}
  \includegraphics[width=2\columnwidth]{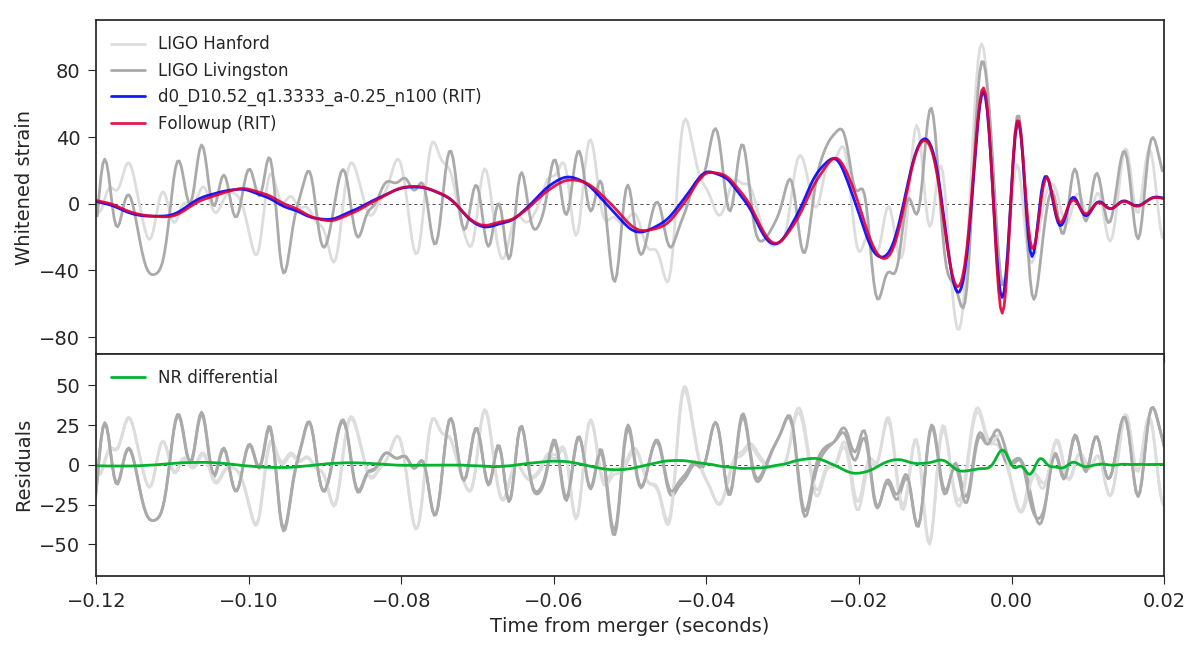}
  \caption{Comparison of the GW170104 signal seen by LIGO detectors
    H1 and L1 (in grey and dark grey) with the computer simulations of
    black hole mergers from RIT at low resolution for the
    nonprecessing case labeled as $\#1$ in Table~\ref{tab:ID} and the highest
    $\ln {\cal L}$ value for an NR simulation given in Table~\ref{tab:lnLmargPeak:OtherSims} (d0\_D10.52\_q1.3333\_a-0.25\_n100).
  \label{fig:UID1_d0}}
\end{figure*}

Figure \ref{fig:UID1_d0} shows an example of these reconstructions for the highest Log-Likelihood NR waveform (top candidate in Table
\ref{tab:lnLmargPeak:OtherSims} and UID\#1.  The top panel of this figure shows the NR 
predicted response in Hanford (blue-red); the Hanford data (grey); and the Livingston data (in dark grey; shifted by -2.93ms and sign
flipped).  
The bottom panel shows the residuals, and the difference between the two simulations in green. Note that the difference between waveforms is small compared to the residuals, but enough to make the simulation in blue (top candidate in Table
\ref{tab:lnLmargPeak:OtherSims}) have a slightly higher Likelihood (63.0 vs. 62.5) over UID\#1 in red, to match the signals over the whole range of frequencies considered. The same simulation resolution have been considered in both cases.

We have also analyzed the finite differences errors produced by fast-response,
low resolution (yet in the convergence regime) simulations of BBH mergers.
The low, medium and high resolutions runs, N100, N118, and N140 respectively,
by the RIT group show a nearly 4th order convergence 
(There are detailed studies of convergence for similar simulations in 
Refs. \cite{Healy:2014yta,Healy:2016lce,Healy:2017mvh})
that allow to extrapolate
to infinite resolution and evaluate the magnitude of the errors in the waveforms as compared to the residuals for this GW170104 event. 
We thus can evaluate the error of the N118 simulation is given by
the (N100-N118) difference, while the error of the N100 waveform is
twice this difference and that of the N140 waveform is half that difference.
This is displayed in the lower half of each panel in Figs.~\ref{fig:UID145} and \ref{fig:UID23}
and provide an alternative evaluation of the errors within a given NR method.

The studies carried out in this paper involving 3-resolutions for each set of parameters well in the convergence regime of the simulations can be very costly from the resources point of view, totaling over 4 million service units (SUs) in computer clusters, as detailed in Table~\ref{tab:SUs}.

\begin{table}
\begin{ruledtabular}
\begin{tabular}{lcccc}
UID & N100 & N118 & N140 & Total \\
\hline
1     & 119 &  184 &  407 &  710 \\
2     & 313 &  451 &  557 & 1321 \\
3     & 145 &  217 &  476 &  838 \\
4     & 130 &  178 &  565 &  873 \\
5     &  67 &  118 &  306 &  491 \\\hline
Total & 774 & 1148 & 2311 & 4233 \\
\end{tabular}
\caption{ kSUs (1000 core-hours) for each RIT run and resolution. \label{tab:SUs}}
  \end{ruledtabular}
\end{table}

According to the variations in the Table \ref{tab:lnLmargPeak}
that evaluates $\ln {\cal L}$ for
the different resolutions we may derive as a rule of thumb that the N100
grid provides a good approximation for the nonprecessing binaries, while
for the precessing ones, N118 is more appropriate. This leads to a
reduction of the SUs needed for these 5 simulations, totaling nearly
1 million SUs, two thirds of which are due to the two precessing cases.
The pseudospectral approach used by the SXS collaboration requires similar
total wallclock times than the above finite differences approach, 
but spends an order of magnitude less resources.
For instance, UID\#1 (SXS:BBH:0626) required 11 kSUs for Lev4, 
7.4 kSUs for Lev3, and 4.7 kSUs hours for Lev2.

\subsection{Likelihood of NR and models}
\label{sec:sub:lnL}

For any proposed coalescing binary, characterized by its outgoing radiation as a function of all directions, we can
compute a single quantity to assess its potential similarity to GW170104, accounting for all possible ways of orienting
the source and placing it in the universe: the marginalized likelihood ($\ln {\cal L}_{\rm marg}$)  \cite{Pankow:2015cra,Abbott:2016apu,Lange:2017wki}.  To provide a sense of scale, the  distribution of $\ln {\cal L}_{\rm marg}$ over the posterior distribution including
all intrinsic parameters is roughly universal
\cite{Abbott:2016apu}, approximately distributed as $\ln {\cal L}_{\rm marg,max} - \chi^2/2$ where $\chi^2$ has $d$ degrees of freedom
(i.e., a mean value of $\ln {\cal L}_{\rm marg,max} - d/2$, and its  90\% credible interval is $\ln {\cal L}_{\rm marg} \ge \ln {\cal L} -
x $, where $x=3.89$ and $x=6.68$ for  $d=4$ and $d=8$, respectively).
For each UID and for each proposed total mass $M$, direct comparison of our simulations to the data allows us to compute
a single number measuring the quality of fit: the marginalized likelihood ${\cal  L}_{\rm marg}$. 
The maximum value of this function (here
denoted by ${\cal L}$) therefore measures the overall quality of fit.    
Table \ref{tab:lnLmargPeak} shows $\ln {\cal L}$ for the five UIDs simulated here.    For comparison, the last column shows
${\cal L}$ calculated using an approximate model for the radiation from a coalescing binary.  Obviously, if these
approximate models and our simulations agree, then we should find the same result for $\ln{\cal L}$
at the same parameters.   Finally, for context, the
peak value of $\ln {\cal L}$ computed using SEOBNRv3 with generic parameters is 63.3.  
%
If our simulation parameters are well-chosen (and if both our simulations and these models are close to true solutions
of Einstein's equations), then this peak value should be in good agreement with the $\ln {\cal L}$ evaluated using our
simulations.

First and foremost, up to Monte Carlo and fitting error, the marginalized likelihoods calculated with NR
agree with each other comparing different resolutions and different approaches to solve the BBH problem, as required given the high degree of similarity between the underlying simulations.   
Second, the marginalized likelihoods computed at these proposed points are substantially \emph{below} the largest ${\cal L}$ found
with approximate models like SEOBNRv3, except for UID3.  Similar to the explanation described in Appendix
\ref{ap:DifferentTargets}, the exception here is due to the differences between the precessing models ($\ln {\cal L}$ was calculated with SEOB but the parameters were suggested with IMRP).  Likewise, the binary parameters at which the peak value of ${\cal L}$
occurs for SEOBNRv3 are substantially different from any of the proposed parameters explored here.   This discrepancy
suggests that the model-based procedure that we adopted to target our followup simulations was not effective at finding
the most likely parameters, as measured with $\ln {\cal L}$.  The poor performance of our targeted followup cannot simply reflect   sampling
error; even though the likelihood surface is nearly flat near the peak, so small errors are amplified in parameter
space, this near-flatness also insures that systematic offsets \emph{should} produce a small change in $\ln {\cal L}$,
\emph{if} the underlying waveform calculations agree; see Appendix \ref{ap:DifferentTargets} for further discussion.
 Instead, we suspect the
biases in ${\cal L}$ arise because the models only approximate the correct solution of Einstein's equations.
Third, we  confirm our hypothesis in Table \ref{tab:lnLmargPeak:OtherSims}  simply by demonstrating that other
simulations (not performed in followup) fit the data substantially better than our targeted parameters.

\begin{figure}
  \includegraphics[angle=0,width=0.85\columnwidth]{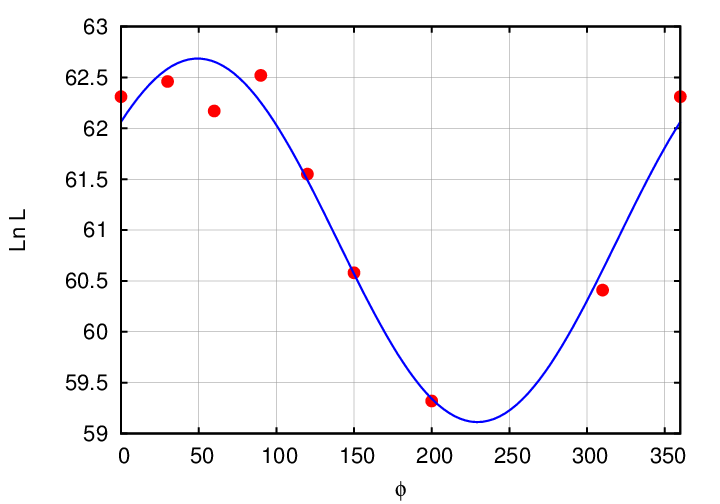}
  \includegraphics[angle=0,width=0.85\columnwidth]{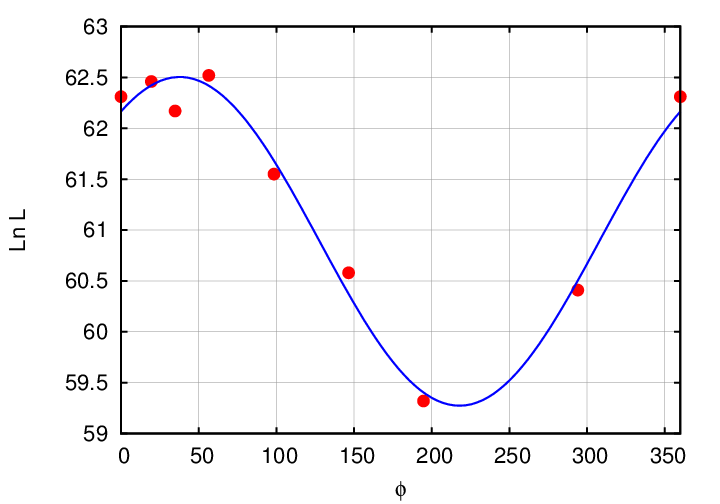}
  \caption{The log-likelihood of the NQ50TH135 series~\cite{Zlochower:2015wga} assuming a period of $2\pi$ versus initial angle
    (top panel) and merger angle (bottom panel.)  Data (red) and fits (blue) are given in Table~\ref{tab:LnL_phi_table}.
  \label{fig:LnL_phi}}
\end{figure}

On the one hand,  NR followup simulations guided by the models (as displayed in Table \ref{tab:lnLmargPeak}) leads to
lower marginalized likelihoods ($\ln {\cal L}$).  Conversely,  other simulations shown in Table
\ref{tab:lnLmargPeak:OtherSims} produce higher
$\ln {\cal L}$, at points in parameter space where the  models predict lower $\ln {\cal L}$.  This discrepancy suggest
the two processes ($\ln {\cal L}$ evaluated with NR and with the models) favor
 different regions of parameter space. In particular, table \ref{tab:lnLmargPeak:OtherSims}, which has one of the
 largest values of $\ln {\cal L}$ among all of the (roughly two thousand) simulations available to us, shows that the top
precessing simulation is 
$\verb|q50_a0_a8_th_135_ph_30|$.
This simulation has a mass ratio of 1:2, i.e. q=1/2, where the smaller
hole is nonspinning and the larger 
hole is spinning with an intrinsic spin magnitude of 0.8 and pointing 
initially in a direction downwards with respect to the orbital angular 
momentum ($\theta$=135 degrees) and an angle of 30 degrees from the line 
joining the two black holes ($\phi$=30 degrees). This simulation belongs
to a family of 6 simulations performed in Ref.~\cite{Zlochower:2015wga} 
labeled as NQ50TH135PH[0,30,60,90,120,150]. Those runs, supplemented by
two control runs with angles $\phi=200,\,310$ we performed for this paper, 
are displayed in Fig.~\ref{fig:LnL_phi} versus
the $\ln {\cal L}$ for this GW170104 event. The lower panels plots all those 
simulation with respect to their $\phi$-angle at merger as defined in
Ref.~\cite{Zlochower:2015wga} and given in table XXI in that paper. 
The continuous
curve provide a fit (detailed in table \ref{tab:LnL_phi_table})
for such values as reference and an estimate of the maximum value located
near the phi=30 simulation.

\begin{table}
\begin{ruledtabular}
\begin{tabular}{cccc}
$\phi$ & $\phi_{merger}$ & $\ln {\cal L}$ & $M_z/M_\odot$ \\
\hline
0      & 0               & 62.3& 54.9\\
30     & 19.5            & 62.5& 55.2\\
60     & 34.8            & 62.2& 54.1\\
90     & 56.5            & 62.5& 54.4\\
120    & 98.5            & 61.6& 54.1\\
150    & 146.5           & 60.6& 54.5\\
210    & 194.7	         & 59.3& 55.1\\
310    & 294.0           & 60.4& 54.6\\
\hline
$A$    & $B$             & $C$ & $\mathrm{RMS}$ \\
\hline
$1.23\pm0.21$ & $-0.75\pm0.15$ & $61.1\pm0.15$ & 0.38 \\
\hline 
$A_{merger}$ & $B_{merger}$ & $C_{merger}$ & $\mathrm{RMS}_{merger}$ \\
\hline
$1.08\pm0.18$ & $-0.47\pm0.19$ & $61.1\pm0.15$ & 0.37 \\
\end{tabular}
\caption{The log-likelihood of the NQ50TH135 series~\cite{Zlochower:2015wga}.  Fittings of the form $\ln {\cal L} = A sin( \pi/180 \phi + B ) + C$ is also given for both the initial $\phi$ and $\phi_{merger}$.
\label{tab:LnL_phi_table}}
  \end{ruledtabular}
\end{table}

\begin{figure}
 \includegraphics[angle=270,width=\columnwidth]{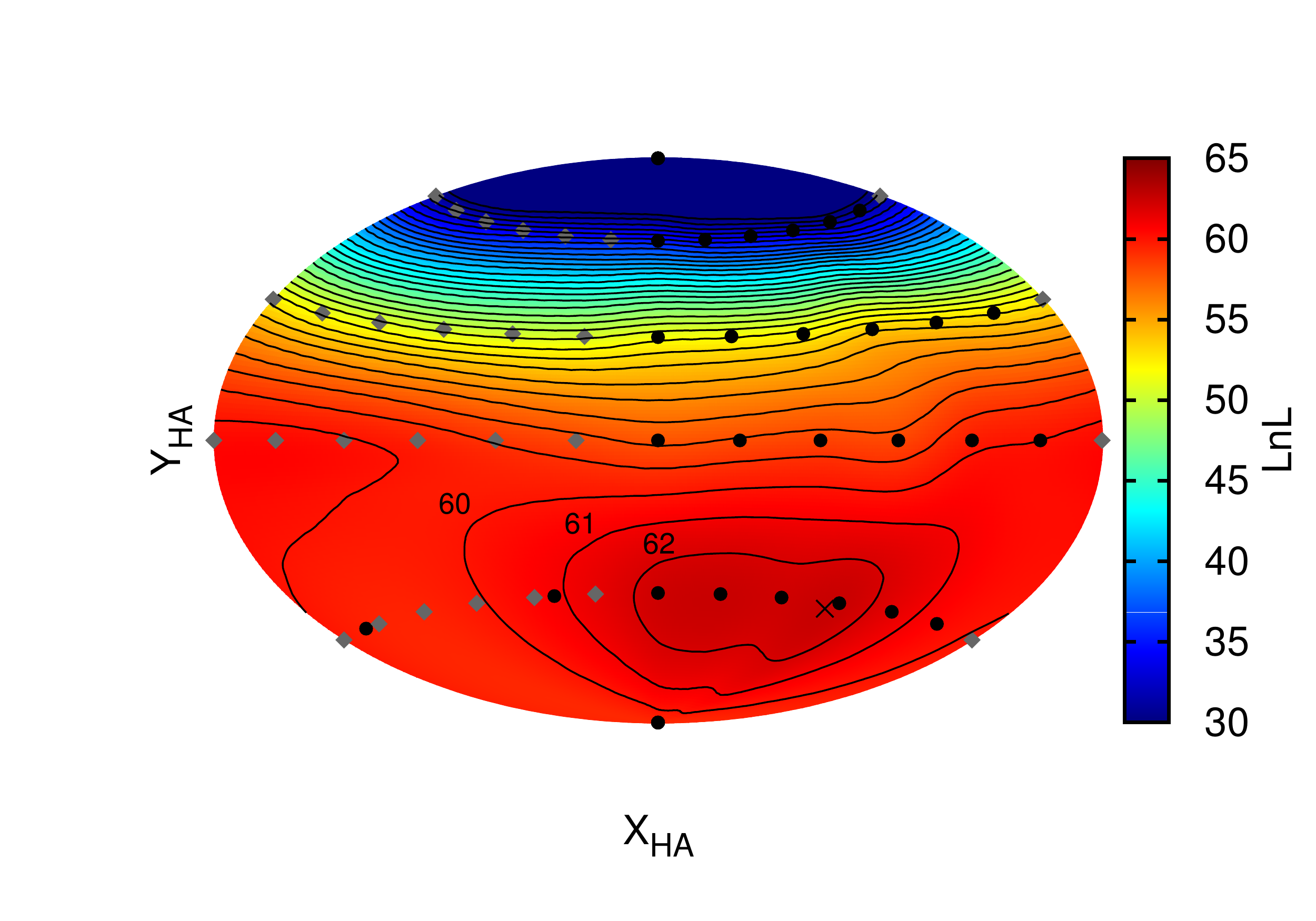}
 \caption{The log-likelihood of the NQ50THPHI series~\cite{Zlochower:2015wga} 
as a color map with red giving the highest $\ln{\cal L}$ and blue the lowest. The
black dots (and grey diamonds, obtained by symmetry) represent the NR simulations and we have used Hammer-Aitoff coordinates
$X_{HA}, Y_{HA}$, to represent the map and level curves with the top values of
$\ln {\cal L}=60,61,62$. The maximum, marked with an X, is 
located at TH=137, PH=87 reaching $\ln {\cal L}=62.6$.
 \label{fig:LnL_map_initial}}
\end{figure}

The notable results displayed in Fig. \ref{fig:LnL_phi},
where $\ln {\cal L}$ seems to be sensitive to the orientation of the 
spin of the larger hole on the orbital plane, are consistent with broader trends that can be extracted using similar
simulations: here, the set of 24 simulations of the family 
NQ50TH[30,60,90,135]PH[0,30,60,90,120,150] given in 
Ref.~\cite{Zlochower:2015wga}
supplemented by the two aligned runs NQ50TH[0,180]PH0 given in 
Ref.~\cite{Healy:2016lce} and two runs specifically performed for this paper,
NQ50TH135PH[200,310].   
These simulations all have $q=1/2$, a nonspinning smaller BH, and a spinning BH with fixed spin magnitude but changing
orientation.   
Figure \ref{fig:LnL_map_initial} shows a color-map derived from the maximum $\ln {\cal L}$ obtained for each of these
simulations, using standard (MatLab) plotting tools. The last surface levels indicates the regions of largest likelihood (60,61,62) and a maximum, marked with an X, is located at TH=137, PH=87 with 
$\ln {\cal L}$ of 62.6.
This results allow us to perform followup simulations seeking for this maximum.
In the plot, the black points are the NR simulations and the black curves are level sets of
the color-map.
Instead of plotting in the angles theta and phi, we plot in the Hammer-Aitoff coordinates
\footnote{Weisstein, Eric W.``Hammer-Aitoff Equal-Area Projection.'' From MathWorld--A Wolfram Web Resource. http://mathworld.wolfram.com/Hammer-AitoffEqual-AreaProjection.html}, which is a
coordinate system where the whole angular space can be viewed as a 2d map.  The points at the top left and bottom left
are the poles, $\theta=0$ at the top, and $\theta=\pi$ at the bottom.  The line connecting the two is the $\phi=0$ line.  As
you move from left to right, $\phi$ increases from 0 to 150 degrees (the maximum value of $\phi$ available in these simulations).

\subsection{Reconstructed NR waveforms}\label{sec:CI}

\begin{figure}
  \includegraphics[width=\columnwidth]{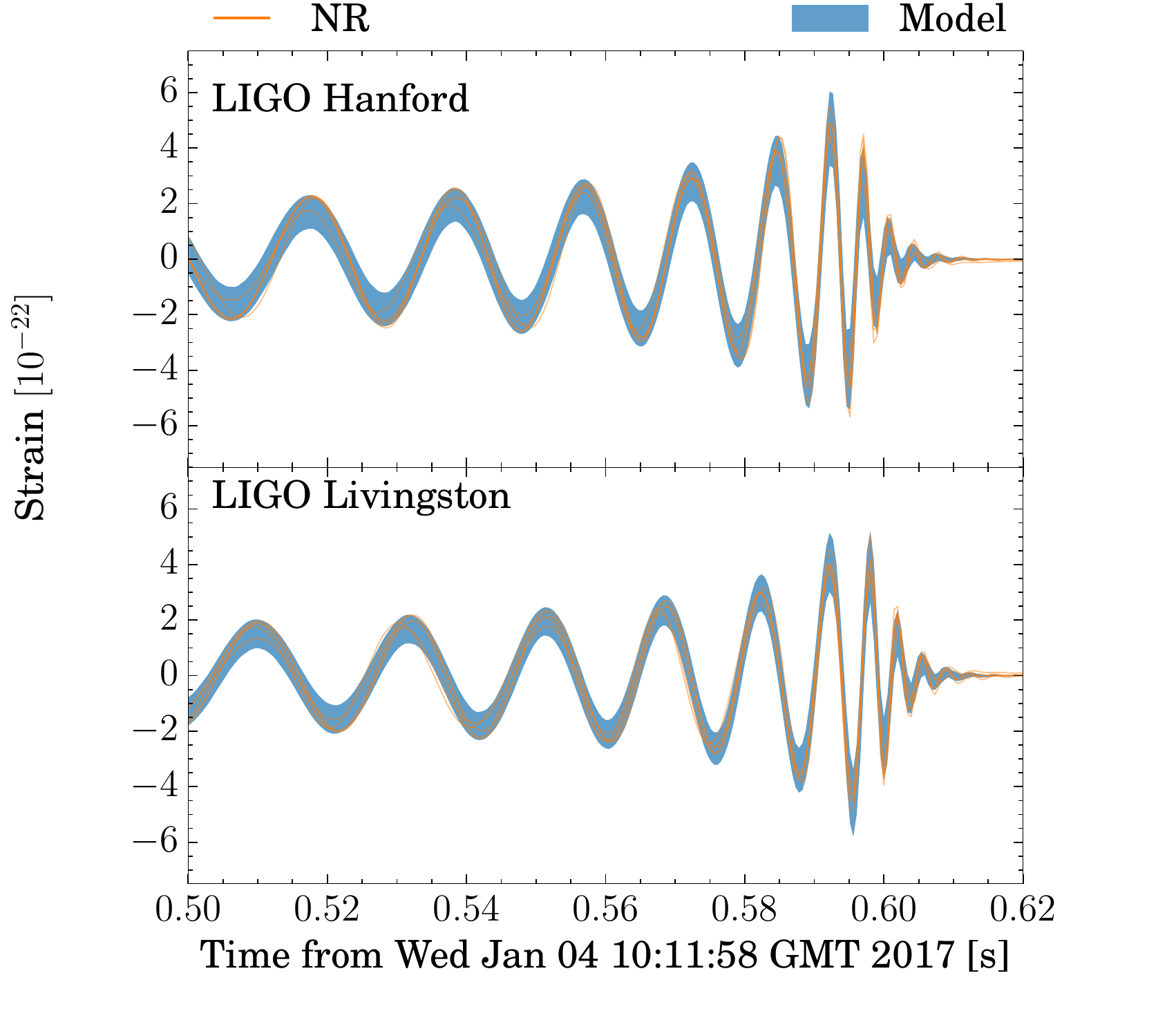}
  \caption{Comparison of the 90\% confidence intervals of GW170104
    from the two precessing models with the computer simulations of
    black hole mergers (in orange) from the best-fitting NR simulations listed in Table \ref{tab:lnLmargPeak:OtherSims}.
  \label{fig:CI}}
\end{figure}

The analysis above -- a difference in $\ln {\cal L}$ for models that should represent the same physical binary which is
comparable to the expected range of $\ln {\cal L}_{\rm marg}$ over the posterior -- suggests modest tension can exist between our NR simulations and the models used to draw inferences
about GW170104.  To illustrate this tension, in Fig.~\ref{fig:CI} we display the 90\% confidence intervals
of the precessing follow up cases (\#2 and \#3) computed by the two
approximate/phenomenological models comparing them with the full numerical simulations
(RIT's with N100 resolutions, note that increasing the numerical resolutions
to N118 and N140 reinforces this point).  
For each simulation, the waveform is generated by first fixing the  total mass --  selected by maximizing ${\cal L}$ --
and then choosing extrinsic parameters which maximize the likelihood.
At merger, these reconstructed waveforms appear to be in modest tension with the confidence interval reported for
$h(t)$; for example, the peaks and troughs of the yellow (NR) curves are consistently at the boundaries of what the 90\%
credible intervals derived from waveforms allow.  This
illustration, however, relies on a non representative metric 
to assess waveform similarity (i.e, differences in the  GW strain as a function of time, without reference to detector
sensitivity, assessed by eye).

\begin{figure}
\includegraphics[width=\columnwidth]{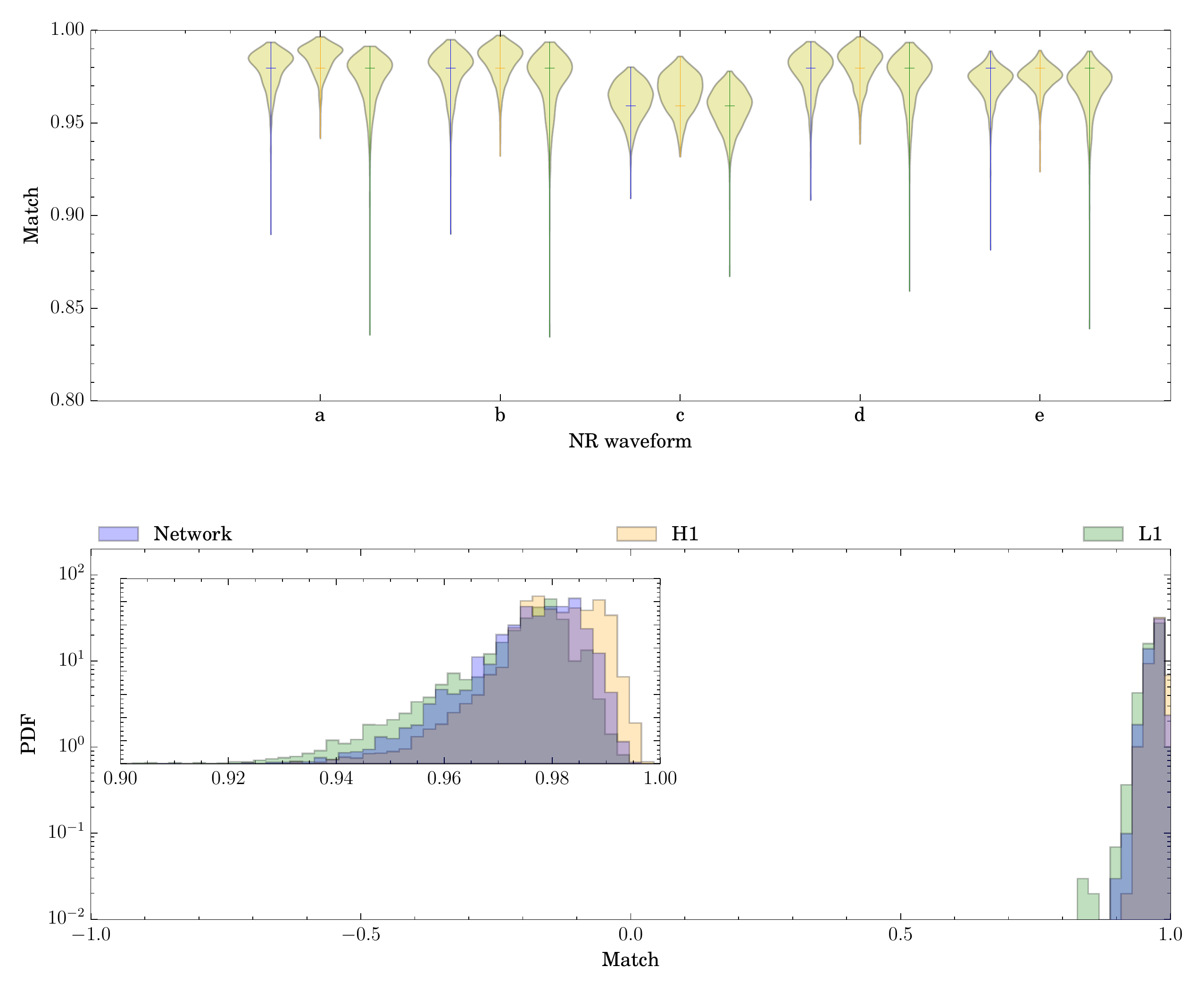}
\caption{\label{fig:OverlapViolinAndCum} Distribution of the overall match between each NR waveform (a,b,c,d,e) listed
  in Table \ref{tab:lnLmargPeak:OtherSims} and the (distribution of) waveforms produced by model-based parameter
  inference, as reported in \cite{Abbott:2017vtc}.  Matches are produced for the signal in H1, L1, and overall.  If an NR signal is perfectly consistent with these models for some
  parameters, then the distribution of matches will be well-approximated by a $\chi^2$ distribution with $d-1$ degrees
  of freedom.  Many of the best-fitting NR simulations have a distribution of matches that is significantly offset
  relative to this expected distribution, reflecting the mild tension shown in Figure \ref{fig:CI}.
}
\end{figure}

To remedy this deficiency, Figure \ref{fig:OverlapViolinAndCum} uses the match to compare our reconstructed NR waveforms with
reconstructed waveforms drawn from the posterior parameter distribution of GW170104.   The top panel uses a violin plot
to illustrate the distribution of matches, with a solid bar showing the median value.  The bottom panel shows a sample
cumulative distribution.   The median and maximum of this distribution provides a measure of how consistent the $h(t)$
estimate via NR is with the distribution provided by the model.  Using the maximum likelihood waveform from the
model and posterior, these distributions should be proportional to a (centrally) $\chi^2$ distributed quantity, with
median mismatch $N/2\rho^2$ for $N$ the number of model degrees of freedom and $\rho$ the signal to noise ratio (SNR),
where the specific choice for $\rho$  depends on the signal and detector/network being studied (e.g., for GW170104, the
network SNR was $\simeq 13$) .  By contrast, in several
of these overlap distributions, the peak and median values are manifestly offset downward, supporting a significant
systematic difference between the radiation predicted from our approximate models and our NR waveforms, each generated
from targeted NR followup simulations using parameters drawn from these selfsame model parameter distributions.

\subsection{Discussion}\label{sec:Discsussion}

Using comparisons to data via $\ln{\cal L}$ as our guide, we found in Section \ref{sec:sub:lnL} that model-based and NR-based analyses seem to have
maxima (in $\ln {\cal L}$) in different parts of parameter space; see Appendix  \ref{ap:DifferentTargets} for greater detail.  In the region identified as a good fit by model-based
analysis, corresponding NR simulations have a low $\ln {\cal L}$.  Conversely, several NR simulations with distinctly
different parameters had a larger $\ln {\cal L}$ than the corresponding targeted NR simulations \emph{and} model-based
comparisons evaluated at the same parameters.   The two functions $\ln{\cal L}$, evaluated using models and NR on
parameters designed to be similar to and representative of plausible parameters for GW170104, do not agree, implying
systematic differences between models and NR (i.e., a change in $\ln {\cal L}\simeq 2$).   
While we have for simplicity adopted one procedure which identifies candidate parameters to select our followup
simulations, we emphasize that the specific procedure is largely arbitrary, as in this work we simply  demonstrate the
two marginalized likelihoods (NR and model-based) disagree somewhere.  Changes in the exact location and value of the marginalized
likelihood are of less interest changes in the full posterior distribution; the latter subject is beyond the scope of
this study.

The NR followup simulations and Bayesian inferences used in this work were performed soon after the
identification of GW170104, and as such did not benefit from recent improvements in waveform modeling.  
Notably, by calibrating to a large suite of numerical relativity simulations, surrogate waveform models have been
generated that, in a suitable part of parameter space, are markedly superior to any of the waveform models used for parameter inference to date
\cite{Blackman:2015pia,Blackman:2017dfb}.     
Parameter inferences performed with these models should be more reliable and (by optimizing $\ln {\cal L}$) enable
better targets for NR followup simulations.

For simplicity and brevity, we have directly compared our nonprecessing and precessing simulations to only one of the
two extant families of phenomenological waveform models (SEOBv3/v4).   While the two models are in good agreement for
nonprecessing binaries, the other model (IMRP) has technical complications that limit its utility for our study.  On the
one hand, we cannot generate a similar
waveform with similar initial conditions,
 preventing us from performing the straightforward comparisons shown in Figure \ref{fig:UID2_match}.  [As a frequency domain model, it did not  adopt the same time conventions as NR and time-domain models
   for the precession phase  (see,e.g., Williamson et al 2017 \cite{Williamson:2017evr}).]    
On the other hand, the implementations available do not provide a spin-weighted
spherical harmonic decomposition, preventing us from performing the mode-by-mode mismatch calculations in Table
\ref{tab:match_uid3}. 

Previous investigations have demonstrated by example that posterior inferences with approximate waveform models can be
biased, even for parameters consistent with observed binary black hole \cite{Williamson:2017evr,LIGO-O1-PENR-Systematics}.
For example, a previous large study using simulations consistent with GW150914 found that, despite the brevity and
relative simplicity of its signal, the inferred parameters could be biased for certain binary configurations relative to
the line of sight
\cite{LIGO-O1-PENR-Systematics}, and much less so for others (e.g., nonprecessing and comparable-mass binaries).   The relevance and frequency of these  configurations is not yet determined and
depends on the binary black hole population which nature provides.

\mysec{Conclusions}\label{sec:Conclusions}

After the detection of GW170104  \cite{Abbott:2017vtc}, we performed several simulations of binary black hole mergers, intending to reproduce
LIGO's observations using simulations with similar parameters.   The parameters used were selected based on LIGO's reported
inferences about GW170104, generated by comparing two approximate models for binary black hole merger to the GW170104
data.     Comparing these  targeted simulations of binary black hole mergers, we find good agreement.
We have shown that the differences among typical numerical simulations, used
as a measure of their error, is much smaller (by over an order of magnitude)
than the residuals of observation versus theory.   
On the other hand, we demonstrate (expected) differences between our numerical solutions to general relativity and the approximate models
used to target our simulations.    Because we used these models to identify candidate parameters for followup, our
followup simulations were systematically biased away from the best-fitting parameters.  
These biases are not surprising, as the models used do not fully incorporate all the physics of binary merger, including
higher modes and all features of precession, and are known to modestly disagree both with one another and with NR simulations.  
This does not mean that the models are not recovering the full signal: 
both models and NR could find similar likelihoods, but for different
parameters. These bias can be particularly large for small mass ratios and
highly spinning precessing binaries.
We demonstrate that other,
pre-existing simulations with different parameters fit the data substantially better than the configurations targeted by
model-based techniques.   

We have shown here (and in previous studies
\cite{Bustillo:2015ova,Lange:2017wki})
that the standard low resolution, fast-response, simulations
provide an accurate description of GW signals, and can improve over 
the parameters determined by the models 
(See Table \ref{tab:lnLmargPeak:OtherSims} and Fig.~\ref{fig:LnL_phi})
for precessing and non-precessing cases (note that while SEOBNRv4 improves
on the inaccurate~\cite{LovelaceLousto2016}
SEOBNRv2~\cite{Taracchini:2013rva}, it is still not at comparable
accuracy to the NR simulations, See 
Figs.~\ref{fig:UID1_match}-\ref{fig:UID2_match}, for instance).
The tension between the models
and the full numerical simulations
(notwithstanding \cite{Abbott:2016wiq}) 
may be crucial in determining parameters
such as individual spin of the holes and tests of general relativity
for the large SNR signals, where the limitations of the models is larger).
Both this study, focused on 
GW170104, and the investigation by  \cite{Williamson:2017evr}, carried out on  GW151226,  point to the limitations of existing models to accurately determine
 binary parameters in the case of precessing BBH.

\begin{figure}
  \includegraphics[width=0.9\columnwidth]{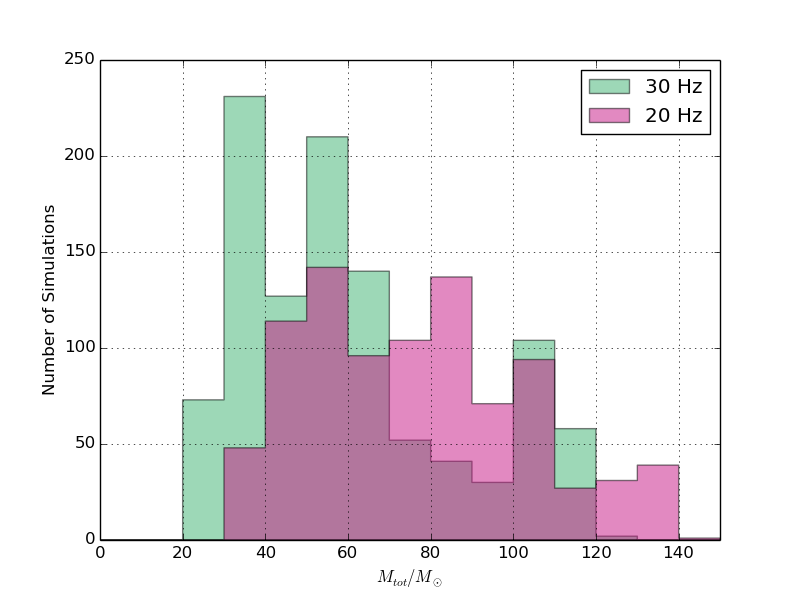}
  \includegraphics[width=0.9\columnwidth]{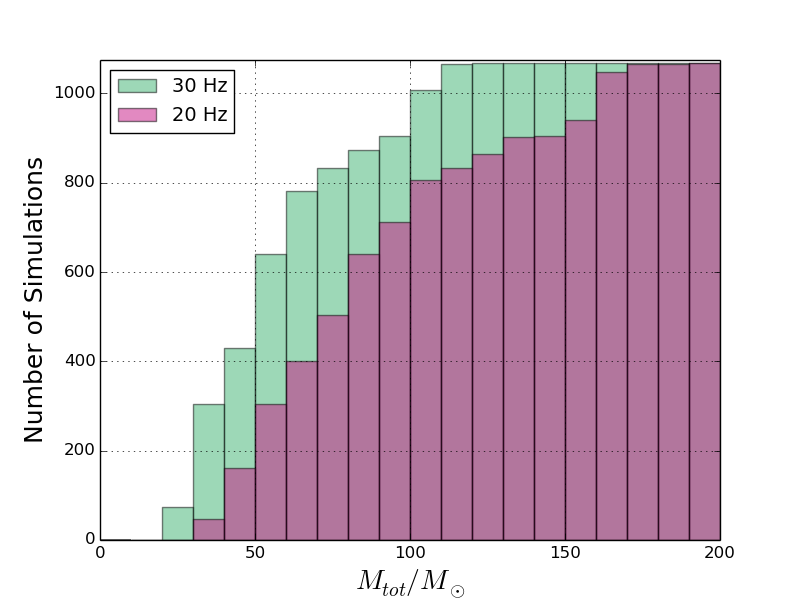}
  \caption{This plot shows histograms of the $\#$ of simulations in a given total mass bin of size $10M_\odot$, for starting at 20Hz and for 
    starting at 30Hz with the given total mass.
    This is for all the runs in the public SXS+GT+RIT Catalogs~\cite{Mroue:2013xna, Jani:2016wkt, Healy:2017psd}.
    The bottom histograms shows how many simulations in the catalogs can be used from 20Hz or 30Hz from a minimal mass on, i.e. the cumulative of the upper
    plot.
  \label{fig:mtotal}}
\end{figure}

Regarding prospects for future followups,
Figure~\ref{fig:mtotal} shows the distributions of the minimal total mass of the 
BBH systems in the NR catalogs \cite{Mroue:2013xna, Jani:2016wkt, Healy:2017psd} given a starting gravitational wave frequency
of 20 or 30 Hz in the source frame and its cumulative.
This provides a coverage for the current events observed by LIGO (redshift
effects improve this coverage by a factor of $(1+z)$, where $z$ is the
redshift). Coverage
of lower total masses would require longer simulations or hybridization
of the current waveforms.

Finally, we demonstrated the power of using purely 
numerical waveforms to determine parameters of a binary black hole merger
as the previous case of GW150914 \cite{Abbott:2016apu} and similarly
in the case of the source GW170104. More work is needed though to systematically
and robustly include hybridization of waveforms and the case of generically
precessing binaries \cite{Lange:2017wki}.

\mysec{Acknowledgments}
The authors thank S. Husa,  C. Berry, K.Chatziioannou, and A. Zimmerman for their feedback on this work. 
The RIT authors gratefully acknowledge the NSF for financial support from Grants
PHY-1607520, No.\ PHY-1707946, No.\ ACI-1550436, No.\ AST-1516150,
No.\ ACI-1516125, No.\ PHY-1726215.  This work used the Extreme Science and Engineering
Discovery Environment (XSEDE) [allocation TG-PHY060027N], which is
supported by NSF grant No. ACI-1548562.
Computational resources were also provided by the NewHorizons and
BlueSky Clusters at the Rochester Institute of Technology, which were
supported by NSF grants No.\ PHY-0722703, No.\ DMS-0820923, No.\
AST-1028087, and No.\ PHY-1229173. 
ROS is supported by NSF AST-1412449, PHY-1505629, and PHY-1607520.
The GT authors gratefully acknowledge NSF support through awards
PHY-1505824, 1505524, and XSEDE TG-PHY120016.
They also gratefully acknowledge support from Cullen-Peck and Dunn Families.
The SXS collaboration authors gratefully acknowledge the NSF for financial support from Grants: No. PHY-1307489, No. PHY-1606522, PHY-1606654, and AST- 1333129. They also  gratefully acknowledge support for this research at CITA from NSERC 
of Canada, the Ontario Early Researcher Awards Program, the Canada
Research Chairs Program, and the Canadian Institute for Advanced 
Research.  Calculations were done on the ORCA computer cluster, supported by NSF grant PHY-1429873, the Research Corporation for Science Advancement, CSU Fullerton, the GPC supercomputer at the SciNet HPC Consortium~\cite{scinet}; SciNet is funded by: the Canada Foundation for Innovation (CFI) under the
auspices of Compute Canada; the Government of Ontario; Ontario Research Fund (ORF) -- Research Excellence; and the University of Toronto. Further calculations were performed on the Briar\'ee cluster at Sherbrooke University, managed by Calcul Qu\'ebec and Compute Canada and with operation funded by the Canada Foundation for Innovation (CFI), Minist\'ere de l'\'Economie, de l'Innovation et des Exportations du Quebec (MEIE), RMGA and the Fonds de recherche du Qu\'ebec - Nature et Technologies (FRQ-NT).
PK gratefully acknowledges support from the Sherman Fairchild Foundation and NSF grants PHY-1606654 and AST- 1333129 at Cornell.
\bibliography{local.bib}

\begin{thebibliography}{110}%
\makeatletter
\providecommand \@ifxundefined [1]{%
 \@ifx{#1\undefined}
}%
\providecommand \@ifnum [1]{%
 \ifnum #1\expandafter \@firstoftwo
 \else \expandafter \@secondoftwo
 \fi
}%
\providecommand \@ifx [1]{%
 \ifx #1\expandafter \@firstoftwo
 \else \expandafter \@secondoftwo
 \fi
}%
\providecommand \natexlab [1]{#1}%
\providecommand \enquote  [1]{``#1''}%
\providecommand \bibnamefont  [1]{#1}%
\providecommand \bibfnamefont [1]{#1}%
\providecommand \citenamefont [1]{#1}%
\providecommand \href@noop [0]{\@secondoftwo}%
\providecommand \href [0]{\begingroup \@sanitize@url \@href}%
\providecommand \@href[1]{\@@startlink{#1}\@@href}%
\providecommand \@@href[1]{\endgroup#1\@@endlink}%
\providecommand \@sanitize@url [0]{\catcode `\\12\catcode `\$12\catcode
  `\&12\catcode `\#12\catcode `\^12\catcode `\_12\catcode `\%12\relax}%
\providecommand \@@startlink[1]{}%
\providecommand \@@endlink[0]{}%
\providecommand \url  [0]{\begingroup\@sanitize@url \@url }%
\providecommand \@url [1]{\endgroup\@href {#1}{\urlprefix }}%
\providecommand \urlprefix  [0]{URL }%
\providecommand \Eprint [0]{\href }%
\providecommand \doibase [0]{http://dx.doi.org/}%
\providecommand \selectlanguage [0]{\@gobble}%
\providecommand \bibinfo  [0]{\@secondoftwo}%
\providecommand \bibfield  [0]{\@secondoftwo}%
\providecommand \translation [1]{[#1]}%
\providecommand \BibitemOpen [0]{}%
\providecommand \bibitemStop [0]{}%
\providecommand \bibitemNoStop [0]{.\EOS\space}%
\providecommand \EOS [0]{\spacefactor3000\relax}%
\providecommand \BibitemShut  [1]{\csname bibitem#1\endcsname}%
\let\auto@bib@innerbib\@empty
\bibitem [{\citenamefont {Abbott}\ \emph
  {et~al.}(2016{\natexlab{a}})\citenamefont {Abbott} \emph
  {et~al.}}]{LIGOVirgo2016a}%
  \BibitemOpen
  \bibfield  {author} {\bibinfo {author} {\bibfnamefont {B.~P.}\ \bibnamefont
  {Abbott}} \emph {et~al.} (\bibinfo {collaboration} {LIGO Scientific
  Collaboration, Virgo Collaboration}),\ }\href {\doibase
  10.1103/PhysRevLett.116.061102} {\bibfield  {journal} {\bibinfo  {journal}
  {Phys.\ Rev.\ Lett.}\ }\textbf {\bibinfo {volume} {116}},\ \bibinfo {pages}
  {061102} (\bibinfo {year} {2016}{\natexlab{a}})},\ \Eprint
  {http://arxiv.org/abs/1602.03837} {arXiv:1602.03837 [gr-qc]} \BibitemShut
  {NoStop}%
\bibitem [{\citenamefont {Abbott}\ \emph
  {et~al.}(2016{\natexlab{b}})\citenamefont {Abbott} \emph
  {et~al.}}]{Abbott:2016nmj}%
  \BibitemOpen
  \bibfield  {author} {\bibinfo {author} {\bibfnamefont {B.~P.}\ \bibnamefont
  {Abbott}} \emph {et~al.} (\bibinfo {collaboration} {LIGO Scientific
  Collaboration, Virgo Collaboration}),\ }\href {\doibase
  10.1103/PhysRevLett.116.241103} {\bibfield  {journal} {\bibinfo  {journal}
  {Phys. Rev. Lett.}\ }\textbf {\bibinfo {volume} {116}},\ \bibinfo {pages}
  {241103} (\bibinfo {year} {2016}{\natexlab{b}})},\ \Eprint
  {http://arxiv.org/abs/1606.04855} {arXiv:1606.04855 [gr-qc]} \BibitemShut
  {NoStop}%
\bibitem [{\citenamefont {Abbott}\ \emph
  {et~al.}(2016{\natexlab{c}})\citenamefont {Abbott} \emph
  {et~al.}}]{TheLIGOScientific:2016pea}%
  \BibitemOpen
  \bibfield  {author} {\bibinfo {author} {\bibfnamefont {B.~P.}\ \bibnamefont
  {Abbott}} \emph {et~al.} (\bibinfo {collaboration} {Virgo, LIGO
  Scientific}),\ }\href {\doibase 10.1103/PhysRevX.6.041015} {\bibfield
  {journal} {\bibinfo  {journal} {Phys. Rev.}\ }\textbf {\bibinfo {volume}
  {X6}},\ \bibinfo {pages} {041015} (\bibinfo {year} {2016}{\natexlab{c}})},\
  \Eprint {http://arxiv.org/abs/1606.04856} {arXiv:1606.04856 [gr-qc]}
  \BibitemShut {NoStop}%
\bibitem [{\citenamefont {{Abbott}}\ \emph {et~al.}(2017)\citenamefont
  {{Abbott}}, \citenamefont {{Abbott}}, \citenamefont {{Abbott}}, \citenamefont
  {{Acernese}}, \citenamefont {{Ackley}}, \citenamefont {{Adams}},
  \citenamefont {{Adams}}, \citenamefont {{Addesso}}, \citenamefont
  {{Adhikari}}, \citenamefont {{Adya}},\ and\ \citenamefont
  {et~al.}}]{2017PhRvL.118v1101A}%
  \BibitemOpen
  \bibfield  {author} {\bibinfo {author} {\bibfnamefont {B.~P.}\ \bibnamefont
  {{Abbott}}}, \bibinfo {author} {\bibfnamefont {R.}~\bibnamefont {{Abbott}}},
  \bibinfo {author} {\bibfnamefont {T.~D.}\ \bibnamefont {{Abbott}}}, \bibinfo
  {author} {\bibfnamefont {F.}~\bibnamefont {{Acernese}}}, \bibinfo {author}
  {\bibfnamefont {K.}~\bibnamefont {{Ackley}}}, \bibinfo {author}
  {\bibfnamefont {C.}~\bibnamefont {{Adams}}}, \bibinfo {author} {\bibfnamefont
  {T.}~\bibnamefont {{Adams}}}, \bibinfo {author} {\bibfnamefont
  {P.}~\bibnamefont {{Addesso}}}, \bibinfo {author} {\bibfnamefont {R.~X.}\
  \bibnamefont {{Adhikari}}}, \bibinfo {author} {\bibfnamefont {V.~B.}\
  \bibnamefont {{Adya}}}, \ and\ \bibinfo {author} {\bibnamefont {et~al.}},\
  }\href {\doibase 10.1103/PhysRevLett.118.221101} {\bibfield  {journal}
  {\bibinfo  {journal} {Physical Review Letters}\ }\textbf {\bibinfo {volume}
  {118}},\ \bibinfo {eid} {221101} (\bibinfo {year} {2017})},\ \Eprint
  {http://arxiv.org/abs/1706.01812} {arXiv:1706.01812 [gr-qc]} \BibitemShut
  {NoStop}%
\bibitem [{\citenamefont {Abbott}\ \emph
  {et~al.}(2017{\natexlab{a}})\citenamefont {Abbott} \emph
  {et~al.}}]{Abbott:2017gyy}%
  \BibitemOpen
  \bibfield  {author} {\bibinfo {author} {\bibfnamefont {B.~P.}\ \bibnamefont
  {Abbott}} \emph {et~al.} (\bibinfo {collaboration} {Virgo, LIGO
  Scientific}),\ }\href@noop {} {\  (\bibinfo {year} {2017}{\natexlab{a}})},\
  \Eprint {http://arxiv.org/abs/1711.05578} {arXiv:1711.05578 [astro-ph.HE]}
  \BibitemShut {NoStop}%
\bibitem [{\citenamefont {Abbott}\ \emph
  {et~al.}(2017{\natexlab{b}})\citenamefont {Abbott} \emph
  {et~al.}}]{Abbott:2017oio}%
  \BibitemOpen
  \bibfield  {author} {\bibinfo {author} {\bibfnamefont {B.~P.}\ \bibnamefont
  {Abbott}} \emph {et~al.} (\bibinfo {collaboration} {Virgo, LIGO
  Scientific}),\ }\href {\doibase 10.1103/PhysRevLett.119.141101} {\bibfield
  {journal} {\bibinfo  {journal} {Phys. Rev. Lett.}\ }\textbf {\bibinfo
  {volume} {119}},\ \bibinfo {pages} {141101} (\bibinfo {year}
  {2017}{\natexlab{b}})},\ \Eprint {http://arxiv.org/abs/1709.09660}
  {arXiv:1709.09660 [gr-qc]} \BibitemShut {NoStop}%
\bibitem [{\citenamefont {{Taracchini}}\ \emph {et~al.}(2014)\citenamefont
  {{Taracchini}}, \citenamefont {{Buonanno}}, \citenamefont {{Pan}},
  \citenamefont {{Hinderer}}, \citenamefont {{Boyle}}, \citenamefont
  {{Hemberger}}, \citenamefont {{Kidder}}, \citenamefont {{Lovelace}},
  \citenamefont {{Mrou{\'e}}}, \citenamefont {{Pfeiffer}}, \citenamefont
  {{Scheel}}, \citenamefont {{Szil{\'a}gyi}}, \citenamefont {{Taylor}},\ and\
  \citenamefont {{Zenginoglu}}}]{2014PhRvD..89f1502T}%
  \BibitemOpen
  \bibfield  {author} {\bibinfo {author} {\bibfnamefont {A.}~\bibnamefont
  {{Taracchini}}}, \bibinfo {author} {\bibfnamefont {A.}~\bibnamefont
  {{Buonanno}}}, \bibinfo {author} {\bibfnamefont {Y.}~\bibnamefont {{Pan}}},
  \bibinfo {author} {\bibfnamefont {T.}~\bibnamefont {{Hinderer}}}, \bibinfo
  {author} {\bibfnamefont {M.}~\bibnamefont {{Boyle}}}, \bibinfo {author}
  {\bibfnamefont {D.~A.}\ \bibnamefont {{Hemberger}}}, \bibinfo {author}
  {\bibfnamefont {L.~E.}\ \bibnamefont {{Kidder}}}, \bibinfo {author}
  {\bibfnamefont {G.}~\bibnamefont {{Lovelace}}}, \bibinfo {author}
  {\bibfnamefont {A.~H.}\ \bibnamefont {{Mrou{\'e}}}}, \bibinfo {author}
  {\bibfnamefont {H.~P.}\ \bibnamefont {{Pfeiffer}}}, \bibinfo {author}
  {\bibfnamefont {M.~A.}\ \bibnamefont {{Scheel}}}, \bibinfo {author}
  {\bibfnamefont {B.}~\bibnamefont {{Szil{\'a}gyi}}}, \bibinfo {author}
  {\bibfnamefont {N.~W.}\ \bibnamefont {{Taylor}}}, \ and\ \bibinfo {author}
  {\bibfnamefont {A.}~\bibnamefont {{Zenginoglu}}},\ }\href {\doibase
  10.1103/PhysRevD.89.061502} {\bibfield  {journal} {\bibinfo  {journal}
  {\prd}\ }\textbf {\bibinfo {volume} {89}},\ \bibinfo {eid} {061502} (\bibinfo
  {year} {2014})},\ \Eprint {http://arxiv.org/abs/1311.2544} {arXiv:1311.2544
  [gr-qc]} \BibitemShut {NoStop}%
\bibitem [{\citenamefont {P{\"u}rrer}(2014)}]{Purrer:2014}%
  \BibitemOpen
  \bibfield  {author} {\bibinfo {author} {\bibfnamefont {M.}~\bibnamefont
  {P{\"u}rrer}},\ }\href {http://stacks.iop.org/0264-9381/31/i=19/a=195010}
  {\bibfield  {journal} {\bibinfo  {journal} {Class.\ Quantum Grav.}\ }\textbf
  {\bibinfo {volume} {31}},\ \bibinfo {pages} {195010} (\bibinfo {year}
  {2014})},\ \Eprint {http://arxiv.org/abs/1402.4146} {arXiv:1402.4146 [gr-qc]}
  \BibitemShut {NoStop}%
\bibitem [{\citenamefont {Hannam}\ \emph {et~al.}(2014)\citenamefont {Hannam},
  \citenamefont {Schmidt}, \citenamefont {Boh{\' e}}, \citenamefont {Haegel},
  \citenamefont {Husa} \emph {et~al.}}]{Hannam:2013oca}%
  \BibitemOpen
  \bibfield  {author} {\bibinfo {author} {\bibfnamefont {M.}~\bibnamefont
  {Hannam}}, \bibinfo {author} {\bibfnamefont {P.}~\bibnamefont {Schmidt}},
  \bibinfo {author} {\bibfnamefont {A.}~\bibnamefont {Boh{\' e}}}, \bibinfo
  {author} {\bibfnamefont {L.}~\bibnamefont {Haegel}}, \bibinfo {author}
  {\bibfnamefont {S.}~\bibnamefont {Husa}},  \emph {et~al.},\ }\href {\doibase
  10.1103/PhysRevLett.113.151101} {\bibfield  {journal} {\bibinfo  {journal}
  {Phys.\ Rev.\ Lett.}\ }\textbf {\bibinfo {volume} {113}},\ \bibinfo {pages}
  {151101} (\bibinfo {year} {2014})},\ \Eprint {http://arxiv.org/abs/1308.3271}
  {arXiv:1308.3271 [gr-qc]} \BibitemShut {NoStop}%
\bibitem [{\citenamefont {{Abbott et al. (The LIGO Scientific Collaboration and
  the Virgo Collaboration)}}(2016)}]{NRPaper}%
  \BibitemOpen
  \bibfield  {author} {\bibinfo {author} {\bibfnamefont {B.}~\bibnamefont
  {{Abbott et al. (The LIGO Scientific Collaboration and the Virgo
  Collaboration)}}},\ }\href {\doibase 10.1103/PhysRevD.94.064035} {\bibfield
  {journal} {\bibinfo  {journal} {\prd}\ }\textbf {\bibinfo {volume} {94}},\
  \bibinfo {pages} {064035} (\bibinfo {year} {2016})}\BibitemShut {NoStop}%
\bibitem [{\citenamefont {{Lange}}\ \emph {et~al.}(2017)\citenamefont
  {{Lange}}, \citenamefont {{O'Shaughnessy}}, \citenamefont {{Boyle}},
  \citenamefont {{Calder{\'o}n Bustillo}}, \citenamefont {{Campanelli}},
  \citenamefont {{Chu}}, \citenamefont {{Clark}}, \citenamefont {{Demos}},
  \citenamefont {{Fong}}, \citenamefont {{Healy}}, \citenamefont {{Hemberger}},
  \citenamefont {{Hinder}}, \citenamefont {{Jani}}, \citenamefont {{Khamesra}},
  \citenamefont {{Kidder}}, \citenamefont {{Kumar}}, \citenamefont {{Laguna}},
  \citenamefont {{Lousto}}, \citenamefont {{Lovelace}}, \citenamefont
  {{Ossokine}}, \citenamefont {{Pfeiffer}}, \citenamefont {{Scheel}},
  \citenamefont {{Shoemaker}}, \citenamefont {{Szilagyi}}, \citenamefont
  {{Teukolsky}},\ and\ \citenamefont {{Zlochower}}}]{Lange2017}%
  \BibitemOpen
  \bibfield  {author} {\bibinfo {author} {\bibfnamefont {J.}~\bibnamefont
  {{Lange}}}, \bibinfo {author} {\bibfnamefont {R.}~\bibnamefont
  {{O'Shaughnessy}}}, \bibinfo {author} {\bibfnamefont {M.}~\bibnamefont
  {{Boyle}}}, \bibinfo {author} {\bibfnamefont {J.}~\bibnamefont {{Calder{\'o}n
  Bustillo}}}, \bibinfo {author} {\bibfnamefont {M.}~\bibnamefont
  {{Campanelli}}}, \bibinfo {author} {\bibfnamefont {T.}~\bibnamefont {{Chu}}},
  \bibinfo {author} {\bibfnamefont {J.~A.}\ \bibnamefont {{Clark}}}, \bibinfo
  {author} {\bibfnamefont {N.}~\bibnamefont {{Demos}}}, \bibinfo {author}
  {\bibfnamefont {H.}~\bibnamefont {{Fong}}}, \bibinfo {author} {\bibfnamefont
  {J.}~\bibnamefont {{Healy}}}, \bibinfo {author} {\bibfnamefont
  {D.}~\bibnamefont {{Hemberger}}}, \bibinfo {author} {\bibfnamefont
  {I.}~\bibnamefont {{Hinder}}}, \bibinfo {author} {\bibfnamefont
  {K.}~\bibnamefont {{Jani}}}, \bibinfo {author} {\bibfnamefont
  {B.}~\bibnamefont {{Khamesra}}}, \bibinfo {author} {\bibfnamefont {L.~E.}\
  \bibnamefont {{Kidder}}}, \bibinfo {author} {\bibfnamefont {P.}~\bibnamefont
  {{Kumar}}}, \bibinfo {author} {\bibfnamefont {P.}~\bibnamefont {{Laguna}}},
  \bibinfo {author} {\bibfnamefont {C.~O.}\ \bibnamefont {{Lousto}}}, \bibinfo
  {author} {\bibfnamefont {G.}~\bibnamefont {{Lovelace}}}, \bibinfo {author}
  {\bibfnamefont {S.}~\bibnamefont {{Ossokine}}}, \bibinfo {author}
  {\bibfnamefont {H.}~\bibnamefont {{Pfeiffer}}}, \bibinfo {author}
  {\bibfnamefont {M.~A.}\ \bibnamefont {{Scheel}}}, \bibinfo {author}
  {\bibfnamefont {D.}~\bibnamefont {{Shoemaker}}}, \bibinfo {author}
  {\bibfnamefont {B.}~\bibnamefont {{Szilagyi}}}, \bibinfo {author}
  {\bibfnamefont {S.}~\bibnamefont {{Teukolsky}}}, \ and\ \bibinfo {author}
  {\bibfnamefont {Y.}~\bibnamefont {{Zlochower}}},\ }\href@noop {} {\bibfield
  {journal} {\bibinfo  {journal} {ArXiv e-prints}\ } (\bibinfo {year}
  {2017})},\ \Eprint {http://arxiv.org/abs/1705.09833} {arXiv:1705.09833
  [gr-qc]} \BibitemShut {NoStop}%
\bibitem [{\citenamefont {Pretorius}(2005)}]{Pretorius:2005gq}%
  \BibitemOpen
  \bibfield  {author} {\bibinfo {author} {\bibfnamefont {F.}~\bibnamefont
  {Pretorius}},\ }\href {\doibase 10.1103/PhysRevLett.95.121101} {\bibfield
  {journal} {\bibinfo  {journal} {Phys. Rev. Lett.}\ }\textbf {\bibinfo
  {volume} {95}},\ \bibinfo {pages} {121101} (\bibinfo {year} {2005})},\
  \Eprint {http://arxiv.org/abs/gr-qc/0507014} {arXiv:gr-qc/0507014 [gr-qc]}
  \BibitemShut {NoStop}%
\bibitem [{\citenamefont {Campanelli}\ \emph {et~al.}(2006)\citenamefont
  {Campanelli}, \citenamefont {Lousto}, \citenamefont {Marronetti},\ and\
  \citenamefont {Zlochower}}]{Campanelli:2005dd}%
  \BibitemOpen
  \bibfield  {author} {\bibinfo {author} {\bibfnamefont {M.}~\bibnamefont
  {Campanelli}}, \bibinfo {author} {\bibfnamefont {C.~O.}\ \bibnamefont
  {Lousto}}, \bibinfo {author} {\bibfnamefont {P.}~\bibnamefont {Marronetti}},
  \ and\ \bibinfo {author} {\bibfnamefont {Y.}~\bibnamefont {Zlochower}},\
  }\href {\doibase 10.1103/PhysRevLett.96.111101} {\bibfield  {journal}
  {\bibinfo  {journal} {Phys. Rev. Lett.}\ }\textbf {\bibinfo {volume} {96}},\
  \bibinfo {pages} {111101} (\bibinfo {year} {2006})},\ \Eprint
  {http://arxiv.org/abs/gr-qc/0511048} {arXiv:gr-qc/0511048 [gr-qc]}
  \BibitemShut {NoStop}%
\bibitem [{\citenamefont {Baker}\ \emph {et~al.}(2006)\citenamefont {Baker},
  \citenamefont {Centrella}, \citenamefont {Choi}, \citenamefont {Koppitz},\
  and\ \citenamefont {van Meter}}]{Baker:2005vv}%
  \BibitemOpen
  \bibfield  {author} {\bibinfo {author} {\bibfnamefont {J.~G.}\ \bibnamefont
  {Baker}}, \bibinfo {author} {\bibfnamefont {J.}~\bibnamefont {Centrella}},
  \bibinfo {author} {\bibfnamefont {D.-I.}\ \bibnamefont {Choi}}, \bibinfo
  {author} {\bibfnamefont {M.}~\bibnamefont {Koppitz}}, \ and\ \bibinfo
  {author} {\bibfnamefont {J.}~\bibnamefont {van Meter}},\ }\href {\doibase
  10.1103/PhysRevLett.96.111102} {\bibfield  {journal} {\bibinfo  {journal}
  {Phys. Rev. Lett.}\ }\textbf {\bibinfo {volume} {96}},\ \bibinfo {pages}
  {111102} (\bibinfo {year} {2006})},\ \Eprint
  {http://arxiv.org/abs/gr-qc/0511103} {arXiv:gr-qc/0511103 [gr-qc]}
  \BibitemShut {NoStop}%
\bibitem [{\citenamefont {Mroue}\ \emph {et~al.}(2013)\citenamefont {Mroue}
  \emph {et~al.}}]{Mroue:2013xna}%
  \BibitemOpen
  \bibfield  {author} {\bibinfo {author} {\bibfnamefont {A.~H.}\ \bibnamefont
  {Mroue}} \emph {et~al.},\ }\href {\doibase 10.1103/PhysRevLett.111.241104}
  {\bibfield  {journal} {\bibinfo  {journal} {Phys. Rev. Lett.}\ }\textbf
  {\bibinfo {volume} {111}},\ \bibinfo {pages} {241104} (\bibinfo {year}
  {2013})},\ \Eprint {http://arxiv.org/abs/1304.6077} {arXiv:1304.6077 [gr-qc]}
  \BibitemShut {NoStop}%
\bibitem [{\citenamefont {Jani}\ \emph {et~al.}(2016)\citenamefont {Jani},
  \citenamefont {Healy}, \citenamefont {Clark}, \citenamefont {London},
  \citenamefont {Laguna},\ and\ \citenamefont {Shoemaker}}]{Jani:2016wkt}%
  \BibitemOpen
  \bibfield  {author} {\bibinfo {author} {\bibfnamefont {K.}~\bibnamefont
  {Jani}}, \bibinfo {author} {\bibfnamefont {J.}~\bibnamefont {Healy}},
  \bibinfo {author} {\bibfnamefont {J.~A.}\ \bibnamefont {Clark}}, \bibinfo
  {author} {\bibfnamefont {L.}~\bibnamefont {London}}, \bibinfo {author}
  {\bibfnamefont {P.}~\bibnamefont {Laguna}}, \ and\ \bibinfo {author}
  {\bibfnamefont {D.}~\bibnamefont {Shoemaker}},\ }\href {\doibase
  10.1088/0264-9381/33/20/204001} {\bibfield  {journal} {\bibinfo  {journal}
  {Class. Quant. Grav.}\ }\textbf {\bibinfo {volume} {33}},\ \bibinfo {pages}
  {204001} (\bibinfo {year} {2016})},\ \Eprint
  {http://arxiv.org/abs/1605.03204} {arXiv:1605.03204 [gr-qc]} \BibitemShut
  {NoStop}%
\bibitem [{\citenamefont {Healy}\ \emph
  {et~al.}(2017{\natexlab{a}})\citenamefont {Healy}, \citenamefont {Lousto},
  \citenamefont {Zlochower},\ and\ \citenamefont {Campanelli}}]{Healy:2017psd}%
  \BibitemOpen
  \bibfield  {author} {\bibinfo {author} {\bibfnamefont {J.}~\bibnamefont
  {Healy}}, \bibinfo {author} {\bibfnamefont {C.~O.}\ \bibnamefont {Lousto}},
  \bibinfo {author} {\bibfnamefont {Y.}~\bibnamefont {Zlochower}}, \ and\
  \bibinfo {author} {\bibfnamefont {M.}~\bibnamefont {Campanelli}},\
  }\href@noop {} {\  (\bibinfo {year} {2017}{\natexlab{a}})},\ \Eprint
  {http://arxiv.org/abs/1703.03423} {arXiv:1703.03423 [gr-qc]} \BibitemShut
  {NoStop}%
\bibitem [{\citenamefont {Lange}\ \emph {et~al.}(2017)\citenamefont {Lange}
  \emph {et~al.}}]{Lange:2017wki}%
  \BibitemOpen
  \bibfield  {author} {\bibinfo {author} {\bibfnamefont {J.}~\bibnamefont
  {Lange}} \emph {et~al.},\ }\href@noop {} {\  (\bibinfo {year} {2017})},\
  \Eprint {http://arxiv.org/abs/1705.09833} {arXiv:1705.09833 [gr-qc]}
  \BibitemShut {NoStop}%
\bibitem [{\citenamefont {Healy}\ \emph {et~al.}(2014)\citenamefont {Healy},
  \citenamefont {Lousto},\ and\ \citenamefont {Zlochower}}]{Healy:2014yta}%
  \BibitemOpen
  \bibfield  {author} {\bibinfo {author} {\bibfnamefont {J.}~\bibnamefont
  {Healy}}, \bibinfo {author} {\bibfnamefont {C.~O.}\ \bibnamefont {Lousto}}, \
  and\ \bibinfo {author} {\bibfnamefont {Y.}~\bibnamefont {Zlochower}},\
  }\href@noop {} {\bibfield  {journal} {\bibinfo  {journal} {Phys.\ Rev.\ D}\
  }\textbf {\bibinfo {volume} {89}},\ \bibinfo {pages} {104052} (\bibinfo
  {year} {2014})},\ \Eprint {http://arxiv.org/abs/1406.7295} {arXiv:1406.7295
  [gr-qc]} \BibitemShut {NoStop}%
\bibitem [{\citenamefont {Ghosh}\ \emph {et~al.}(2016)\citenamefont {Ghosh}
  \emph {et~al.}}]{Ghosh:2016qgn}%
  \BibitemOpen
  \bibfield  {author} {\bibinfo {author} {\bibfnamefont {A.}~\bibnamefont
  {Ghosh}} \emph {et~al.},\ }\href {\doibase 10.1103/PhysRevD.94.021101}
  {\bibfield  {journal} {\bibinfo  {journal} {Phys. Rev.}\ }\textbf {\bibinfo
  {volume} {D94}},\ \bibinfo {pages} {021101} (\bibinfo {year} {2016})},\
  \Eprint {http://arxiv.org/abs/1602.02453} {arXiv:1602.02453 [gr-qc]}
  \BibitemShut {NoStop}%
\bibitem [{\citenamefont {Abbott}\ \emph
  {et~al.}(2016{\natexlab{d}})\citenamefont {Abbott} \emph
  {et~al.}}]{TheLIGOScientific:2016src}%
  \BibitemOpen
  \bibfield  {author} {\bibinfo {author} {\bibfnamefont {B.~P.}\ \bibnamefont
  {Abbott}} \emph {et~al.} (\bibinfo {collaboration} {LIGO Scientific
  Collaboration, Virgo Collaboration}),\ }\href@noop {} {\bibfield  {journal}
  {\bibinfo  {journal} {Phys.~Rev.~Lett.}\ }\textbf {\bibinfo {volume} {116}},\
  \bibinfo {pages} {221101} (\bibinfo {year} {2016}{\natexlab{d}})},\ \Eprint
  {http://arxiv.org/abs/1602.03841} {arXiv:1602.03841 [gr-qc]} \BibitemShut
  {NoStop}%
\bibitem [{\citenamefont {Zlochower}\ \emph {et~al.}(2005)\citenamefont
  {Zlochower}, \citenamefont {Baker}, \citenamefont {Campanelli},\ and\
  \citenamefont {Lousto}}]{Zlochower:2005bj}%
  \BibitemOpen
  \bibfield  {author} {\bibinfo {author} {\bibfnamefont {Y.}~\bibnamefont
  {Zlochower}}, \bibinfo {author} {\bibfnamefont {J.}~\bibnamefont {Baker}},
  \bibinfo {author} {\bibfnamefont {M.}~\bibnamefont {Campanelli}}, \ and\
  \bibinfo {author} {\bibfnamefont {C.}~\bibnamefont {Lousto}},\ }\href
  {\doibase 10.1103/PhysRevD.72.024021} {\bibfield  {journal} {\bibinfo
  {journal} {Phys.\ Rev.\ D}\ }\textbf {\bibinfo {volume} {72}},\ \bibinfo
  {pages} {024021} (\bibinfo {year} {2005})},\ \Eprint
  {http://arxiv.org/abs/gr-qc/0505055} {arXiv:gr-qc/0505055 [gr-qc]}
  \BibitemShut {NoStop}%
\bibitem [{\citenamefont {Marronetti}\ \emph {et~al.}(2008)\citenamefont
  {Marronetti}, \citenamefont {Tichy}, \citenamefont {Bruegmann}, \citenamefont
  {Gonzalez},\ and\ \citenamefont {Sperhake}}]{Marronetti:2007wz}%
  \BibitemOpen
  \bibfield  {author} {\bibinfo {author} {\bibfnamefont {P.}~\bibnamefont
  {Marronetti}}, \bibinfo {author} {\bibfnamefont {W.}~\bibnamefont {Tichy}},
  \bibinfo {author} {\bibfnamefont {B.}~\bibnamefont {Bruegmann}}, \bibinfo
  {author} {\bibfnamefont {J.}~\bibnamefont {Gonzalez}}, \ and\ \bibinfo
  {author} {\bibfnamefont {U.}~\bibnamefont {Sperhake}},\ }\href {\doibase
  10.1103/PhysRevD.77.064010} {\bibfield  {journal} {\bibinfo  {journal}
  {Phys.\ Rev.\ D}\ }\textbf {\bibinfo {volume} {77}},\ \bibinfo {pages}
  {064010} (\bibinfo {year} {2008})},\ \Eprint {http://arxiv.org/abs/0709.2160}
  {arXiv:0709.2160 [gr-qc]} \BibitemShut {NoStop}%
\bibitem [{\citenamefont {Lousto}\ and\ \citenamefont
  {Zlochower}(2008)}]{Lousto:2007rj}%
  \BibitemOpen
  \bibfield  {author} {\bibinfo {author} {\bibfnamefont {C.~O.}\ \bibnamefont
  {Lousto}}\ and\ \bibinfo {author} {\bibfnamefont {Y.}~\bibnamefont
  {Zlochower}},\ }\href {\doibase 10.1103/PhysRevD.77.024034} {\bibfield
  {journal} {\bibinfo  {journal} {Phys.\ Rev.\ D}\ }\textbf {\bibinfo {volume}
  {77}},\ \bibinfo {pages} {024034} (\bibinfo {year} {2008})},\ \Eprint
  {http://arxiv.org/abs/0711.1165} {arXiv:0711.1165 [gr-qc]} \BibitemShut
  {NoStop}%
\bibitem [{\citenamefont {{L{\"o}ffler}}\ \emph {et~al.}(2012)\citenamefont
  {{L{\"o}ffler}}, \citenamefont {{Faber}}, \citenamefont {{Bentivegna}},
  \citenamefont {{Bode}}, \citenamefont {{Diener}}, \citenamefont {{Haas}},
  \citenamefont {{Hinder}}, \citenamefont {{Mundim}}, \citenamefont {{Ott}},
  \citenamefont {{Schnetter}}, \citenamefont {{Allen}}, \citenamefont
  {{Campanelli}},\ and\ \citenamefont {{Laguna}}}]{Loffler:2011ay}%
  \BibitemOpen
  \bibfield  {author} {\bibinfo {author} {\bibfnamefont {F.}~\bibnamefont
  {{L{\"o}ffler}}}, \bibinfo {author} {\bibfnamefont {J.}~\bibnamefont
  {{Faber}}}, \bibinfo {author} {\bibfnamefont {E.}~\bibnamefont
  {{Bentivegna}}}, \bibinfo {author} {\bibfnamefont {T.}~\bibnamefont
  {{Bode}}}, \bibinfo {author} {\bibfnamefont {P.}~\bibnamefont {{Diener}}},
  \bibinfo {author} {\bibfnamefont {R.}~\bibnamefont {{Haas}}}, \bibinfo
  {author} {\bibfnamefont {I.}~\bibnamefont {{Hinder}}}, \bibinfo {author}
  {\bibfnamefont {B.~C.}\ \bibnamefont {{Mundim}}}, \bibinfo {author}
  {\bibfnamefont {C.~D.}\ \bibnamefont {{Ott}}}, \bibinfo {author}
  {\bibfnamefont {E.}~\bibnamefont {{Schnetter}}}, \bibinfo {author}
  {\bibfnamefont {G.}~\bibnamefont {{Allen}}}, \bibinfo {author} {\bibfnamefont
  {M.}~\bibnamefont {{Campanelli}}}, \ and\ \bibinfo {author} {\bibfnamefont
  {P.}~\bibnamefont {{Laguna}}},\ }\href {\doibase
  10.1088/0264-9381/29/11/115001} {\bibfield  {journal} {\bibinfo  {journal}
  {Class.\ Quantum Grav.}\ }\textbf {\bibinfo {volume} {29}},\ \bibinfo {eid}
  {115001} (\bibinfo {year} {2012})},\ \Eprint {http://arxiv.org/abs/1111.3344}
  {arXiv:1111.3344 [gr-qc]} \BibitemShut {NoStop}%
\bibitem [{ein()}]{einsteintoolkit}%
  \BibitemOpen
  \href@noop {} {}\bibinfo {note} {Einstein Toolkit home page:
  \url{http://einsteintoolkit.org}}\BibitemShut {NoStop}%
\bibitem [{cac()}]{cactus_web}%
  \BibitemOpen
  \href@noop {} {}\bibinfo {note} {Cactus Computational Toolkit home page: {\tt
  http://cactuscode.org}}\BibitemShut {NoStop}%
\bibitem [{\citenamefont {Schnetter}\ \emph {et~al.}(2004)\citenamefont
  {Schnetter}, \citenamefont {Hawley},\ and\ \citenamefont
  {Hawke}}]{Schnetter-etal-03b}%
  \BibitemOpen
  \bibfield  {author} {\bibinfo {author} {\bibfnamefont {E.}~\bibnamefont
  {Schnetter}}, \bibinfo {author} {\bibfnamefont {S.~H.}\ \bibnamefont
  {Hawley}}, \ and\ \bibinfo {author} {\bibfnamefont {I.}~\bibnamefont
  {Hawke}},\ }\href@noop {} {\bibfield  {journal} {\bibinfo  {journal} {Class.
  Quant. Grav.}\ }\textbf {\bibinfo {volume} {21}},\ \bibinfo {pages} {1465}
  (\bibinfo {year} {2004})},\ \Eprint {http://arxiv.org/abs/gr-qc/0310042}
  {gr-qc/0310042} \BibitemShut {NoStop}%
\bibitem [{\citenamefont {Healy}\ \emph
  {et~al.}(2017{\natexlab{b}})\citenamefont {Healy}, \citenamefont {Lousto},
  \citenamefont {Nakano},\ and\ \citenamefont {Zlochower}}]{Healy:2017zqj}%
  \BibitemOpen
  \bibfield  {author} {\bibinfo {author} {\bibfnamefont {J.}~\bibnamefont
  {Healy}}, \bibinfo {author} {\bibfnamefont {C.~O.}\ \bibnamefont {Lousto}},
  \bibinfo {author} {\bibfnamefont {H.}~\bibnamefont {Nakano}}, \ and\ \bibinfo
  {author} {\bibfnamefont {Y.}~\bibnamefont {Zlochower}},\ }\href {\doibase
  10.1088/1361-6382/aa7929} {\bibfield  {journal} {\bibinfo  {journal} {Class.
  Quant. Grav.}\ }\textbf {\bibinfo {volume} {34}},\ \bibinfo {pages} {145011}
  (\bibinfo {year} {2017}{\natexlab{b}})},\ \Eprint
  {http://arxiv.org/abs/1702.00872} {arXiv:1702.00872 [gr-qc]} \BibitemShut
  {NoStop}%
\bibitem [{\citenamefont {Brandt}\ and\ \citenamefont
  {Br{\"u}gmann}(1997)}]{Brandt97b}%
  \BibitemOpen
  \bibfield  {author} {\bibinfo {author} {\bibfnamefont {S.}~\bibnamefont
  {Brandt}}\ and\ \bibinfo {author} {\bibfnamefont {B.}~\bibnamefont
  {Br{\"u}gmann}},\ }\href@noop {} {\bibfield  {journal} {\bibinfo  {journal}
  {Phys. Rev. Lett.}\ }\textbf {\bibinfo {volume} {78}},\ \bibinfo {pages}
  {3606} (\bibinfo {year} {1997})},\ \Eprint
  {http://arxiv.org/abs/gr-qc/9703066} {gr-qc/9703066} \BibitemShut {NoStop}%
\bibitem [{\citenamefont {Ansorg}\ \emph {et~al.}(2004)\citenamefont {Ansorg},
  \citenamefont {Bruegmann},\ and\ \citenamefont {Tichy}}]{Ansorg:2004ds}%
  \BibitemOpen
  \bibfield  {author} {\bibinfo {author} {\bibfnamefont {M.}~\bibnamefont
  {Ansorg}}, \bibinfo {author} {\bibfnamefont {B.}~\bibnamefont {Bruegmann}}, \
  and\ \bibinfo {author} {\bibfnamefont {W.}~\bibnamefont {Tichy}},\ }\href
  {\doibase 10.1103/PhysRevD.70.064011} {\bibfield  {journal} {\bibinfo
  {journal} {Phys. Rev.}\ }\textbf {\bibinfo {volume} {D70}},\ \bibinfo {pages}
  {064011} (\bibinfo {year} {2004})},\ \Eprint
  {http://arxiv.org/abs/gr-qc/0404056} {arXiv:gr-qc/0404056 [gr-qc]}
  \BibitemShut {NoStop}%
\bibitem [{\citenamefont {Thornburg}(2004)}]{Thornburg2003:AH-finding}%
  \BibitemOpen
  \bibfield  {author} {\bibinfo {author} {\bibfnamefont {J.}~\bibnamefont
  {Thornburg}},\ }\href {\doibase 10.1088/0264-9381/21/2/026} {\bibfield
  {journal} {\bibinfo  {journal} {Class. Quant. Grav.}\ }\textbf {\bibinfo
  {volume} {21}},\ \bibinfo {pages} {743} (\bibinfo {year} {2004})},\ \Eprint
  {http://arxiv.org/abs/gr-qc/0306056} {gr-qc/0306056} \BibitemShut {NoStop}%
\bibitem [{\citenamefont {Dreyer}\ \emph {et~al.}(2003)\citenamefont {Dreyer},
  \citenamefont {Krishnan}, \citenamefont {Shoemaker},\ and\ \citenamefont
  {Schnetter}}]{Dreyer02a}%
  \BibitemOpen
  \bibfield  {author} {\bibinfo {author} {\bibfnamefont {O.}~\bibnamefont
  {Dreyer}}, \bibinfo {author} {\bibfnamefont {B.}~\bibnamefont {Krishnan}},
  \bibinfo {author} {\bibfnamefont {D.}~\bibnamefont {Shoemaker}}, \ and\
  \bibinfo {author} {\bibfnamefont {E.}~\bibnamefont {Schnetter}},\ }\href@noop
  {} {\bibfield  {journal} {\bibinfo  {journal} {Phys. Rev.}\ }\textbf
  {\bibinfo {volume} {D67}},\ \bibinfo {pages} {024018} (\bibinfo {year}
  {2003})},\ \Eprint {http://arxiv.org/abs/gr-qc/0206008} {gr-qc/0206008}
  \BibitemShut {NoStop}%
\bibitem [{\citenamefont {Campanelli}\ \emph {et~al.}(2007)\citenamefont
  {Campanelli}, \citenamefont {Lousto}, \citenamefont {Zlochower},
  \citenamefont {Krishnan},\ and\ \citenamefont {Merritt}}]{Campanelli:2006fy}%
  \BibitemOpen
  \bibfield  {author} {\bibinfo {author} {\bibfnamefont {M.}~\bibnamefont
  {Campanelli}}, \bibinfo {author} {\bibfnamefont {C.~O.}\ \bibnamefont
  {Lousto}}, \bibinfo {author} {\bibfnamefont {Y.}~\bibnamefont {Zlochower}},
  \bibinfo {author} {\bibfnamefont {B.}~\bibnamefont {Krishnan}}, \ and\
  \bibinfo {author} {\bibfnamefont {D.}~\bibnamefont {Merritt}},\ }\href
  {\doibase 10.1103/PhysRevD.75.064030} {\bibfield  {journal} {\bibinfo
  {journal} {Phys.\ Rev.\ D}\ }\textbf {\bibinfo {volume} {75}},\ \bibinfo
  {pages} {064030} (\bibinfo {year} {2007})},\ \Eprint
  {http://arxiv.org/abs/gr-qc/0612076} {arXiv:gr-qc/0612076 [gr-qc]}
  \BibitemShut {NoStop}%
\bibitem [{\citenamefont {Campanelli}\ and\ \citenamefont
  {Lousto}(1999)}]{Campanelli:1998jv}%
  \BibitemOpen
  \bibfield  {author} {\bibinfo {author} {\bibfnamefont {M.}~\bibnamefont
  {Campanelli}}\ and\ \bibinfo {author} {\bibfnamefont {C.~O.}\ \bibnamefont
  {Lousto}},\ }\href {\doibase 10.1103/PhysRevD.59.124022} {\bibfield
  {journal} {\bibinfo  {journal} {Phys.\ Rev.\ D}\ }\textbf {\bibinfo {volume}
  {59}},\ \bibinfo {pages} {124022} (\bibinfo {year} {1999})},\ \Eprint
  {http://arxiv.org/abs/gr-qc/9811019} {arXiv:gr-qc/9811019 [gr-qc]}
  \BibitemShut {NoStop}%
\bibitem [{\citenamefont {Lousto}\ and\ \citenamefont
  {Zlochower}(2007)}]{Lousto:2007mh}%
  \BibitemOpen
  \bibfield  {author} {\bibinfo {author} {\bibfnamefont {C.~O.}\ \bibnamefont
  {Lousto}}\ and\ \bibinfo {author} {\bibfnamefont {Y.}~\bibnamefont
  {Zlochower}},\ }\href {\doibase 10.1103/PhysRevD.76.041502} {\bibfield
  {journal} {\bibinfo  {journal} {Phys.\ Rev.\ D}\ }\textbf {\bibinfo {volume}
  {76}},\ \bibinfo {pages} {041502} (\bibinfo {year} {2007})},\ \Eprint
  {http://arxiv.org/abs/gr-qc/0703061} {arXiv:gr-qc/0703061 [GR-QC]}
  \BibitemShut {NoStop}%
\bibitem [{\citenamefont {Nakano}\ \emph {et~al.}(2015)\citenamefont {Nakano},
  \citenamefont {Healy}, \citenamefont {Lousto},\ and\ \citenamefont
  {Zlochower}}]{Nakano:2015pta}%
  \BibitemOpen
  \bibfield  {author} {\bibinfo {author} {\bibfnamefont {H.}~\bibnamefont
  {Nakano}}, \bibinfo {author} {\bibfnamefont {J.}~\bibnamefont {Healy}},
  \bibinfo {author} {\bibfnamefont {C.~O.}\ \bibnamefont {Lousto}}, \ and\
  \bibinfo {author} {\bibfnamefont {Y.}~\bibnamefont {Zlochower}},\ }\href
  {\doibase 10.1103/PhysRevD.91.104022} {\bibfield  {journal} {\bibinfo
  {journal} {Phys.\ Rev.\ D}\ }\textbf {\bibinfo {volume} {91}},\ \bibinfo
  {pages} {104022} (\bibinfo {year} {2015})},\ \Eprint
  {http://arxiv.org/abs/1503.00718} {arXiv:1503.00718 [gr-qc]} \BibitemShut
  {NoStop}%
\bibitem [{\citenamefont {Healy}\ \emph
  {et~al.}(2017{\natexlab{c}})\citenamefont {Healy}, \citenamefont {Lousto},\
  and\ \citenamefont {Zlochower}}]{Healy:2017mvh}%
  \BibitemOpen
  \bibfield  {author} {\bibinfo {author} {\bibfnamefont {J.}~\bibnamefont
  {Healy}}, \bibinfo {author} {\bibfnamefont {C.~O.}\ \bibnamefont {Lousto}}, \
  and\ \bibinfo {author} {\bibfnamefont {Y.}~\bibnamefont {Zlochower}},\ }\href
  {\doibase 10.1103/PhysRevD.96.024031} {\bibfield  {journal} {\bibinfo
  {journal} {Phys. Rev.}\ }\textbf {\bibinfo {volume} {D96}},\ \bibinfo {pages}
  {024031} (\bibinfo {year} {2017}{\natexlab{c}})},\ \Eprint
  {http://arxiv.org/abs/1705.07034} {arXiv:1705.07034 [gr-qc]} \BibitemShut
  {NoStop}%
\bibitem [{\citenamefont {Br{\"u}gmann}\ \emph {et~al.}(2008)\citenamefont
  {Br{\"u}gmann}, \citenamefont {Gonz\'{a}lez}, \citenamefont {Hannam},
  \citenamefont {Husa}, \citenamefont {Sperhake},\ and\ \citenamefont
  {Tichy}}]{Bruegmann2006}%
  \BibitemOpen
  \bibfield  {author} {\bibinfo {author} {\bibfnamefont {B.}~\bibnamefont
  {Br{\"u}gmann}}, \bibinfo {author} {\bibfnamefont {J.~A.}\ \bibnamefont
  {Gonz\'{a}lez}}, \bibinfo {author} {\bibfnamefont {M.}~\bibnamefont
  {Hannam}}, \bibinfo {author} {\bibfnamefont {S.}~\bibnamefont {Husa}},
  \bibinfo {author} {\bibfnamefont {U.}~\bibnamefont {Sperhake}}, \ and\
  \bibinfo {author} {\bibfnamefont {W.}~\bibnamefont {Tichy}},\ }\href
  {\doibase 10.1103/PhysRevD.77.024027} {\bibfield  {journal} {\bibinfo
  {journal} {Phys.\ Rev.\ D}\ }\textbf {\bibinfo {volume} {77}},\ \bibinfo
  {eid} {024027} (\bibinfo {year} {2008})},\ \Eprint
  {http://arxiv.org/abs/gr-qc/0610128} {gr-qc/0610128} \BibitemShut {NoStop}%
\bibitem [{\citenamefont {Herrmann}\ \emph
  {et~al.}(2007{\natexlab{a}})\citenamefont {Herrmann}, \citenamefont {Hinder},
  \citenamefont {Shoemaker}, \citenamefont {Laguna},\ and\ \citenamefont
  {Matzner}}]{Herrmann:2007ex}%
  \BibitemOpen
  \bibfield  {author} {\bibinfo {author} {\bibfnamefont {F.}~\bibnamefont
  {Herrmann}}, \bibinfo {author} {\bibfnamefont {I.}~\bibnamefont {Hinder}},
  \bibinfo {author} {\bibfnamefont {D.~M.}\ \bibnamefont {Shoemaker}}, \bibinfo
  {author} {\bibfnamefont {P.}~\bibnamefont {Laguna}}, \ and\ \bibinfo {author}
  {\bibfnamefont {R.~A.}\ \bibnamefont {Matzner}},\ }\href {\doibase
  10.1103/PhysRevD.76.084032} {\bibfield  {journal} {\bibinfo  {journal} {Phys.
  Rev.}\ }\textbf {\bibinfo {volume} {D76}},\ \bibinfo {pages} {084032}
  (\bibinfo {year} {2007}{\natexlab{a}})},\ \Eprint
  {http://arxiv.org/abs/0706.2541} {arXiv:0706.2541 [gr-qc]} \BibitemShut
  {NoStop}%
\bibitem [{\citenamefont {Herrmann}\ \emph
  {et~al.}(2007{\natexlab{b}})\citenamefont {Herrmann}, \citenamefont {Hinder},
  \citenamefont {Shoemaker}, \citenamefont {Laguna},\ and\ \citenamefont
  {Matzner}}]{Herrmann:2007ac}%
  \BibitemOpen
  \bibfield  {author} {\bibinfo {author} {\bibfnamefont {F.}~\bibnamefont
  {Herrmann}}, \bibinfo {author} {\bibfnamefont {I.}~\bibnamefont {Hinder}},
  \bibinfo {author} {\bibfnamefont {D.}~\bibnamefont {Shoemaker}}, \bibinfo
  {author} {\bibfnamefont {P.}~\bibnamefont {Laguna}}, \ and\ \bibinfo {author}
  {\bibfnamefont {R.~A.}\ \bibnamefont {Matzner}},\ }\href {\doibase
  10.1086/513603} {\bibfield  {journal} {\bibinfo  {journal} {Astrophys. J.}\
  }\textbf {\bibinfo {volume} {661}},\ \bibinfo {pages} {430} (\bibinfo {year}
  {2007}{\natexlab{b}})},\ \Eprint {http://arxiv.org/abs/gr-qc/0701143}
  {arXiv:gr-qc/0701143 [GR-QC]} \BibitemShut {NoStop}%
\bibitem [{\citenamefont {Hinder}\ \emph {et~al.}(2008)\citenamefont {Hinder},
  \citenamefont {Vaishnav}, \citenamefont {Herrmann}, \citenamefont
  {Shoemaker},\ and\ \citenamefont {Laguna}}]{Hinder:2007qu}%
  \BibitemOpen
  \bibfield  {author} {\bibinfo {author} {\bibfnamefont {I.}~\bibnamefont
  {Hinder}}, \bibinfo {author} {\bibfnamefont {B.}~\bibnamefont {Vaishnav}},
  \bibinfo {author} {\bibfnamefont {F.}~\bibnamefont {Herrmann}}, \bibinfo
  {author} {\bibfnamefont {D.}~\bibnamefont {Shoemaker}}, \ and\ \bibinfo
  {author} {\bibfnamefont {P.}~\bibnamefont {Laguna}},\ }\href {\doibase
  10.1103/PhysRevD.77.081502} {\bibfield  {journal} {\bibinfo  {journal} {Phys.
  Rev.}\ }\textbf {\bibinfo {volume} {D77}},\ \bibinfo {pages} {081502}
  (\bibinfo {year} {2008})},\ \Eprint {http://arxiv.org/abs/0710.5167}
  {arXiv:0710.5167 [gr-qc]} \BibitemShut {NoStop}%
\bibitem [{\citenamefont {Healy}\ \emph
  {et~al.}(2009{\natexlab{a}})\citenamefont {Healy}, \citenamefont {Herrmann},
  \citenamefont {Hinder}, \citenamefont {Shoemaker}, \citenamefont {Laguna},\
  and\ \citenamefont {Matzner}}]{Healy:2008js}%
  \BibitemOpen
  \bibfield  {author} {\bibinfo {author} {\bibfnamefont {J.}~\bibnamefont
  {Healy}}, \bibinfo {author} {\bibfnamefont {F.}~\bibnamefont {Herrmann}},
  \bibinfo {author} {\bibfnamefont {I.}~\bibnamefont {Hinder}}, \bibinfo
  {author} {\bibfnamefont {D.~M.}\ \bibnamefont {Shoemaker}}, \bibinfo {author}
  {\bibfnamefont {P.}~\bibnamefont {Laguna}}, \ and\ \bibinfo {author}
  {\bibfnamefont {R.~A.}\ \bibnamefont {Matzner}},\ }\href {\doibase
  10.1103/PhysRevLett.102.041101} {\bibfield  {journal} {\bibinfo  {journal}
  {Phys. Rev. Lett.}\ }\textbf {\bibinfo {volume} {102}},\ \bibinfo {pages}
  {041101} (\bibinfo {year} {2009}{\natexlab{a}})},\ \Eprint
  {http://arxiv.org/abs/0807.3292} {arXiv:0807.3292 [gr-qc]} \BibitemShut
  {NoStop}%
\bibitem [{\citenamefont {Hinder}\ \emph {et~al.}(2010)\citenamefont {Hinder},
  \citenamefont {Herrmann}, \citenamefont {Laguna},\ and\ \citenamefont
  {Shoemaker}}]{Hinder:2008kv}%
  \BibitemOpen
  \bibfield  {author} {\bibinfo {author} {\bibfnamefont {I.}~\bibnamefont
  {Hinder}}, \bibinfo {author} {\bibfnamefont {F.}~\bibnamefont {Herrmann}},
  \bibinfo {author} {\bibfnamefont {P.}~\bibnamefont {Laguna}}, \ and\ \bibinfo
  {author} {\bibfnamefont {D.}~\bibnamefont {Shoemaker}},\ }\href {\doibase
  10.1103/PhysRevD.82.024033} {\bibfield  {journal} {\bibinfo  {journal}
  {Phys.\ Rev.\ D}\ }\textbf {\bibinfo {volume} {82}},\ \bibinfo {pages}
  {024033} (\bibinfo {year} {2010})},\ \Eprint {http://arxiv.org/abs/0806.1037}
  {arXiv:0806.1037 [gr-qc]} \BibitemShut {NoStop}%
\bibitem [{\citenamefont {Healy}\ \emph
  {et~al.}(2009{\natexlab{b}})\citenamefont {Healy}, \citenamefont {Levin},\
  and\ \citenamefont {Shoemaker}}]{Healy:2009zm}%
  \BibitemOpen
  \bibfield  {author} {\bibinfo {author} {\bibfnamefont {J.}~\bibnamefont
  {Healy}}, \bibinfo {author} {\bibfnamefont {J.}~\bibnamefont {Levin}}, \ and\
  \bibinfo {author} {\bibfnamefont {D.}~\bibnamefont {Shoemaker}},\ }\href
  {\doibase 10.1103/PhysRevLett.103.131101} {\bibfield  {journal} {\bibinfo
  {journal} {Phys. Rev. Lett.}\ }\textbf {\bibinfo {volume} {103}},\ \bibinfo
  {pages} {131101} (\bibinfo {year} {2009}{\natexlab{b}})},\ \Eprint
  {http://arxiv.org/abs/0907.0671} {arXiv:0907.0671 [gr-qc]} \BibitemShut
  {NoStop}%
\bibitem [{\citenamefont {Healy}\ \emph {et~al.}(2010)\citenamefont {Healy},
  \citenamefont {Laguna}, \citenamefont {Matzner},\ and\ \citenamefont
  {Shoemaker}}]{Healy:2009ir}%
  \BibitemOpen
  \bibfield  {author} {\bibinfo {author} {\bibfnamefont {J.}~\bibnamefont
  {Healy}}, \bibinfo {author} {\bibfnamefont {P.}~\bibnamefont {Laguna}},
  \bibinfo {author} {\bibfnamefont {R.~A.}\ \bibnamefont {Matzner}}, \ and\
  \bibinfo {author} {\bibfnamefont {D.~M.}\ \bibnamefont {Shoemaker}},\ }\href
  {\doibase 10.1103/PhysRevD.81.081501} {\bibfield  {journal} {\bibinfo
  {journal} {Phys.\ Rev.\ D}\ }\textbf {\bibinfo {volume} {81}},\ \bibinfo
  {pages} {081501} (\bibinfo {year} {2010})},\ \Eprint
  {http://arxiv.org/abs/0905.3914} {arXiv:0905.3914 [gr-qc]} \BibitemShut
  {NoStop}%
\bibitem [{\citenamefont {Bode}\ \emph {et~al.}(2010)\citenamefont {Bode},
  \citenamefont {Haas}, \citenamefont {Bogdanovic}, \citenamefont {Laguna},\
  and\ \citenamefont {Shoemaker}}]{Bode:2009mt}%
  \BibitemOpen
  \bibfield  {author} {\bibinfo {author} {\bibfnamefont {T.}~\bibnamefont
  {Bode}}, \bibinfo {author} {\bibfnamefont {R.}~\bibnamefont {Haas}}, \bibinfo
  {author} {\bibfnamefont {T.}~\bibnamefont {Bogdanovic}}, \bibinfo {author}
  {\bibfnamefont {P.}~\bibnamefont {Laguna}}, \ and\ \bibinfo {author}
  {\bibfnamefont {D.}~\bibnamefont {Shoemaker}},\ }\href {\doibase
  10.1088/0004-637X/715/2/1117} {\bibfield  {journal} {\bibinfo  {journal}
  {Astrophys. J.}\ }\textbf {\bibinfo {volume} {715}},\ \bibinfo {pages} {1117}
  (\bibinfo {year} {2010})},\ \Eprint {http://arxiv.org/abs/0912.0087}
  {arXiv:0912.0087 [gr-qc]} \BibitemShut {NoStop}%
\bibitem [{\citenamefont {Husa}\ \emph {et~al.}(2006)\citenamefont {Husa},
  \citenamefont {Hinder},\ and\ \citenamefont {Lechner}}]{Husa:2004ip}%
  \BibitemOpen
  \bibfield  {author} {\bibinfo {author} {\bibfnamefont {S.}~\bibnamefont
  {Husa}}, \bibinfo {author} {\bibfnamefont {I.}~\bibnamefont {Hinder}}, \ and\
  \bibinfo {author} {\bibfnamefont {C.}~\bibnamefont {Lechner}},\ }\href
  {\doibase 10.1016/j.cpc.2006.02.002} {\bibfield  {journal} {\bibinfo
  {journal} {Comput. Phys. Commun.}\ }\textbf {\bibinfo {volume} {174}},\
  \bibinfo {pages} {983} (\bibinfo {year} {2006})},\ \Eprint
  {http://arxiv.org/abs/gr-qc/0404023} {arXiv:gr-qc/0404023 [gr-qc]}
  \BibitemShut {NoStop}%
\bibitem [{SpE()}]{SpECwebsite}%
  \BibitemOpen
  \href@noop {} {}\bibinfo {howpublished}
  {\url{http://www.black-holes.org/SpEC.html}}\BibitemShut {NoStop}%
\bibitem [{\citenamefont {Lovelace}\ \emph {et~al.}(2008)\citenamefont
  {Lovelace}, \citenamefont {Owen}, \citenamefont {Pfeiffer},\ and\
  \citenamefont {Chu}}]{Lovelace2008}%
  \BibitemOpen
  \bibfield  {author} {\bibinfo {author} {\bibfnamefont {G.}~\bibnamefont
  {Lovelace}}, \bibinfo {author} {\bibfnamefont {R.}~\bibnamefont {Owen}},
  \bibinfo {author} {\bibfnamefont {H.~P.}\ \bibnamefont {Pfeiffer}}, \ and\
  \bibinfo {author} {\bibfnamefont {T.}~\bibnamefont {Chu}},\ }\href {\doibase
  10.1103/PhysRevD.78.084017} {\bibfield  {journal} {\bibinfo  {journal}
  {Phys.\ Rev.\ D}\ }\textbf {\bibinfo {volume} {78}},\ \bibinfo {pages}
  {084017} (\bibinfo {year} {2008})}\BibitemShut {NoStop}%
\bibitem [{\citenamefont {York}(1999)}]{York1999}%
  \BibitemOpen
  \bibfield  {author} {\bibinfo {author} {\bibfnamefont {J.~W.}\ \bibnamefont
  {York}},\ }\href {\doibase 10.1103/PhysRevLett.82.1350} {\bibfield  {journal}
  {\bibinfo  {journal} {Phys.\ Rev.\ Lett.}\ }\textbf {\bibinfo {volume}
  {82}},\ \bibinfo {pages} {1350} (\bibinfo {year} {1999})}\BibitemShut
  {NoStop}%
\bibitem [{\citenamefont {Pfeiffer}\ \emph {et~al.}(2003)\citenamefont
  {Pfeiffer}, \citenamefont {Kidder}, \citenamefont {Scheel},\ and\
  \citenamefont {Teukolsky}}]{Pfeiffer2003}%
  \BibitemOpen
  \bibfield  {author} {\bibinfo {author} {\bibfnamefont {H.~P.}\ \bibnamefont
  {Pfeiffer}}, \bibinfo {author} {\bibfnamefont {L.~E.}\ \bibnamefont
  {Kidder}}, \bibinfo {author} {\bibfnamefont {M.~A.}\ \bibnamefont {Scheel}},
  \ and\ \bibinfo {author} {\bibfnamefont {S.~A.}\ \bibnamefont {Teukolsky}},\
  }\href {\doibase 10.1016/S0010-4655(02)00847-0} {\bibfield  {journal}
  {\bibinfo  {journal} {Comput.\ Phys.\ Commun.}\ }\textbf {\bibinfo {volume}
  {152}},\ \bibinfo {pages} {253} (\bibinfo {year} {2003})},\ \Eprint
  {http://arxiv.org/abs/gr-qc/0202096} {gr-qc/0202096} \BibitemShut {NoStop}%
\bibitem [{\citenamefont {{Ossokine}}\ \emph {et~al.}(2015)\citenamefont
  {{Ossokine}}, \citenamefont {{Foucart}}, \citenamefont {{Pfeiffer}},
  \citenamefont {{Boyle}},\ and\ \citenamefont
  {{Szil{\'a}gyi}}}]{Ossokine:2015yla}%
  \BibitemOpen
  \bibfield  {author} {\bibinfo {author} {\bibfnamefont {S.}~\bibnamefont
  {{Ossokine}}}, \bibinfo {author} {\bibfnamefont {F.}~\bibnamefont
  {{Foucart}}}, \bibinfo {author} {\bibfnamefont {H.~P.}\ \bibnamefont
  {{Pfeiffer}}}, \bibinfo {author} {\bibfnamefont {M.}~\bibnamefont {{Boyle}}},
  \ and\ \bibinfo {author} {\bibfnamefont {B.}~\bibnamefont {{Szil{\'a}gyi}}},\
  }\href {\doibase 10.1088/0264-9381/32/24/245010} {\bibfield  {journal}
  {\bibinfo  {journal} {Class.\ Quantum Grav.}\ }\textbf {\bibinfo {volume}
  {32}},\ \bibinfo {eid} {245010} (\bibinfo {year} {2015})},\ \Eprint
  {http://arxiv.org/abs/1506.01689} {arXiv:1506.01689 [gr-qc]} \BibitemShut
  {NoStop}%
\bibitem [{\citenamefont {{Caudill}}\ \emph {et~al.}(2006)\citenamefont
  {{Caudill}}, \citenamefont {{Cook}}, \citenamefont {{Grigsby}},\ and\
  \citenamefont {{Pfeiffer}}}]{Caudill-etal:2006}%
  \BibitemOpen
  \bibfield  {author} {\bibinfo {author} {\bibfnamefont {M.}~\bibnamefont
  {{Caudill}}}, \bibinfo {author} {\bibfnamefont {G.~B.}\ \bibnamefont
  {{Cook}}}, \bibinfo {author} {\bibfnamefont {J.~D.}\ \bibnamefont
  {{Grigsby}}}, \ and\ \bibinfo {author} {\bibfnamefont {H.~P.}\ \bibnamefont
  {{Pfeiffer}}},\ }\href@noop {} {\bibfield  {journal} {\bibinfo  {journal}
  {Phys.\ Rev.\ D}\ }\textbf {\bibinfo {volume} {74}},\ \bibinfo {pages}
  {064011} (\bibinfo {year} {2006})},\ \Eprint
  {http://arxiv.org/abs/gr-qc/0605053} {gr-qc/0605053} \BibitemShut {NoStop}%
\bibitem [{\citenamefont {Pfeiffer}\ \emph {et~al.}(2007)\citenamefont
  {Pfeiffer}, \citenamefont {Brown}, \citenamefont {Kidder}, \citenamefont
  {Lindblom}, \citenamefont {Lovelace},\ and\ \citenamefont
  {Scheel}}]{Pfeiffer-Brown-etal:2007}%
  \BibitemOpen
  \bibfield  {author} {\bibinfo {author} {\bibfnamefont {H.~P.}\ \bibnamefont
  {Pfeiffer}}, \bibinfo {author} {\bibfnamefont {D.~A.}\ \bibnamefont {Brown}},
  \bibinfo {author} {\bibfnamefont {L.~E.}\ \bibnamefont {Kidder}}, \bibinfo
  {author} {\bibfnamefont {L.}~\bibnamefont {Lindblom}}, \bibinfo {author}
  {\bibfnamefont {G.}~\bibnamefont {Lovelace}}, \ and\ \bibinfo {author}
  {\bibfnamefont {M.~A.}\ \bibnamefont {Scheel}},\ }\href@noop {} {\bibfield
  {journal} {\bibinfo  {journal} {Class.\ Quantum Grav.}\ }\textbf {\bibinfo
  {volume} {24}},\ \bibinfo {pages} {S59} (\bibinfo {year} {2007})},\ \Eprint
  {http://arxiv.org/abs/gr-qc/0702106} {gr-qc/0702106} \BibitemShut {NoStop}%
\bibitem [{\citenamefont {Buonanno}\ \emph {et~al.}(2011)\citenamefont
  {Buonanno}, \citenamefont {Kidder}, \citenamefont {Mrou\'{e}}, \citenamefont
  {Pfeiffer},\ and\ \citenamefont {Taracchini}}]{Buonanno:2010yk}%
  \BibitemOpen
  \bibfield  {author} {\bibinfo {author} {\bibfnamefont {A.}~\bibnamefont
  {Buonanno}}, \bibinfo {author} {\bibfnamefont {L.~E.}\ \bibnamefont
  {Kidder}}, \bibinfo {author} {\bibfnamefont {A.~H.}\ \bibnamefont
  {Mrou\'{e}}}, \bibinfo {author} {\bibfnamefont {H.~P.}\ \bibnamefont
  {Pfeiffer}}, \ and\ \bibinfo {author} {\bibfnamefont {A.}~\bibnamefont
  {Taracchini}},\ }\href {\doibase 10.1103/PhysRevD.83.104034} {\bibfield
  {journal} {\bibinfo  {journal} {Phys.\ Rev.\ D}\ }\textbf {\bibinfo {volume}
  {83}},\ \bibinfo {pages} {104034} (\bibinfo {year} {2011})},\ \Eprint
  {http://arxiv.org/abs/1012.1549} {arXiv:1012.1549 [gr-qc]} \BibitemShut
  {NoStop}%
\bibitem [{\citenamefont {Mrou\'e}\ and\ \citenamefont
  {Pfeiffer}(2012)}]{Mroue:2012kv}%
  \BibitemOpen
  \bibfield  {author} {\bibinfo {author} {\bibfnamefont {A.~H.}\ \bibnamefont
  {Mrou\'e}}\ and\ \bibinfo {author} {\bibfnamefont {H.~P.}\ \bibnamefont
  {Pfeiffer}},\ }\href@noop {} {\  (\bibinfo {year} {2012})},\ \Eprint
  {http://arxiv.org/abs/1210.2958} {arXiv:1210.2958 [gr-qc]} \BibitemShut
  {NoStop}%
\bibitem [{\citenamefont {Lindblom}\ \emph {et~al.}(2006)\citenamefont
  {Lindblom}, \citenamefont {Scheel}, \citenamefont {Kidder}, \citenamefont
  {Owen},\ and\ \citenamefont {Rinne}}]{Lindblom2006}%
  \BibitemOpen
  \bibfield  {author} {\bibinfo {author} {\bibfnamefont {L.}~\bibnamefont
  {Lindblom}}, \bibinfo {author} {\bibfnamefont {M.~A.}\ \bibnamefont
  {Scheel}}, \bibinfo {author} {\bibfnamefont {L.~E.}\ \bibnamefont {Kidder}},
  \bibinfo {author} {\bibfnamefont {R.}~\bibnamefont {Owen}}, \ and\ \bibinfo
  {author} {\bibfnamefont {O.}~\bibnamefont {Rinne}},\ }\href {\doibase
  10.1088/0264-9381/23/16/S09} {\bibfield  {journal} {\bibinfo  {journal}
  {Class.\ Quantum Grav.}\ }\textbf {\bibinfo {volume} {23}},\ \bibinfo {pages}
  {S447} (\bibinfo {year} {2006})},\ \Eprint
  {http://arxiv.org/abs/gr-qc/0512093} {gr-qc/0512093} \BibitemShut {NoStop}%
\bibitem [{\citenamefont {Friedrich}(1985)}]{Friedrich1985}%
  \BibitemOpen
  \bibfield  {author} {\bibinfo {author} {\bibfnamefont {H.}~\bibnamefont
  {Friedrich}},\ }\href {\doibase 10.1007/BF01217728} {\bibfield  {journal}
  {\bibinfo  {journal} {Commun.\ Math.\ Phys.}\ }\textbf {\bibinfo {volume}
  {100}},\ \bibinfo {pages} {525} (\bibinfo {year} {1985})}\BibitemShut
  {NoStop}%
\bibitem [{\citenamefont {Garfinkle}(2002)}]{Garfinkle2002}%
  \BibitemOpen
  \bibfield  {author} {\bibinfo {author} {\bibfnamefont {D.}~\bibnamefont
  {Garfinkle}},\ }\href@noop {} {\bibfield  {journal} {\bibinfo  {journal}
  {Phys.\ Rev.\ D}\ }\textbf {\bibinfo {volume} {65}},\ \bibinfo {pages}
  {044029} (\bibinfo {year} {2002})}\BibitemShut {NoStop}%
\bibitem [{\citenamefont {{Pretorius}}(2005)}]{Pretorius2005c}%
  \BibitemOpen
  \bibfield  {author} {\bibinfo {author} {\bibfnamefont {F.}~\bibnamefont
  {{Pretorius}}},\ }\href {\doibase 10.1088/0264-9381/22/2/014} {\bibfield
  {journal} {\bibinfo  {journal} {Class.\ Quantum Grav.}\ }\textbf {\bibinfo
  {volume} {22}},\ \bibinfo {pages} {425} (\bibinfo {year} {2005})},\ \Eprint
  {http://arxiv.org/abs/gr-qc/0407110} {gr-qc/0407110} \BibitemShut {NoStop}%
\bibitem [{\citenamefont {Lindblom}\ and\ \citenamefont
  {Szil\'agyi}(2009)}]{Lindblom2009c}%
  \BibitemOpen
  \bibfield  {author} {\bibinfo {author} {\bibfnamefont {L.}~\bibnamefont
  {Lindblom}}\ and\ \bibinfo {author} {\bibfnamefont {B.}~\bibnamefont
  {Szil\'agyi}},\ }\href@noop {} {\bibfield  {journal} {\bibinfo  {journal}
  {Phys.\ Rev.\ D}\ }\textbf {\bibinfo {volume} {80}},\ \bibinfo {pages}
  {084019} (\bibinfo {year} {2009})},\ \Eprint
  {http://arxiv.org/abs/arXiv:0904.4873} {arXiv:0904.4873} \BibitemShut
  {NoStop}%
\bibitem [{\citenamefont {Choptuik}\ and\ \citenamefont
  {Pretorius}(2010)}]{Choptuik:2009ww}%
  \BibitemOpen
  \bibfield  {author} {\bibinfo {author} {\bibfnamefont {M.~W.}\ \bibnamefont
  {Choptuik}}\ and\ \bibinfo {author} {\bibfnamefont {F.}~\bibnamefont
  {Pretorius}},\ }\href {\doibase 10.1103/PhysRevLett.104.111101} {\bibfield
  {journal} {\bibinfo  {journal} {Phys.\ Rev.\ Lett.}\ }\textbf {\bibinfo
  {volume} {104}},\ \bibinfo {pages} {111101} (\bibinfo {year} {2010})},\
  \Eprint {http://arxiv.org/abs/0908.1780} {arXiv:0908.1780 [gr-qc]}
  \BibitemShut {NoStop}%
\bibitem [{\citenamefont {{Szil{\'a}gyi}}\ \emph {et~al.}(2009)\citenamefont
  {{Szil{\'a}gyi}}, \citenamefont {{Lindblom}},\ and\ \citenamefont
  {{Scheel}}}]{Szilagyi:2009qz}%
  \BibitemOpen
  \bibfield  {author} {\bibinfo {author} {\bibfnamefont {B.}~\bibnamefont
  {{Szil{\'a}gyi}}}, \bibinfo {author} {\bibfnamefont {L.}~\bibnamefont
  {{Lindblom}}}, \ and\ \bibinfo {author} {\bibfnamefont {M.~A.}\ \bibnamefont
  {{Scheel}}},\ }\href {\doibase 10.1103/PhysRevD.80.124010} {\bibfield
  {journal} {\bibinfo  {journal} {Phys.\ Rev.\ D}\ }\textbf {\bibinfo {volume}
  {80}},\ \bibinfo {eid} {124010} (\bibinfo {year} {2009})},\ \Eprint
  {http://arxiv.org/abs/0909.3557} {arXiv:0909.3557 [gr-qc]} \BibitemShut
  {NoStop}%
\bibitem [{\citenamefont {Lovelace}\ \emph {et~al.}(2011)\citenamefont
  {Lovelace}, \citenamefont {Scheel},\ and\ \citenamefont
  {Szil{\'a}gyi}}]{Lovelace:2010ne}%
  \BibitemOpen
  \bibfield  {author} {\bibinfo {author} {\bibfnamefont {G.}~\bibnamefont
  {Lovelace}}, \bibinfo {author} {\bibfnamefont {M.~A.}\ \bibnamefont
  {Scheel}}, \ and\ \bibinfo {author} {\bibfnamefont {B.}~\bibnamefont
  {Szil{\'a}gyi}},\ }\href {\doibase 10.1103/PhysRevD.83.024010} {\bibfield
  {journal} {\bibinfo  {journal} {Phys.\ Rev.\ D}\ }\textbf {\bibinfo {volume}
  {83}},\ \bibinfo {pages} {024010} (\bibinfo {year} {2011})},\ \Eprint
  {http://arxiv.org/abs/1010.2777} {arXiv:1010.2777 [gr-qc]} \BibitemShut
  {NoStop}%
\bibitem [{\citenamefont {{Szil{\'a}gyi}}(2014)}]{Szilagyi:2014fna}%
  \BibitemOpen
  \bibfield  {author} {\bibinfo {author} {\bibfnamefont {B.}~\bibnamefont
  {{Szil{\'a}gyi}}},\ }\href {\doibase 10.1142/S0218271814300146} {\bibfield
  {journal} {\bibinfo  {journal} {{Int. J. Mod. Phys. D}}\ }\textbf {\bibinfo
  {volume} {23}},\ \bibinfo {eid} {1430014} (\bibinfo {year} {2014})},\ \Eprint
  {http://arxiv.org/abs/1405.3693} {arXiv:1405.3693 [gr-qc]} \BibitemShut
  {NoStop}%
\bibitem [{\citenamefont {{M. A. Scheel, M. Boyle, T. Chu, L. E. Kidder, K. D.
  Matthews and H. P. Pfeiffer}}(2009)}]{Scheel2009}%
  \BibitemOpen
  \bibfield  {author} {\bibinfo {author} {\bibnamefont {{M. A. Scheel, M.
  Boyle, T. Chu, L. E. Kidder, K. D. Matthews and H. P. Pfeiffer}}},\
  }\href@noop {} {\bibfield  {journal} {\bibinfo  {journal} {Phys.\ Rev.\ D}\
  }\textbf {\bibinfo {volume} {79}},\ \bibinfo {pages} {024003} (\bibinfo
  {year} {2009})},\ \Eprint {http://arxiv.org/abs/arXiv:gr-qc/0810.1767}
  {arXiv:gr-qc/0810.1767} \BibitemShut {NoStop}%
\bibitem [{\citenamefont {{Hemberger}}\ \emph {et~al.}(2013)\citenamefont
  {{Hemberger}}, \citenamefont {{Scheel}}, \citenamefont {{Kidder}},
  \citenamefont {{Szil{\'a}gyi}}, \citenamefont {{Lovelace}}, \citenamefont
  {{Taylor}},\ and\ \citenamefont {{Teukolsky}}}]{Hemberger:2012jz}%
  \BibitemOpen
  \bibfield  {author} {\bibinfo {author} {\bibfnamefont {D.~A.}\ \bibnamefont
  {{Hemberger}}}, \bibinfo {author} {\bibfnamefont {M.~A.}\ \bibnamefont
  {{Scheel}}}, \bibinfo {author} {\bibfnamefont {L.~E.}\ \bibnamefont
  {{Kidder}}}, \bibinfo {author} {\bibfnamefont {B.}~\bibnamefont
  {{Szil{\'a}gyi}}}, \bibinfo {author} {\bibfnamefont {G.}~\bibnamefont
  {{Lovelace}}}, \bibinfo {author} {\bibfnamefont {N.~W.}\ \bibnamefont
  {{Taylor}}}, \ and\ \bibinfo {author} {\bibfnamefont {S.~A.}\ \bibnamefont
  {{Teukolsky}}},\ }\href {\doibase 10.1088/0264-9381/30/11/115001} {\bibfield
  {journal} {\bibinfo  {journal} {Class.\ Quantum Grav.}\ }\textbf {\bibinfo
  {volume} {30}},\ \bibinfo {eid} {115001} (\bibinfo {year} {2013})},\ \Eprint
  {http://arxiv.org/abs/1211.6079} {arXiv:1211.6079 [gr-qc]} \BibitemShut
  {NoStop}%
\bibitem [{\citenamefont {Ossokine}\ \emph {et~al.}(2013)\citenamefont
  {Ossokine}, \citenamefont {Kidder},\ and\ \citenamefont
  {Pfeiffer}}]{Ossokine:2013zga}%
  \BibitemOpen
  \bibfield  {author} {\bibinfo {author} {\bibfnamefont {S.}~\bibnamefont
  {Ossokine}}, \bibinfo {author} {\bibfnamefont {L.~E.}\ \bibnamefont
  {Kidder}}, \ and\ \bibinfo {author} {\bibfnamefont {H.~P.}\ \bibnamefont
  {Pfeiffer}},\ }\href {\doibase 10.1103/PhysRevD.88.084031} {\bibfield
  {journal} {\bibinfo  {journal} {Phys.\ Rev.\ D}\ }\textbf {\bibinfo {volume}
  {88}},\ \bibinfo {pages} {084031} (\bibinfo {year} {2013})},\ \Eprint
  {http://arxiv.org/abs/1304.3067} {arXiv:1304.3067 [gr-qc]} \BibitemShut
  {NoStop}%
\bibitem [{\citenamefont {{Scheel}}\ \emph {et~al.}(2015)\citenamefont
  {{Scheel}}, \citenamefont {{Giesler}}, \citenamefont {{Hemberger}},
  \citenamefont {{Lovelace}}, \citenamefont {{Kuper}}, \citenamefont {{Boyle}},
  \citenamefont {{Szil{\'a}gyi}},\ and\ \citenamefont {{Kidder}}}]{Scheel2014}%
  \BibitemOpen
  \bibfield  {author} {\bibinfo {author} {\bibfnamefont {M.~A.}\ \bibnamefont
  {{Scheel}}}, \bibinfo {author} {\bibfnamefont {M.}~\bibnamefont {{Giesler}}},
  \bibinfo {author} {\bibfnamefont {D.~A.}\ \bibnamefont {{Hemberger}}},
  \bibinfo {author} {\bibfnamefont {G.}~\bibnamefont {{Lovelace}}}, \bibinfo
  {author} {\bibfnamefont {K.}~\bibnamefont {{Kuper}}}, \bibinfo {author}
  {\bibfnamefont {M.}~\bibnamefont {{Boyle}}}, \bibinfo {author} {\bibfnamefont
  {B.}~\bibnamefont {{Szil{\'a}gyi}}}, \ and\ \bibinfo {author} {\bibfnamefont
  {L.~E.}\ \bibnamefont {{Kidder}}},\ }\href {\doibase
  10.1088/0264-9381/32/10/105009} {\bibfield  {journal} {\bibinfo  {journal}
  {Class.\ Quantum Grav.}\ }\textbf {\bibinfo {volume} {32}},\ \bibinfo {eid}
  {105009} (\bibinfo {year} {2015})},\ \Eprint {http://arxiv.org/abs/1412.1803}
  {arXiv:1412.1803 [gr-qc]} \BibitemShut {NoStop}%
\bibitem [{\citenamefont {Rinne}(2006)}]{Rinne2006}%
  \BibitemOpen
  \bibfield  {author} {\bibinfo {author} {\bibfnamefont {O.}~\bibnamefont
  {Rinne}},\ }\href {http://stacks.iop.org/0264-9381/23/6275} {\bibfield
  {journal} {\bibinfo  {journal} {Class.\ Quantum Grav.}\ }\textbf {\bibinfo
  {volume} {23}},\ \bibinfo {pages} {6275} (\bibinfo {year}
  {2006})}\BibitemShut {NoStop}%
\bibitem [{\citenamefont {Rinne}\ \emph {et~al.}(2007)\citenamefont {Rinne},
  \citenamefont {Lindblom},\ and\ \citenamefont {Scheel}}]{Rinne2007}%
  \BibitemOpen
  \bibfield  {author} {\bibinfo {author} {\bibfnamefont {O.}~\bibnamefont
  {Rinne}}, \bibinfo {author} {\bibfnamefont {L.}~\bibnamefont {Lindblom}}, \
  and\ \bibinfo {author} {\bibfnamefont {M.~A.}\ \bibnamefont {Scheel}},\
  }\href {http://stacks.iop.org/0264-9381/24/4053} {\bibfield  {journal}
  {\bibinfo  {journal} {Class.\ Quantum Grav.}\ }\textbf {\bibinfo {volume}
  {24}},\ \bibinfo {pages} {4053} (\bibinfo {year} {2007})}\BibitemShut
  {NoStop}%
\bibitem [{\citenamefont {Gundlach}(1998)}]{Gundlach1998}%
  \BibitemOpen
  \bibfield  {author} {\bibinfo {author} {\bibfnamefont {C.}~\bibnamefont
  {Gundlach}},\ }\href {\doibase 10.1103/PhysRevD.57.863} {\bibfield  {journal}
  {\bibinfo  {journal} {Phys.\ Rev.\ D}\ }\textbf {\bibinfo {volume} {57}},\
  \bibinfo {pages} {863} (\bibinfo {year} {1998})}\BibitemShut {NoStop}%
\bibitem [{\citenamefont {Cook}\ and\ \citenamefont
  {Whiting}(2007)}]{Cook2007}%
  \BibitemOpen
  \bibfield  {author} {\bibinfo {author} {\bibfnamefont {G.~B.}\ \bibnamefont
  {Cook}}\ and\ \bibinfo {author} {\bibfnamefont {B.~F.}\ \bibnamefont
  {Whiting}},\ }\href {\doibase 10.1103/PhysRevD.76.041501} {\bibfield
  {journal} {\bibinfo  {journal} {Phys.\ Rev.\ D}\ }\textbf {\bibinfo {volume}
  {76}},\ \bibinfo {eid} {041501(R)} (\bibinfo {year} {2007})}\BibitemShut
  {NoStop}%
\bibitem [{\citenamefont {Owen}(2007)}]{OwenThesis}%
  \BibitemOpen
  \bibfield  {author} {\bibinfo {author} {\bibfnamefont {R.}~\bibnamefont
  {Owen}},\ }\emph {\bibinfo {title} {Topics in Numerical Relativity: {T}he
  periodic standing-wave approximation, the stability of constraints in free
  evolution, and the spin of dynamical black holes}},\ \href
  {http://resolver.caltech.edu/CaltechETD:etd-05252007-143511} {Ph.D. thesis},\
  \bibinfo  {school} {California Institute of Technology} (\bibinfo {year}
  {2007})\BibitemShut {NoStop}%
\bibitem [{\citenamefont {Boyle}\ \emph {et~al.}(2007)\citenamefont {Boyle},
  \citenamefont {Brown}, \citenamefont {Kidder}, \citenamefont {Mrou{\'e}},
  \citenamefont {Pfeiffer}, \citenamefont {Scheel}, \citenamefont {Cook},\ and\
  \citenamefont {Teukolsky}}]{Boyle2007}%
  \BibitemOpen
  \bibfield  {author} {\bibinfo {author} {\bibfnamefont {M.}~\bibnamefont
  {Boyle}}, \bibinfo {author} {\bibfnamefont {D.~A.}\ \bibnamefont {Brown}},
  \bibinfo {author} {\bibfnamefont {L.~E.}\ \bibnamefont {Kidder}}, \bibinfo
  {author} {\bibfnamefont {A.~H.}\ \bibnamefont {Mrou{\'e}}}, \bibinfo {author}
  {\bibfnamefont {H.~P.}\ \bibnamefont {Pfeiffer}}, \bibinfo {author}
  {\bibfnamefont {M.~A.}\ \bibnamefont {Scheel}}, \bibinfo {author}
  {\bibfnamefont {G.~B.}\ \bibnamefont {Cook}}, \ and\ \bibinfo {author}
  {\bibfnamefont {S.~A.}\ \bibnamefont {Teukolsky}},\ }\href@noop {} {\bibfield
   {journal} {\bibinfo  {journal} {Phys.\ Rev.\ D}\ }\textbf {\bibinfo {volume}
  {76}},\ \bibinfo {pages} {124038} (\bibinfo {year} {2007})},\ \Eprint
  {http://arxiv.org/abs/0710.0158} {arXiv:0710.0158 [gr-qc]} \BibitemShut
  {NoStop}%
\bibitem [{\citenamefont {Buchman}\ and\ \citenamefont
  {Sarbach}(2007)}]{Buchman2007}%
  \BibitemOpen
  \bibfield  {author} {\bibinfo {author} {\bibfnamefont {L.~T.}\ \bibnamefont
  {Buchman}}\ and\ \bibinfo {author} {\bibfnamefont {O.~C.~A.}\ \bibnamefont
  {Sarbach}},\ }\href@noop {} {\bibfield  {journal} {\bibinfo  {journal}
  {Class.\ Quantum Grav.}\ }\textbf {\bibinfo {volume} {24}},\ \bibinfo {pages}
  {S307} (\bibinfo {year} {2007})}\BibitemShut {NoStop}%
\bibitem [{\citenamefont {Rinne}\ \emph {et~al.}(2009)\citenamefont {Rinne},
  \citenamefont {Buchman}, \citenamefont {Scheel},\ and\ \citenamefont
  {Pfeiffer}}]{Rinne2008b}%
  \BibitemOpen
  \bibfield  {author} {\bibinfo {author} {\bibfnamefont {O.}~\bibnamefont
  {Rinne}}, \bibinfo {author} {\bibfnamefont {L.~T.}\ \bibnamefont {Buchman}},
  \bibinfo {author} {\bibfnamefont {M.~A.}\ \bibnamefont {Scheel}}, \ and\
  \bibinfo {author} {\bibfnamefont {H.~P.}\ \bibnamefont {Pfeiffer}},\
  }\href@noop {} {\bibfield  {journal} {\bibinfo  {journal} {Class.\ Quantum
  Grav.}\ }\textbf {\bibinfo {volume} {26}},\ \bibinfo {pages} {075009}
  (\bibinfo {year} {2009})}\BibitemShut {NoStop}%
\bibitem [{\citenamefont {Bishop}\ \emph {et~al.}(1997)\citenamefont {Bishop},
  \citenamefont {Gomez}, \citenamefont {Lehner}, \citenamefont {Maharaj},\ and\
  \citenamefont {Winicour}}]{Bishop:1997ik}%
  \BibitemOpen
  \bibfield  {author} {\bibinfo {author} {\bibfnamefont {N.~T.}\ \bibnamefont
  {Bishop}}, \bibinfo {author} {\bibfnamefont {R.}~\bibnamefont {Gomez}},
  \bibinfo {author} {\bibfnamefont {L.}~\bibnamefont {Lehner}}, \bibinfo
  {author} {\bibfnamefont {M.}~\bibnamefont {Maharaj}}, \ and\ \bibinfo
  {author} {\bibfnamefont {J.}~\bibnamefont {Winicour}},\ }\href {\doibase
  10.1103/PhysRevD.56.6298} {\bibfield  {journal} {\bibinfo  {journal} {Phys.\
  Rev. D}\ }\textbf {\bibinfo {volume} {56}},\ \bibinfo {pages} {6298}
  (\bibinfo {year} {1997})},\ \Eprint {http://arxiv.org/abs/gr-qc/9708065}
  {arXiv:gr-qc/9708065} \BibitemShut {NoStop}%
\bibitem [{\citenamefont {Winicour}(2005)}]{Winicour2005}%
  \BibitemOpen
  \bibfield  {author} {\bibinfo {author} {\bibfnamefont {J.}~\bibnamefont
  {Winicour}},\ }\href {http://www.livingreviews.org/lrr-2005-10} {\bibfield
  {journal} {\bibinfo  {journal} {Living Rev.\ Rel.}\ }\textbf {\bibinfo
  {volume} {8}} (\bibinfo {year} {2005})}\BibitemShut {NoStop}%
\bibitem [{\citenamefont {Gomez}\ \emph {et~al.}(2007)\citenamefont {Gomez},
  \citenamefont {Barreto},\ and\ \citenamefont {Frittelli}}]{Gomez:2007cj}%
  \BibitemOpen
  \bibfield  {author} {\bibinfo {author} {\bibfnamefont {R.}~\bibnamefont
  {Gomez}}, \bibinfo {author} {\bibfnamefont {W.}~\bibnamefont {Barreto}}, \
  and\ \bibinfo {author} {\bibfnamefont {S.}~\bibnamefont {Frittelli}},\ }\href
  {\doibase 10.1103/PhysRevD.76.124029} {\bibfield  {journal} {\bibinfo
  {journal} {Phys.\ Rev.\ D}\ }\textbf {\bibinfo {volume} {76}},\ \bibinfo
  {pages} {124029} (\bibinfo {year} {2007})},\ \Eprint
  {http://arxiv.org/abs/0711.0564} {arXiv:0711.0564 [gr-qc]} \BibitemShut
  {NoStop}%
\bibitem [{\citenamefont {Reisswig}\ \emph {et~al.}(2013)\citenamefont
  {Reisswig}, \citenamefont {Bishop},\ and\ \citenamefont
  {Pollney}}]{Reisswig:2012ka}%
  \BibitemOpen
  \bibfield  {author} {\bibinfo {author} {\bibfnamefont {C.}~\bibnamefont
  {Reisswig}}, \bibinfo {author} {\bibfnamefont {N.~T.}\ \bibnamefont
  {Bishop}}, \ and\ \bibinfo {author} {\bibfnamefont {D.}~\bibnamefont
  {Pollney}},\ }\href@noop {} {\bibfield  {journal} {\bibinfo  {journal} {{Gen.
  Rel. Grav.}}\ }\textbf {\bibinfo {volume} {45}},\ \bibinfo {pages} {1069}
  (\bibinfo {year} {2013})},\ \Eprint {http://arxiv.org/abs/1208.3891}
  {arXiv:1208.3891 [gr-qc]} \BibitemShut {NoStop}%
\bibitem [{\citenamefont {Handmer}\ and\ \citenamefont
  {Szil\'{a}gyi}(2015)}]{Handmer:2014}%
  \BibitemOpen
  \bibfield  {author} {\bibinfo {author} {\bibfnamefont {C.~J.}\ \bibnamefont
  {Handmer}}\ and\ \bibinfo {author} {\bibfnamefont {B.}~\bibnamefont
  {Szil\'{a}gyi}},\ }\href@noop {} {\bibfield  {journal} {\bibinfo  {journal}
  {Class.\ Quantum Grav.}\ }\textbf {\bibinfo {volume} {32}},\ \bibinfo {pages}
  {025008} (\bibinfo {year} {2015})},\ \Eprint {http://arxiv.org/abs/1406.7029}
  {arXiv:1406.7029} \BibitemShut {NoStop}%
\bibitem [{\citenamefont {Boyle}\ and\ \citenamefont
  {Mrou{\'{e}}}(2009)}]{Boyle-Mroue:2008}%
  \BibitemOpen
  \bibfield  {author} {\bibinfo {author} {\bibfnamefont {M.}~\bibnamefont
  {Boyle}}\ and\ \bibinfo {author} {\bibfnamefont {A.~H.}\ \bibnamefont
  {Mrou{\'{e}}}},\ }\href {\doibase 10.1103/PhysRevD.80.124045} {\bibfield
  {journal} {\bibinfo  {journal} {Phys.\ Rev.\ D}\ }\textbf {\bibinfo {volume}
  {80}},\ \bibinfo {pages} {124045} (\bibinfo {year} {2009})},\ \Eprint
  {http://arxiv.org/abs/0905.3177} {arXiv:0905.3177 [gr-qc]} \BibitemShut
  {NoStop}%
\bibitem [{\citenamefont {{Taylor}}\ \emph {et~al.}(2013)\citenamefont
  {{Taylor}}, \citenamefont {{Boyle}}, \citenamefont {{Reisswig}},
  \citenamefont {{Scheel}}, \citenamefont {{Chu}}, \citenamefont {{Kidder}},\
  and\ \citenamefont {{Szil{\'a}gyi}}}]{Taylor:2013zia}%
  \BibitemOpen
  \bibfield  {author} {\bibinfo {author} {\bibfnamefont {N.~W.}\ \bibnamefont
  {{Taylor}}}, \bibinfo {author} {\bibfnamefont {M.}~\bibnamefont {{Boyle}}},
  \bibinfo {author} {\bibfnamefont {C.}~\bibnamefont {{Reisswig}}}, \bibinfo
  {author} {\bibfnamefont {M.~A.}\ \bibnamefont {{Scheel}}}, \bibinfo {author}
  {\bibfnamefont {T.}~\bibnamefont {{Chu}}}, \bibinfo {author} {\bibfnamefont
  {L.~E.}\ \bibnamefont {{Kidder}}}, \ and\ \bibinfo {author} {\bibfnamefont
  {B.}~\bibnamefont {{Szil{\'a}gyi}}},\ }\href {\doibase
  10.1103/PhysRevD.88.124010} {\bibfield  {journal} {\bibinfo  {journal}
  {Phys.\ Rev.\ D}\ }\textbf {\bibinfo {volume} {88}},\ \bibinfo {eid} {124010}
  (\bibinfo {year} {2013})},\ \Eprint {http://arxiv.org/abs/1309.3605}
  {arXiv:1309.3605 [gr-qc]} \BibitemShut {NoStop}%
\bibitem [{\citenamefont {{Chu}}\ \emph {et~al.}(2016)\citenamefont {{Chu}},
  \citenamefont {{Fong}}, \citenamefont {{Kumar}}, \citenamefont {{Pfeiffer}},
  \citenamefont {{Boyle}}, \citenamefont {{Hemberger}}, \citenamefont
  {{Kidder}}, \citenamefont {{Scheel}},\ and\ \citenamefont
  {{Szil{\'a}gyi}}}]{Chu:2015kft}%
  \BibitemOpen
  \bibfield  {author} {\bibinfo {author} {\bibfnamefont {T.}~\bibnamefont
  {{Chu}}}, \bibinfo {author} {\bibfnamefont {H.}~\bibnamefont {{Fong}}},
  \bibinfo {author} {\bibfnamefont {P.}~\bibnamefont {{Kumar}}}, \bibinfo
  {author} {\bibfnamefont {H.~P.}\ \bibnamefont {{Pfeiffer}}}, \bibinfo
  {author} {\bibfnamefont {M.}~\bibnamefont {{Boyle}}}, \bibinfo {author}
  {\bibfnamefont {D.~A.}\ \bibnamefont {{Hemberger}}}, \bibinfo {author}
  {\bibfnamefont {L.~E.}\ \bibnamefont {{Kidder}}}, \bibinfo {author}
  {\bibfnamefont {M.~A.}\ \bibnamefont {{Scheel}}}, \ and\ \bibinfo {author}
  {\bibfnamefont {B.}~\bibnamefont {{Szil{\'a}gyi}}},\ }\href {\doibase
  10.1088/0264-9381/33/16/165001} {\bibfield  {journal} {\bibinfo  {journal}
  {Class.\ Quantum Grav.}\ }\textbf {\bibinfo {volume} {33}},\ \bibinfo {eid}
  {165001} (\bibinfo {year} {2016})},\ \Eprint
  {http://arxiv.org/abs/1512.06800} {arXiv:1512.06800 [gr-qc]} \BibitemShut
  {NoStop}%
\bibitem [{\citenamefont {Boyle}(2016{\natexlab{a}})}]{Boyle2015a}%
  \BibitemOpen
  \bibfield  {author} {\bibinfo {author} {\bibfnamefont {M.}~\bibnamefont
  {Boyle}},\ }\href {\doibase 10.1103/PhysRevD.93.084031} {\bibfield  {journal}
  {\bibinfo  {journal} {Phys.\ Rev.\ D}\ }\textbf {\bibinfo {volume} {93}},\
  \bibinfo {pages} {084031} (\bibinfo {year} {2016}{\natexlab{a}})}\BibitemShut
  {NoStop}%
\bibitem [{\citenamefont {Abbott}\ \emph
  {et~al.}(2017{\natexlab{c}})\citenamefont {Abbott} \emph
  {et~al.}}]{Abbott:2017vtc}%
  \BibitemOpen
  \bibfield  {author} {\bibinfo {author} {\bibfnamefont {B.~P.}\ \bibnamefont
  {Abbott}} \emph {et~al.} (\bibinfo {collaboration} {VIRGO, LIGO
  Scientific}),\ }\href {\doibase 10.1103/PhysRevLett.118.221101} {\bibfield
  {journal} {\bibinfo  {journal} {Phys. Rev. Lett.}\ }\textbf {\bibinfo
  {volume} {118}},\ \bibinfo {pages} {221101} (\bibinfo {year}
  {2017}{\natexlab{c}})},\ \Eprint {http://arxiv.org/abs/1706.01812}
  {arXiv:1706.01812 [gr-qc]} \BibitemShut {NoStop}%
\bibitem [{\citenamefont {Veitch}\ \emph {et~al.}(2015)\citenamefont {Veitch},
  \citenamefont {Raymond}, \citenamefont {Farr}, \citenamefont {Farr},
  \citenamefont {Graff}, \citenamefont {Vitale}, \citenamefont {Aylott},
  \citenamefont {Blackburn}, \citenamefont {Christensen}, \citenamefont
  {Coughlin}, \citenamefont {Del~Pozzo}, \citenamefont {Feroz}, \citenamefont
  {Gair}, \citenamefont {Haster}, \citenamefont {Kalogera}, \citenamefont
  {Littenberg}, \citenamefont {Mandel}, \citenamefont {O'Shaughnessy},
  \citenamefont {Pitkin}, \citenamefont {Rodriguez}, \citenamefont {R\"over},
  \citenamefont {Sidery}, \citenamefont {Smith}, \citenamefont {Van Der~Sluys},
  \citenamefont {Vecchio}, \citenamefont {Vousden},\ and\ \citenamefont
  {Wade}}]{Veitch:2015}%
  \BibitemOpen
  \bibfield  {author} {\bibinfo {author} {\bibfnamefont {J.}~\bibnamefont
  {Veitch}}, \bibinfo {author} {\bibfnamefont {V.}~\bibnamefont {Raymond}},
  \bibinfo {author} {\bibfnamefont {B.}~\bibnamefont {Farr}}, \bibinfo {author}
  {\bibfnamefont {W.}~\bibnamefont {Farr}}, \bibinfo {author} {\bibfnamefont
  {P.}~\bibnamefont {Graff}}, \bibinfo {author} {\bibfnamefont
  {S.}~\bibnamefont {Vitale}}, \bibinfo {author} {\bibfnamefont
  {B.}~\bibnamefont {Aylott}}, \bibinfo {author} {\bibfnamefont
  {K.}~\bibnamefont {Blackburn}}, \bibinfo {author} {\bibfnamefont
  {N.}~\bibnamefont {Christensen}}, \bibinfo {author} {\bibfnamefont
  {M.}~\bibnamefont {Coughlin}}, \bibinfo {author} {\bibfnamefont
  {W.}~\bibnamefont {Del~Pozzo}}, \bibinfo {author} {\bibfnamefont
  {F.}~\bibnamefont {Feroz}}, \bibinfo {author} {\bibfnamefont
  {J.}~\bibnamefont {Gair}}, \bibinfo {author} {\bibfnamefont {C.-J.}\
  \bibnamefont {Haster}}, \bibinfo {author} {\bibfnamefont {V.}~\bibnamefont
  {Kalogera}}, \bibinfo {author} {\bibfnamefont {T.}~\bibnamefont
  {Littenberg}}, \bibinfo {author} {\bibfnamefont {I.}~\bibnamefont {Mandel}},
  \bibinfo {author} {\bibfnamefont {R.}~\bibnamefont {O'Shaughnessy}}, \bibinfo
  {author} {\bibfnamefont {M.}~\bibnamefont {Pitkin}}, \bibinfo {author}
  {\bibfnamefont {C.}~\bibnamefont {Rodriguez}}, \bibinfo {author}
  {\bibfnamefont {C.}~\bibnamefont {R\"over}}, \bibinfo {author} {\bibfnamefont
  {T.}~\bibnamefont {Sidery}}, \bibinfo {author} {\bibfnamefont
  {R.}~\bibnamefont {Smith}}, \bibinfo {author} {\bibfnamefont
  {M.}~\bibnamefont {Van Der~Sluys}}, \bibinfo {author} {\bibfnamefont
  {A.}~\bibnamefont {Vecchio}}, \bibinfo {author} {\bibfnamefont
  {W.}~\bibnamefont {Vousden}}, \ and\ \bibinfo {author} {\bibfnamefont
  {L.}~\bibnamefont {Wade}},\ }\href {\doibase 10.1103/PhysRevD.91.042003}
  {\bibfield  {journal} {\bibinfo  {journal} {Phys.\ Rev.\ D}\ }\textbf
  {\bibinfo {volume} {91}},\ \bibinfo {pages} {042003} (\bibinfo {year}
  {2015})}\BibitemShut {NoStop}%
\bibitem [{Note1()}]{Note1}%
  \BibitemOpen
  \bibinfo {note} {Https://dcc.ligo.org/DocDB/0123/T1500606/006/\\
  NRInjectionInfrastructure.pdf}\BibitemShut {NoStop}%
\bibitem [{\citenamefont {Boyle}(2016{\natexlab{b}})}]{Boyle:2015nqa}%
  \BibitemOpen
  \bibfield  {author} {\bibinfo {author} {\bibfnamefont {M.}~\bibnamefont
  {Boyle}},\ }\href {\doibase 10.1103/PhysRevD.93.084031} {\bibfield  {journal}
  {\bibinfo  {journal} {Phys. Rev.}\ }\textbf {\bibinfo {volume} {D93}},\
  \bibinfo {pages} {084031} (\bibinfo {year} {2016}{\natexlab{b}})},\ \Eprint
  {http://arxiv.org/abs/1509.00862} {arXiv:1509.00862 [gr-qc]} \BibitemShut
  {NoStop}%
\bibitem [{\citenamefont {Kelly}\ and\ \citenamefont
  {Baker}(2013)}]{Kelly:2012nd}%
  \BibitemOpen
  \bibfield  {author} {\bibinfo {author} {\bibfnamefont {B.~J.}\ \bibnamefont
  {Kelly}}\ and\ \bibinfo {author} {\bibfnamefont {J.~G.}\ \bibnamefont
  {Baker}},\ }\href {\doibase 10.1103/PhysRevD.87.084004} {\bibfield  {journal}
  {\bibinfo  {journal} {Phys. Rev.}\ }\textbf {\bibinfo {volume} {D87}},\
  \bibinfo {pages} {084004} (\bibinfo {year} {2013})},\ \Eprint
  {http://arxiv.org/abs/1212.5553} {arXiv:1212.5553 [gr-qc]} \BibitemShut
  {NoStop}%
\bibitem [{\citenamefont {{Lovelace}}\ \emph {et~al.}(2016)\citenamefont
  {{Lovelace}}, \citenamefont {{Lousto}}, \citenamefont {{Healy}},
  \citenamefont {{Scheel}}, \citenamefont {{Garcia}}, \citenamefont
  {{O'Shaughnessy}}, \citenamefont {{Boyle}}, \citenamefont {{Campanelli}},
  \citenamefont {{Hemberger}}, \citenamefont {{Kidder}}, \citenamefont
  {{Pfeiffer}}, \citenamefont {{Szil{\'a}gyi}}, \citenamefont {{Teukolsky}},\
  and\ \citenamefont {{Zlochower}}}]{LovelaceLousto2016}%
  \BibitemOpen
  \bibfield  {author} {\bibinfo {author} {\bibfnamefont {G.}~\bibnamefont
  {{Lovelace}}}, \bibinfo {author} {\bibfnamefont {C.~O.}\ \bibnamefont
  {{Lousto}}}, \bibinfo {author} {\bibfnamefont {J.}~\bibnamefont {{Healy}}},
  \bibinfo {author} {\bibfnamefont {M.~A.}\ \bibnamefont {{Scheel}}}, \bibinfo
  {author} {\bibfnamefont {A.}~\bibnamefont {{Garcia}}}, \bibinfo {author}
  {\bibfnamefont {R.}~\bibnamefont {{O'Shaughnessy}}}, \bibinfo {author}
  {\bibfnamefont {M.}~\bibnamefont {{Boyle}}}, \bibinfo {author} {\bibfnamefont
  {M.}~\bibnamefont {{Campanelli}}}, \bibinfo {author} {\bibfnamefont {D.~A.}\
  \bibnamefont {{Hemberger}}}, \bibinfo {author} {\bibfnamefont {L.~E.}\
  \bibnamefont {{Kidder}}}, \bibinfo {author} {\bibfnamefont {H.~P.}\
  \bibnamefont {{Pfeiffer}}}, \bibinfo {author} {\bibfnamefont
  {B.}~\bibnamefont {{Szil{\'a}gyi}}}, \bibinfo {author} {\bibfnamefont
  {S.~A.}\ \bibnamefont {{Teukolsky}}}, \ and\ \bibinfo {author} {\bibfnamefont
  {Y.}~\bibnamefont {{Zlochower}}},\ }\href@noop {} {\bibfield  {journal}
  {\bibinfo  {journal} {Submitted to Class.~Quantum~Grav.; arXiv:1607.05377}\ }
  (\bibinfo {year} {2016})},\ \Eprint {http://arxiv.org/abs/1607.05377}
  {arXiv:1607.05377 [gr-qc]} \BibitemShut {NoStop}%
\bibitem [{\citenamefont {Boh\'e}\ \emph {et~al.}(2017)\citenamefont {Boh\'e},
  \citenamefont {Shao}, \citenamefont {Taracchini}, \citenamefont {Buonanno},
  \citenamefont {Babak}, \citenamefont {Harry}, \citenamefont {Hinder},
  \citenamefont {Ossokine}, \citenamefont {P\"urrer}, \citenamefont {Raymond},
  \citenamefont {Chu}, \citenamefont {Fong}, \citenamefont {Kumar},
  \citenamefont {Pfeiffer}, \citenamefont {Boyle}, \citenamefont {Hemberger},
  \citenamefont {Kidder}, \citenamefont {Lovelace}, \citenamefont {Scheel},\
  and\ \citenamefont {Szil\'agyi}}]{Bohe:2016gbl}%
  \BibitemOpen
  \bibfield  {author} {\bibinfo {author} {\bibfnamefont {A.}~\bibnamefont
  {Boh\'e}}, \bibinfo {author} {\bibfnamefont {L.}~\bibnamefont {Shao}},
  \bibinfo {author} {\bibfnamefont {A.}~\bibnamefont {Taracchini}}, \bibinfo
  {author} {\bibfnamefont {A.}~\bibnamefont {Buonanno}}, \bibinfo {author}
  {\bibfnamefont {S.}~\bibnamefont {Babak}}, \bibinfo {author} {\bibfnamefont
  {I.~W.}\ \bibnamefont {Harry}}, \bibinfo {author} {\bibfnamefont
  {I.}~\bibnamefont {Hinder}}, \bibinfo {author} {\bibfnamefont
  {S.}~\bibnamefont {Ossokine}}, \bibinfo {author} {\bibfnamefont
  {M.}~\bibnamefont {P\"urrer}}, \bibinfo {author} {\bibfnamefont
  {V.}~\bibnamefont {Raymond}}, \bibinfo {author} {\bibfnamefont
  {T.}~\bibnamefont {Chu}}, \bibinfo {author} {\bibfnamefont {H.}~\bibnamefont
  {Fong}}, \bibinfo {author} {\bibfnamefont {P.}~\bibnamefont {Kumar}},
  \bibinfo {author} {\bibfnamefont {H.~P.}\ \bibnamefont {Pfeiffer}}, \bibinfo
  {author} {\bibfnamefont {M.}~\bibnamefont {Boyle}}, \bibinfo {author}
  {\bibfnamefont {D.~A.}\ \bibnamefont {Hemberger}}, \bibinfo {author}
  {\bibfnamefont {L.~E.}\ \bibnamefont {Kidder}}, \bibinfo {author}
  {\bibfnamefont {G.}~\bibnamefont {Lovelace}}, \bibinfo {author}
  {\bibfnamefont {M.~A.}\ \bibnamefont {Scheel}}, \ and\ \bibinfo {author}
  {\bibfnamefont {B.}~\bibnamefont {Szil\'agyi}},\ }\href {\doibase
  10.1103/PhysRevD.95.044028} {\bibfield  {journal} {\bibinfo  {journal} {Phys.
  Rev. D}\ }\textbf {\bibinfo {volume} {95}},\ \bibinfo {pages} {044028}
  (\bibinfo {year} {2017})},\ \Eprint {http://arxiv.org/abs/1611.03703}
  {arXiv:1611.03703 [gr-qc]} \BibitemShut {NoStop}%
\bibitem [{\citenamefont {Healy}\ and\ \citenamefont
  {Lousto}(2017)}]{Healy:2016lce}%
  \BibitemOpen
  \bibfield  {author} {\bibinfo {author} {\bibfnamefont {J.}~\bibnamefont
  {Healy}}\ and\ \bibinfo {author} {\bibfnamefont {C.~O.}\ \bibnamefont
  {Lousto}},\ }\href {\doibase 10.1103/PhysRevD.95.024037} {\bibfield
  {journal} {\bibinfo  {journal} {Phys. Rev.}\ }\textbf {\bibinfo {volume}
  {D95}},\ \bibinfo {pages} {024037} (\bibinfo {year} {2017})},\ \Eprint
  {http://arxiv.org/abs/1610.09713} {arXiv:1610.09713 [gr-qc]} \BibitemShut
  {NoStop}%
\bibitem [{\citenamefont {Jim\'nez-Forteza}\ \emph {et~al.}(2017)\citenamefont
  {Jim\'nez-Forteza}, \citenamefont {Keitel}, \citenamefont {Husa},
  \citenamefont {Hannam}, \citenamefont {Khan},\ and\ \citenamefont
  {Pürrer}}]{Jimenez-Forteza:2016oae}%
  \BibitemOpen
  \bibfield  {author} {\bibinfo {author} {\bibfnamefont {X.}~\bibnamefont
  {Jim\'nez-Forteza}}, \bibinfo {author} {\bibfnamefont {D.}~\bibnamefont
  {Keitel}}, \bibinfo {author} {\bibfnamefont {S.}~\bibnamefont {Husa}},
  \bibinfo {author} {\bibfnamefont {M.}~\bibnamefont {Hannam}}, \bibinfo
  {author} {\bibfnamefont {S.}~\bibnamefont {Khan}}, \ and\ \bibinfo {author}
  {\bibfnamefont {M.}~\bibnamefont {Pürrer}},\ }\href {\doibase
  10.1103/PhysRevD.95.064024} {\bibfield  {journal} {\bibinfo  {journal} {Phys.
  Rev.}\ }\textbf {\bibinfo {volume} {D95}},\ \bibinfo {pages} {064024}
  (\bibinfo {year} {2017})},\ \Eprint {http://arxiv.org/abs/1611.00332}
  {arXiv:1611.00332 [gr-qc]} \BibitemShut {NoStop}%
\bibitem [{\citenamefont {Blackman}\ \emph
  {et~al.}(2017{\natexlab{a}})\citenamefont {Blackman}, \citenamefont {Field},
  \citenamefont {Scheel}, \citenamefont {Galley}, \citenamefont {Ott},
  \citenamefont {Boyle}, \citenamefont {Kidder}, \citenamefont {Pfeiffer},\
  and\ \citenamefont {Szilágyi}}]{Blackman:2017pcm}%
  \BibitemOpen
  \bibfield  {author} {\bibinfo {author} {\bibfnamefont {J.}~\bibnamefont
  {Blackman}}, \bibinfo {author} {\bibfnamefont {S.~E.}\ \bibnamefont {Field}},
  \bibinfo {author} {\bibfnamefont {M.~A.}\ \bibnamefont {Scheel}}, \bibinfo
  {author} {\bibfnamefont {C.~R.}\ \bibnamefont {Galley}}, \bibinfo {author}
  {\bibfnamefont {C.~D.}\ \bibnamefont {Ott}}, \bibinfo {author} {\bibfnamefont
  {M.}~\bibnamefont {Boyle}}, \bibinfo {author} {\bibfnamefont {L.~E.}\
  \bibnamefont {Kidder}}, \bibinfo {author} {\bibfnamefont {H.~P.}\
  \bibnamefont {Pfeiffer}}, \ and\ \bibinfo {author} {\bibfnamefont
  {B.}~\bibnamefont {Szilágyi}},\ }\href {\doibase 10.1103/PhysRevD.96.024058}
  {\bibfield  {journal} {\bibinfo  {journal} {Phys. Rev.}\ }\textbf {\bibinfo
  {volume} {D96}},\ \bibinfo {pages} {024058} (\bibinfo {year}
  {2017}{\natexlab{a}})},\ \Eprint {http://arxiv.org/abs/1705.07089}
  {arXiv:1705.07089 [gr-qc]} \BibitemShut {NoStop}%
\bibitem [{\citenamefont {London}\ \emph {et~al.}(2017)\citenamefont {London},
  \citenamefont {Khan}, \citenamefont {Fauchon-Jones}, \citenamefont {Forteza},
  \citenamefont {Hannam}, \citenamefont {Husa}, \citenamefont {Kalaghatgi},
  \citenamefont {Ohme},\ and\ \citenamefont {Pannarale}}]{London:2017bcn}%
  \BibitemOpen
  \bibfield  {author} {\bibinfo {author} {\bibfnamefont {L.}~\bibnamefont
  {London}}, \bibinfo {author} {\bibfnamefont {S.}~\bibnamefont {Khan}},
  \bibinfo {author} {\bibfnamefont {E.}~\bibnamefont {Fauchon-Jones}}, \bibinfo
  {author} {\bibfnamefont {X.~J.}\ \bibnamefont {Forteza}}, \bibinfo {author}
  {\bibfnamefont {M.}~\bibnamefont {Hannam}}, \bibinfo {author} {\bibfnamefont
  {S.}~\bibnamefont {Husa}}, \bibinfo {author} {\bibfnamefont {C.}~\bibnamefont
  {Kalaghatgi}}, \bibinfo {author} {\bibfnamefont {F.}~\bibnamefont {Ohme}}, \
  and\ \bibinfo {author} {\bibfnamefont {F.}~\bibnamefont {Pannarale}},\
  }\href@noop {} {\  (\bibinfo {year} {2017})},\ \Eprint
  {http://arxiv.org/abs/1708.00404} {arXiv:1708.00404 [gr-qc]} \BibitemShut
  {NoStop}%
\bibitem [{\citenamefont {Pankow}\ \emph {et~al.}(2015)\citenamefont {Pankow},
  \citenamefont {Brady}, \citenamefont {Ochsner},\ and\ \citenamefont
  {O'Shaughnessy}}]{Pankow:2015cra}%
  \BibitemOpen
  \bibfield  {author} {\bibinfo {author} {\bibfnamefont {C.}~\bibnamefont
  {Pankow}}, \bibinfo {author} {\bibfnamefont {P.}~\bibnamefont {Brady}},
  \bibinfo {author} {\bibfnamefont {E.}~\bibnamefont {Ochsner}}, \ and\
  \bibinfo {author} {\bibfnamefont {R.}~\bibnamefont {O'Shaughnessy}},\ }\href
  {\doibase 10.1103/PhysRevD.92.023002} {\bibfield  {journal} {\bibinfo
  {journal} {Phys. Rev.}\ }\textbf {\bibinfo {volume} {D92}},\ \bibinfo {pages}
  {023002} (\bibinfo {year} {2015})},\ \Eprint
  {http://arxiv.org/abs/1502.04370} {arXiv:1502.04370 [gr-qc]} \BibitemShut
  {NoStop}%
\bibitem [{\citenamefont {Abbott}\ \emph
  {et~al.}(2016{\natexlab{e}})\citenamefont {Abbott} \emph
  {et~al.}}]{Abbott:2016apu}%
  \BibitemOpen
  \bibfield  {author} {\bibinfo {author} {\bibfnamefont {B.~P.}\ \bibnamefont
  {Abbott}} \emph {et~al.} (\bibinfo {collaboration} {LIGO Scientific
  Collaboration, Virgo Collaboration}),\ }\href@noop {} {\bibfield  {journal}
  {\bibinfo  {journal} {Phys.~Rev.~D}\ }\textbf {\bibinfo {volume} {94}},\
  \bibinfo {pages} {064035} (\bibinfo {year} {2016}{\natexlab{e}})},\ \Eprint
  {http://arxiv.org/abs/1606.01262} {arXiv:1606.01262 [gr-qc]} \BibitemShut
  {NoStop}%
\bibitem [{\citenamefont {Zlochower}\ and\ \citenamefont
  {Lousto}(2015)}]{Zlochower:2015wga}%
  \BibitemOpen
  \bibfield  {author} {\bibinfo {author} {\bibfnamefont {Y.}~\bibnamefont
  {Zlochower}}\ and\ \bibinfo {author} {\bibfnamefont {C.~O.}\ \bibnamefont
  {Lousto}},\ }\href {\doibase 10.1103/PhysRevD.92.024022} {\bibfield
  {journal} {\bibinfo  {journal} {Phys. Rev.}\ }\textbf {\bibinfo {volume}
  {D92}},\ \bibinfo {pages} {024022} (\bibinfo {year} {2015})},\ \Eprint
  {http://arxiv.org/abs/1503.07536} {arXiv:1503.07536 [gr-qc]} \BibitemShut
  {NoStop}%
\bibitem [{Note2()}]{Note2}%
  \BibitemOpen
  \bibinfo {note} {Weisstein, Eric W.``Hammer-Aitoff Equal-Area Projection.''
  From MathWorld--A Wolfram Web Resource.
  http://mathworld.wolfram.com/Hammer-AitoffEqual-AreaProjection.html}\BibitemShut
  {NoStop}%
\bibitem [{\citenamefont {{Blackman}}\ \emph {et~al.}(2015)\citenamefont
  {{Blackman}}, \citenamefont {{Field}}, \citenamefont {{Galley}},
  \citenamefont {{Szil{\'a}gyi}}, \citenamefont {{Scheel}}, \citenamefont
  {{Tiglio}},\ and\ \citenamefont {{Hemberger}}}]{Blackman:2015pia}%
  \BibitemOpen
  \bibfield  {author} {\bibinfo {author} {\bibfnamefont {J.}~\bibnamefont
  {{Blackman}}}, \bibinfo {author} {\bibfnamefont {S.~E.}\ \bibnamefont
  {{Field}}}, \bibinfo {author} {\bibfnamefont {C.~R.}\ \bibnamefont
  {{Galley}}}, \bibinfo {author} {\bibfnamefont {B.}~\bibnamefont
  {{Szil{\'a}gyi}}}, \bibinfo {author} {\bibfnamefont {M.~A.}\ \bibnamefont
  {{Scheel}}}, \bibinfo {author} {\bibfnamefont {M.}~\bibnamefont {{Tiglio}}},
  \ and\ \bibinfo {author} {\bibfnamefont {D.~A.}\ \bibnamefont
  {{Hemberger}}},\ }\href {\doibase 10.1103/PhysRevLett.115.121102} {\bibfield
  {journal} {\bibinfo  {journal} {Phys.\ Rev.\ Lett.}\ }\textbf {\bibinfo
  {volume} {115}},\ \bibinfo {eid} {121102} (\bibinfo {year} {2015})},\ \Eprint
  {http://arxiv.org/abs/1502.07758} {arXiv:1502.07758 [gr-qc]} \BibitemShut
  {NoStop}%
\bibitem [{\citenamefont {Blackman}\ \emph
  {et~al.}(2017{\natexlab{b}})\citenamefont {Blackman}, \citenamefont {Field},
  \citenamefont {Scheel}, \citenamefont {Galley}, \citenamefont {Hemberger},
  \citenamefont {Schmidt},\ and\ \citenamefont {Smith}}]{Blackman:2017dfb}%
  \BibitemOpen
  \bibfield  {author} {\bibinfo {author} {\bibfnamefont {J.}~\bibnamefont
  {Blackman}}, \bibinfo {author} {\bibfnamefont {S.~E.}\ \bibnamefont {Field}},
  \bibinfo {author} {\bibfnamefont {M.~A.}\ \bibnamefont {Scheel}}, \bibinfo
  {author} {\bibfnamefont {C.~R.}\ \bibnamefont {Galley}}, \bibinfo {author}
  {\bibfnamefont {D.~A.}\ \bibnamefont {Hemberger}}, \bibinfo {author}
  {\bibfnamefont {P.}~\bibnamefont {Schmidt}}, \ and\ \bibinfo {author}
  {\bibfnamefont {R.}~\bibnamefont {Smith}},\ }\href@noop {} {\  (\bibinfo
  {year} {2017}{\natexlab{b}})},\ \Eprint {http://arxiv.org/abs/1701.00550}
  {arXiv:1701.00550 [gr-qc]} \BibitemShut {NoStop}%
\bibitem [{\citenamefont {Williamson}\ \emph {et~al.}(2017)\citenamefont
  {Williamson}, \citenamefont {Lange}, \citenamefont {O'Shaughnessy},
  \citenamefont {Clark}, \citenamefont {Kumar}, \citenamefont
  {Calderón~Bustillo},\ and\ \citenamefont {Veitch}}]{Williamson:2017evr}%
  \BibitemOpen
  \bibfield  {author} {\bibinfo {author} {\bibfnamefont {A.~R.}\ \bibnamefont
  {Williamson}}, \bibinfo {author} {\bibfnamefont {J.}~\bibnamefont {Lange}},
  \bibinfo {author} {\bibfnamefont {R.}~\bibnamefont {O'Shaughnessy}}, \bibinfo
  {author} {\bibfnamefont {J.~A.}\ \bibnamefont {Clark}}, \bibinfo {author}
  {\bibfnamefont {P.}~\bibnamefont {Kumar}}, \bibinfo {author} {\bibfnamefont
  {J.}~\bibnamefont {Calderón~Bustillo}}, \ and\ \bibinfo {author}
  {\bibfnamefont {J.}~\bibnamefont {Veitch}},\ }\href@noop {} {\  (\bibinfo
  {year} {2017})},\ \Eprint {http://arxiv.org/abs/1709.03095} {arXiv:1709.03095
  [gr-qc]} \BibitemShut {NoStop}%
\bibitem [{\citenamefont {{Abbott}}\ \emph {et~al.}(2017)\citenamefont
  {{Abbott}}, \citenamefont {{Abbott}}, \citenamefont {{Abbott}}, \citenamefont
  {{Abernathy}}, \citenamefont {{Acernese}}, \citenamefont {{Ackley}},
  \citenamefont {{Adams}}, \citenamefont {{Adams}}, \citenamefont {{Addesso}},
  \citenamefont {{Adhikari}},\ and\ \citenamefont
  {et~al.}}]{LIGO-O1-PENR-Systematics}%
  \BibitemOpen
  \bibfield  {author} {\bibinfo {author} {\bibfnamefont {B.~P.}\ \bibnamefont
  {{Abbott}}}, \bibinfo {author} {\bibfnamefont {R.}~\bibnamefont {{Abbott}}},
  \bibinfo {author} {\bibfnamefont {T.~D.}\ \bibnamefont {{Abbott}}}, \bibinfo
  {author} {\bibfnamefont {M.~R.}\ \bibnamefont {{Abernathy}}}, \bibinfo
  {author} {\bibfnamefont {F.}~\bibnamefont {{Acernese}}}, \bibinfo {author}
  {\bibfnamefont {K.}~\bibnamefont {{Ackley}}}, \bibinfo {author}
  {\bibfnamefont {C.}~\bibnamefont {{Adams}}}, \bibinfo {author} {\bibfnamefont
  {T.}~\bibnamefont {{Adams}}}, \bibinfo {author} {\bibfnamefont
  {P.}~\bibnamefont {{Addesso}}}, \bibinfo {author} {\bibfnamefont {R.~X.}\
  \bibnamefont {{Adhikari}}}, \ and\ \bibinfo {author} {\bibnamefont
  {et~al.}},\ }\href {\doibase 10.1088/1361-6382/aa6854} {\bibfield  {journal}
  {\bibinfo  {journal} {Classical and Quantum Gravity}\ }\textbf {\bibinfo
  {volume} {34}},\ \bibinfo {eid} {104002} (\bibinfo {year} {2017})},\ \Eprint
  {http://arxiv.org/abs/1611.07531} {arXiv:1611.07531 [gr-qc]} \BibitemShut
  {NoStop}%
\bibitem [{\citenamefont {Bustillo}\ \emph {et~al.}(2015)\citenamefont
  {Bustillo}, \citenamefont {Bohé}, \citenamefont {Husa}, \citenamefont
  {Sintes}, \citenamefont {Hannam} \emph {et~al.}}]{Bustillo:2015ova}%
  \BibitemOpen
  \bibfield  {author} {\bibinfo {author} {\bibfnamefont {J.~C.}\ \bibnamefont
  {Bustillo}}, \bibinfo {author} {\bibfnamefont {A.}~\bibnamefont {Bohé}},
  \bibinfo {author} {\bibfnamefont {S.}~\bibnamefont {Husa}}, \bibinfo {author}
  {\bibfnamefont {A.~M.}\ \bibnamefont {Sintes}}, \bibinfo {author}
  {\bibfnamefont {M.}~\bibnamefont {Hannam}},  \emph {et~al.},\ }\href@noop {}
  {\  (\bibinfo {year} {2015})},\ \Eprint {http://arxiv.org/abs/1501.00918}
  {arXiv:1501.00918 [gr-qc]} \BibitemShut {NoStop}%
\bibitem [{\citenamefont {Taracchini}\ \emph {et~al.}(2014)\citenamefont
  {Taracchini}, \citenamefont {Buonanno}, \citenamefont {Pan}, \citenamefont
  {Hinderer}, \citenamefont {Boyle}, \citenamefont {Hemberger}, \citenamefont
  {Kidder}, \citenamefont {Lovelace}, \citenamefont {Mroue}, \citenamefont
  {Pfeiffer}, \citenamefont {Scheel}, \citenamefont {Szil{\'a}gyi},
  \citenamefont {Taylor},\ and\ \citenamefont
  {Zenginoglu}}]{Taracchini:2013rva}%
  \BibitemOpen
  \bibfield  {author} {\bibinfo {author} {\bibfnamefont {A.}~\bibnamefont
  {Taracchini}}, \bibinfo {author} {\bibfnamefont {A.}~\bibnamefont
  {Buonanno}}, \bibinfo {author} {\bibfnamefont {Y.}~\bibnamefont {Pan}},
  \bibinfo {author} {\bibfnamefont {T.}~\bibnamefont {Hinderer}}, \bibinfo
  {author} {\bibfnamefont {M.}~\bibnamefont {Boyle}}, \bibinfo {author}
  {\bibfnamefont {D.~A.}\ \bibnamefont {Hemberger}}, \bibinfo {author}
  {\bibfnamefont {L.~E.}\ \bibnamefont {Kidder}}, \bibinfo {author}
  {\bibfnamefont {G.}~\bibnamefont {Lovelace}}, \bibinfo {author}
  {\bibfnamefont {A.~H.}\ \bibnamefont {Mroue}}, \bibinfo {author}
  {\bibfnamefont {H.~P.}\ \bibnamefont {Pfeiffer}}, \bibinfo {author}
  {\bibfnamefont {M.~A.}\ \bibnamefont {Scheel}}, \bibinfo {author}
  {\bibfnamefont {B.}~\bibnamefont {Szil{\'a}gyi}}, \bibinfo {author}
  {\bibfnamefont {N.~W.}\ \bibnamefont {Taylor}}, \ and\ \bibinfo {author}
  {\bibfnamefont {A.}~\bibnamefont {Zenginoglu}},\ }\href@noop {} {\bibfield
  {journal} {\bibinfo  {journal} {Phys.\ Rev.\ D}\ }\textbf {\bibinfo {volume}
  {89 (R)}},\ \bibinfo {pages} {061502} (\bibinfo {year} {2014})},\ \Eprint
  {http://arxiv.org/abs/1311.2544} {arXiv:1311.2544 [gr-qc]} \BibitemShut
  {NoStop}%
\bibitem [{\citenamefont {Abbott}\ \emph
  {et~al.}(2016{\natexlab{f}})\citenamefont {Abbott} \emph
  {et~al.}}]{Abbott:2016wiq}%
  \BibitemOpen
  \bibfield  {author} {\bibinfo {author} {\bibfnamefont {B.~P.}\ \bibnamefont
  {Abbott}} \emph {et~al.} (\bibinfo {collaboration} {Virgo, LIGO
  Scientific}),\ }\href@noop {} {\  (\bibinfo {year} {2016}{\natexlab{f}})},\
  \Eprint {http://arxiv.org/abs/1611.07531} {arXiv:1611.07531 [gr-qc]}
  \BibitemShut {NoStop}%
\bibitem [{\citenamefont {Loken}\ \emph {et~al.}(2010)\citenamefont {Loken},
  \citenamefont {Gruner}, \citenamefont {Groer}, \citenamefont {Peltier},
  \citenamefont {Bunn}, \citenamefont {Craig}, \citenamefont {Henriques},
  \citenamefont {Dempsey}, \citenamefont {Yu}, \citenamefont {Chen},
  \citenamefont {Dursi}, \citenamefont {Chong}, \citenamefont {Northrup},
  \citenamefont {Pinto}, \citenamefont {Knecht},\ and\ \citenamefont
  {Zon}}]{scinet}%
  \BibitemOpen
  \bibfield  {author} {\bibinfo {author} {\bibfnamefont {C.}~\bibnamefont
  {Loken}}, \bibinfo {author} {\bibfnamefont {D.}~\bibnamefont {Gruner}},
  \bibinfo {author} {\bibfnamefont {L.}~\bibnamefont {Groer}}, \bibinfo
  {author} {\bibfnamefont {R.}~\bibnamefont {Peltier}}, \bibinfo {author}
  {\bibfnamefont {N.}~\bibnamefont {Bunn}}, \bibinfo {author} {\bibfnamefont
  {M.}~\bibnamefont {Craig}}, \bibinfo {author} {\bibfnamefont
  {T.}~\bibnamefont {Henriques}}, \bibinfo {author} {\bibfnamefont
  {J.}~\bibnamefont {Dempsey}}, \bibinfo {author} {\bibfnamefont {C.-H.}\
  \bibnamefont {Yu}}, \bibinfo {author} {\bibfnamefont {J.}~\bibnamefont
  {Chen}}, \bibinfo {author} {\bibfnamefont {L.~J.}\ \bibnamefont {Dursi}},
  \bibinfo {author} {\bibfnamefont {J.}~\bibnamefont {Chong}}, \bibinfo
  {author} {\bibfnamefont {S.}~\bibnamefont {Northrup}}, \bibinfo {author}
  {\bibfnamefont {J.}~\bibnamefont {Pinto}}, \bibinfo {author} {\bibfnamefont
  {N.}~\bibnamefont {Knecht}}, \ and\ \bibinfo {author} {\bibfnamefont {R.~V.}\
  \bibnamefont {Zon}},\ }\href {\doibase 10.1088/1742-6596/256/1/012026}
  {\bibfield  {journal} {\bibinfo  {journal} {J. Phys.: Conf. Ser.}\ }\textbf
  {\bibinfo {volume} {256}},\ \bibinfo {pages} {012026} (\bibinfo {year}
  {2010})}\BibitemShut {NoStop}%
\end{thebibliography}%
\bibliographystyle{apsrev4-1}

\appendix

\section{Impact of discrete posterior sampling}
In the text, we performed several calculations that depend on on an inferred posterior distribution: identifying parameters for NR
followup calculations; maximizing the marginalized likelihood $\ln{\cal L}$; and calculating mismatches.   These
calculations are performed using a finite-size collection of approximately independent, identically-distributed samples
from a posterior distribution \cite{Veitch:2015}.    
In this appendix we briefly quantify the (small) effects our finite sample size has on our conclusions and comparisons.
For simplicity, we will conservatively standardize our calculations to $N=3000$ posterior samples; in practice, usually many more were used.

The match and marginalized likelihood distributions are well-described by a $\chi^2$ distribution with a suitable number
of degrees of freedom, corresponding to the model dimension of the intrinsic parameter space (i.e., $d=4$ for
calculations which omit precession,  and $d=8$ for calculations which include it).     For example if $\ln {\cal L}_*$
is the true maximum marginalized likelihood, then the marginalized likelihood distribution over the posterior is
well-approximated by the distribution of $\ln {\cal L} = \ln {\cal L}_{\rm max} - x/2$ where $x$ is $\chi^2$ distributed
with $d$ degrees of freedom.  If we have $N$ independent draws from the $\chi^2$ distribution, the smallest value of $x$
will be distributed according to $P(>x|d)^N = (1-P(<x|d))^N$, where $P(<x|d)$ is the cumulative  distribution for the
$\chi^2$ distribution.  At 95\% confidence, the maximum value of $x$ over the $N$ samples is therefore
$P^{-1}(0.05^{1/N})$.   As a result, if we estimate the maximum value of $\ln {\cal L}$ with the maximum over our
posterior samples, we find am estimate which is smaller than the true maximum value $\ln {\cal L}_{\rm max}$ by $0.5P^{-1}(0.05^{1/N})$.
Evaluating this expression for $d=4$ and $d=8$ in the conservative limit of only $N=3000$ samples, we find a systematic
sampling error of $0.045$ ($0.31$) in $d=4$ ($d=8$), respectively, in our estimate of the peak marginalized likelihood.
This systematic sampling error in our estimate of the peak marginalized likelihood is  smaller than the differences in
marginalized likelihoods discussed in the text and figures.

Likewise, given the number of samples, the targeted parameters should be very close to the true maximum a posteriori
values.  Qualitatively speaking, due to finite sample size effects, our estimate of  each parameter $z$ has an
uncertainty of roughly $\sigma_z/\sqrt{N}$, or roughly $2\%$ of the width of the distribution using our fiducial sample
size.  

\section{Mixed messages: Maximum likelihood, $\ln {\cal L}$, and  \emph{a posteriori}}
\label{ap:DifferentTargets}
One goal of this work is to demonstrate, by a concrete counterexample, that NR followup must be targeted and assessed
self-consistently.

One source of inconsistency in our original NR followup strategy was the algorithm by which NR simulations were
selected from model-based inference.  
Our  NR followup simulations were selected by (approximately) maximizing the a posteriori probability, proportional to the (15-dimensional)
likelihood $L$; the (7-dimensional) prior $p(\theta)$ for extrinsic parameters $\theta$; and the (8-dimensional) prior
for intrinsic parameters
$p(\lambda)$.    This maximum a posteriori (MaP) location does not generally correspond to the
parameters which maximize the likelihood (maxL).    The intrinsic parameters selected by both approaches also do not
cause the marginalized likelihood ${\cal L}_{\rm marg}(\lambda) = \int d\theta p(\theta) L(\lambda,\theta)$ to take on
its largest value.   
In principle, to avoid introducing artificial inconsistencies simply due to the choice of point estimate, we should have
targeted followup simulations using $\ln {\cal L}$.  
To assess how much our choice of targeting impacted our estimate of $\ln {\cal L}$, we evaluated the
marginalized likelihood at our estimates of all three locations.  Each location was approximated by our (finite-size) set of posterior
samples.    
For the posterior distribution adopted to generate UID4 -- a nonprecessing production-quality analysis where  SEOBNRv4 was both used to
generate the reference posterior used to find the MaP and maxL parameters and to compute a model-based $\ln {\cal L}$--
we find that the model-based $\ln {\cal L}$ values at the MaP and
maxL points to be effectively indistinguishable due to Monte Carlo error (61.4 and 61.2 respectively, with an estimated
Monte Carlo error of $0.1$).   
This similarity suggests that, when a fully self-consistent analysis is performed, then even if the MaP and maxL
parameters differ slightly, they  will
produce similar values of $\ln{\cal L}$, with differences far smaller than the differences between NR and model-based analysis.

%
For the reasons described in Section \ref{sec:Discsussion}, we consistently adopt SEOB-based models  to evaluate our
model-based $\ln {\cal L}$.   Because different phenomenological approximants do not agree, the posterior distributions
used to identify the parameters for  UID3 and 5 used an IMRD/P-based approximant produce different MaP and maxL
parameters.   Conversely, to the extent these models agree, they should estimate model parameters corresponding to the
same values of $\ln
{\cal L}$.   In fact, however, when we evaluate $\ln {\cal L}$ with SEOBNRv4 on the MaP and maxL parameters of the
posterior used to find UID5, we find both values disagree with the values seen for UID4, being lower (60.8) and higher
(62) respectively.   These differences in $\ln {\cal L}$ clearly indicate small differences between the two
model-based analyses, comparable to (but smaller than) the differences seen between model-based analyses and NR. 

\end{document}